\documentclass[aps,pre,amsmath,amsfonts,amssymb,9pt,nofootinbib,showpacs]{revtex4}

\usepackage{graphicx}
\usepackage{epsfig}

\newtheorem{propo}{Proposition}

\def\prooft{\hspace{0.5cm}{\bf Proof:}\hspace{0.1cm}}
\def\endproof{\hfill$\Box$\vspace{0.4cm}}

\def\g{\alpha}
\def\a{\alpha}
\def\b{\beta}

\def\ve{\varepsilon}
\def\cG{{\cal G}}
\def\cF{{\cal F}}

\def\<{\langle}
\def\>{\rangle}

\def\prob{{\mathbb P}}
\def\E{{\mathbb E}}
\def\entro{{\cal H}}
\def\R{{\mathbb R}}
\def\ve{\varepsilon}
\def\Qh{\widehat{Q}}
\def\Ph{\widehat{P}}
\def\Mh{\widehat{M}}
\def\Nh{\widehat{N}}
\def\Th{\widehat{T}}
\def\Sh{\widehat{S}}
\def\Rh{\widehat{R}}
\def\Tt{\widetilde{T}}
\def\Ft{\widetilde{F}}
\def\Gt{\widetilde{G}}
\def\gt{\tilde{g}}
\def\atanh{{\rm atanh}}
\def\qh{\hat{q}}
\def\sh{\hat{s}}
\def\fh{\hat{f}}
\def\phh{\hat{\phi}}
\def\eph{\hat{\epsilon}}

\def\rh{\hat{\rho}}
\def\Qc{{\cal Q}}
\def\Qch{\widehat{\cal Q}}

\def\te{\gamma}

\def\de{{\rm d}}
\def\ind{{\mathbb I}}

\def\R{{R}}
\def\Re{{\cal R}}

\def\cT{{\cal T}}

\def\o{{\vartheta}}
\def\bb{b_{\tiny{<}}}
\def\ba{b_{\tiny{>}}}

\def\Maj{{\sf M}}

\def\Lap{\, {\sf L}}
\def\gf{\, {\sf gf}}

\begin{document}

\date{\today}

\title{On the dynamics of the glass transition on Bethe lattices}
\author{ Andrea~Montanari$^{\,1}$ and  Guilhem~Semerjian$^{\,2}$}
\affiliation{$^{1\,}$ Laboratoire de Physique Th\'{e}orique de l'Ecole 
Normale Sup\'{e}rieure\footnote{UMR 8549, Unit{\'e}   Mixte de Recherche du 
Centre National de la Recherche Scientifique  et de 
l' Ecole Normale Sup{\'e}rieure.}}
\affiliation{$^{2\,}$
Dipartimento di Fisica, INFM (UdR and SMC centre),
Universit\`{a} di Roma ``La Sapienza''
}

\begin{abstract} 
The Glauber dynamics of  disordered spin models with multi-spin
interactions on sparse random graphs (Bethe lattices) is investigated. 
Such models undergo a dynamical glass transition upon decreasing the
temperature or increasing the degree of constrainedness.
Our analysis is based upon a detailed study of large scale rearrangements 
which control the slow dynamics of the system close to
the dynamical transition. Particular attention is devoted to the neighborhood
of a zero temperature tricritical point.

Both the approach and several key results are conjectured
to be valid in a considerably more general context.
\end{abstract}

\maketitle

\section{Introduction}
%
\label{IntroductionSection}
%
%
\subsection{Motivation}

Bethe lattices\footnote{We denote by this term random graphs with arbitrary
degree distribution (but bounded average degree). A more precise definition 
is given in Section \ref{DefinitionSection} for the particular case studied 
here.} can be used to define a large 
variety of analytically tractable statistical mechanics models.
A sophisticated (and still improving) theory has been developed for
computing the thermodynamic properties of such systems in great generality
(cf. for instance \cite{MezardParisiBethe} and references therein).
Among the recurring predictions of this theory is the occurrence 
(in models with frustration or self-induced frustration) of 
a {\em dynamical phase transition} (DPT) \cite{FranzFerro}. 
While the system free energy remains analytic at the DPT, this 
is revealed by more refined thermodynamic potentials \cite{FranzPotential}
(such as the configurational entropy) which indicate the appearence of a 
broken-ergodicity phase.

Despite its eminently dynamical character, the dynamics itself at 
the DPT is poorly understood.
For fully-connected (FC) mean field models, in which each degree
of freedom (spin, particle, etc) interacts with all the others, 
one can usually solve  the dynamics in terms of a closed set of equations
for the two points correlation and response functions. 
This is no longer true for models on  Bethe lattices, which have finite 
connectivity (each degree of freedom interacts with a finite number 
of neighbors). Correlation and response functions involving
an arbitrary number of times are relevant in this case~\cite{SeCuMo}.
This is somehow healthy: there is no hope to obtain a finite set of closed 
equations describing the dynamics of realistic (finite-dimensional)
systems. Alternative approaches developed for Bethe lattice models
may, on the other hand, prove more general.

Consider the relaxation of a particular degree of freedom, for instance 
a spin. As the DPT is approached,
dynamics slows down because, in order for the spin to decorrelate,
a larger and larger number of degrees of freedom 
have to be rearranged coherently.
Our approach consists in computing some detailed properties 
(e.g. their minimal size or the minimal energy barrier to be overcome
in order to realize them) of such rearrangements and then using these 
properties to estimate relaxation times.
While we shall develop these ideas on a specific Bethe-lattice model, 
they are susceptible of generalization to a variety of other examples.
The cases of $k$-core percolation and kinetically constrained models 
\cite{RitortReview} are, for instance, quite straightforward, and 
will be treated in  a related publication~\cite{NostroKCM} 
(the model treated in the present paper is in fact strictly related to 
$2$-core percolation on random hypergraphs).

Due to the lack of a simple order parameter and the
extensive configurational entropy, the broken ergodicity phase 
is often referred to as a ``glassy" phase. 
In the 80's, Kirkpatrick, Wolynes and Thirumalai~\cite{KiTh,KiWo} stressed
the analogy between the DPT (in the context of fully-connected models) and 
the mode coupling transition \cite{MCT,MCA,DynamicsReview}, 
thus motivating a considerable effort aimed at a theoretical 
understanding of structural glasses. One recent advance has been the 
definition a ``dynamical length scale" associated
to the dynamical slowing down~\cite{Donati,Bennemann}. 
Within mean-field models and mode coupling
theory (MCT) this length scale diverges at the 
DPT~\cite{FranzChi4A,FranzChi4B,BiroliBouchaud}.
However, the relation between this phenomenon and the diverging
relaxation time is poorly understood. Also, the basic physical mechanism
underlying this divergence in glasses is presumably different from 
the one occurring in simpler system, and is still 
misterious~\cite{JPB}.
Finally, it would be useful to establish a quantitative relation between
the dynamical slowing down and some purely static observable.

In order to clarify these points, let us contrast our understanding of the DPT
with the one we have of the paramagnetic--to--ferromagnetic transition 
in the Ising model. In this case a typical configuration close to the 
transition looks like a patchwork of positively and negatively magnetized 
regions. The typical linear size of each region is the correlation length 
$\xi$, and total magnetization is $\sqrt{\chi}$ 
($\chi$ being the susceptibility).
Assuming that the magnetization performs a random walk, we can estimate the
relaxation time as $\tau\gtrsim \chi$. Since standard scaling
theory predicts $\chi\sim \xi^{\gamma/\nu}$, we get 
$\tau\gtrsim \xi^{2\gamma/\nu}$. In fact
the dynamical scaling hypothesis states that $\tau\sim \xi^z$ with exponent
$z\ge 2\gamma/\nu$~\cite{Halperin}.

A further source of motivation is to understand the role of activation 
at the DPT. Much of our understanding of this phase transition comes from
the study of a  particular model: the $p$-spin spherical spin 
glass~\cite{CrisantiSommers_Statics,CrisantiSommers_Dynamics,Leticia_Houches}.
MCT has been shown to be exact for this model~\cite{MCA}.
Moreover, the energy landscape and its relation to dynamics
are relatively simple~\cite{KurchanTAP}. Above the transition temperature, the 
system never visits minima and moves among saddles. As temperature 
is lowered the number of energy-decreasing directions decreases, slowing down 
the dynamics~\cite{Cavagna}. Below the critical point, the system is 
typically found near a local minimum and is separated by diverging 
(in the thermodynamic limit) energy barriers from other minima.  
Activation is therefore irrelevant both above 
and below the DPT.  It is not necessary in the first case, and it cannot
take place within physical time-scales in the second. 
This has lead to question whether activated processes could spoil the 
predictions of MCT for more realistic (finite-dimensional) models. 
In this case a standard nucleation argument suggests that ergodicity is
restored within a finite (in the thermodynamic limit) time~\cite{Kob}.

Like other mean-field systems, models on Bethe lattices are characterized by
diverging energy barriers within the broken ergodicity phase.
Therefore, they cannot help understanding the role of activation in this regime.
On the other hand, finite energy barriers are present in the ergodic phase 
and one may ask whether they modify the MCT critical behavior upon 
approaching the DPT. Notice that such a question is in general not
well-defined because the distinction between activated and non-activated
processes becomes sharp only in the zero-temperature limit. 
In the following we shall study in detail a zero-temperature
multi-critical point where the distinction makes sense.
%
%
\subsection{Overview of the paper}

The model and some basic notations are introduced in Section 
\ref{DefinitionSection}. The phase diagram is described  in terms of 
two basic parameters: the temperature $T$, and $\g$ which
quantify the degree of constrainedness of the system.
We then discuss the critical properties of the DPT in Sections 
\ref{OriginSection} and \ref{RearrSection}. Let $T_{\rm d}$ be the phase 
transition temperature. We will show that $n$-points susceptibilities
stay finite at $T_{\rm d}$ for any given $n$. On the other hand, we  
introduce a correlation length $\ell$ using point-to-set correlations
and show that $\ell\sim (T-T_{\rm d})^{-1/2}$. 
This correlation length is connected to relaxation times using a 
disagreement percolation argument which implies a bound of the 
form $\tau\gtrsim (T-T_{\rm d})^{-1/2}$.

We then consider the $T\to 0$, finite $\g$ limit (still within
the ergodic phase). In this regime, the dynamics is dominated
by activated processes and the relaxation time can be
derived from a standard Arrhenius argument. The relevant large scale
rearrangements minimize the energy barrier to be overcome
in order to realize them, cf.~Sec.~\ref{sec_minbarr}.
The barriers height diverges as $\g$ increases and approach 
the DPT point $\g_{\rm d}$ (at which $T_{\rm d}=0$).
In Sec.~\ref{GeometricalSection}, we compute several properties of 
these rearrangements, such as their size, depth and cooperativity. 
We also consider minimal size rearrangements (cf. Sec.~\ref{sec_minsize}), 
which share several properties with
minimal barrier ones, and are somewhat easier to analyze.
We compare the critical behavior as $\g\to\g_{\rm d}$
with standard MCT predictions. We find that several of these predictions 
are violated.
More surprisingly  some universal properties of MCT
(such as the divergence of the four point susceptibility) 
hold even in this extremely activated regime.

The point $(\g = \g_{\rm d},T=0)$ is a tricritical point,
analogous to the percolation point in diluted ferromagnets. 
In Sec.~\ref{ScalingSection}, we use standard scaling theory 
in order to connect the $T\to 0$, $\g<\g_{\rm d}$ 
and $T\to T_{\rm d}(\g)>0$, $\g>\g_{\rm d}$ regimes.
The scaling hypothesis is verified through numerical simulations.
Under this hypothesis, temperature is a relevant variable and 
a crossover from an activation-dominated to a thermal regime takes place.
We argue that the thermal regime is controlled by usual MCT critical 
behavior at any finite temperature.

Some conclusive remarks are put forward in Sec.~\ref{ConclusionSection},
while most of technical calculations are collected in the Appendices.

A synthetic account of our results was published in Ref.~\cite{NostroLettera}.
%
\section{Definitions and notations}
%
\label{DefinitionSection}

%
%
\subsection{Model}
\label{ModelSection}

In this paper we focus on Ising models with $p$-spin ($p\ge 3$) interactions:
\begin{equation}
H(\sigma)=
\sum_{(i_1\dots i_p)\in\cG} (1-J_{i_1 \dots i_p} \sigma_{i_1} 
\cdots \sigma_{i_p}) \, ,\label{Hamiltonian}
\end{equation}
Here $\sigma_i=\pm 1$ are $N$ Ising spins, $\cG$ is a set of
$M=N\g$ $p$-uples of indices, and $J_{i_1 \dots i_p}$ 
are quenched couplings taking values $\pm 1$ with equal probability.
The above Hamiltonian is twice the number of violated
constraints $ \sigma_{i_1} \cdots \sigma_{i_p} =J_{i_1\dots i_p}$. 
Since this condition can be written as the exclusive OR of
$p$ boolean variables, the problem is known in
computer science as XORSAT~\cite{XOR_CS}.
Any configuration satisfying all these constraints has zero energy (is 
unfrustrated).

In order to complete our definition, we need to 
specify the (hyper)graph of interactions $\cG$. In this paper we consider
$\cG$ to be a random graph constructed by taking the $M$
interacting $p$-uples of sites $\{(i_1\dots i_p)\}$ to be quenched 
random variables uniformly distributed in  the set of 
$N \choose p$ possible $p$-uples~\cite{XOR}.
In the thermodynamic limit $N\to\infty$, the number of interactions a given 
spin belongs to (its degree) is a Poisson random variable with parameter 
$p\g$. Moreover, the shortest loop through such a spin is (typically)
of order $\log_{c} N$ with $c=p(p-1)\g$ \cite{Bollobas}. 

\begin{figure}
\includegraphics[width=10.5cm]{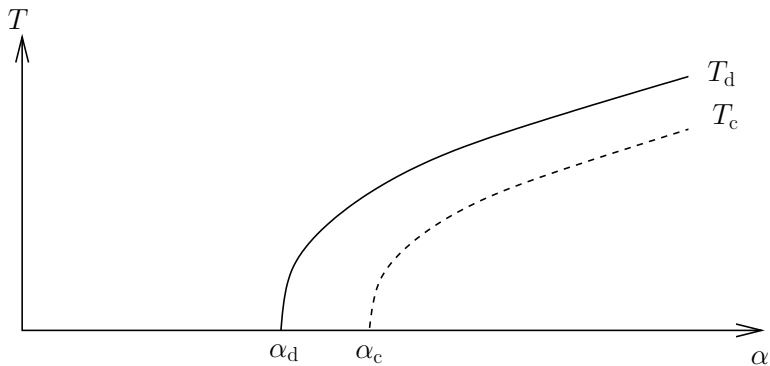}
\caption{Schematic view of the phase diagram of the diluted $p$-spin model.} 
\label{phasediag}
\end{figure}

This property allows (at least in principle) for an exact treatment
of the thermodynamics.
The phase diagram is sketched in Fig.~\ref{phasediag}.
For $\g < \g_{\rm d}$ and above the critical line $T_{\rm d}(\g)$ the system is ergodic 
and the free energy density is given by the usual paramagnetic
expression $-\beta f = \log 2+\g((\log\cosh\beta)-1)$. The free energy 
density remains analytic across the line $T_{\rm d}(\g)$, 
but ergodicity is broken in the whole region $T<T_{\rm d}(\g)$.
Finally, a true thermodynamic transition of the random-first
order type \cite{KiTh,KiWo} takes place at $T_{\rm c}(\g)$.

Two regimes have attracted particular attention. In the ``fully-connected"
limit $\g\to\infty$, $T\propto \sqrt{\g}$, both statics and
dynamics can be treated analytically, showing a typical MCT 
transition~\cite{Gardner,KiTh}. 
The relaxation time diverges as $|T-T_{\rm d}|^{-\te_{\rm mct}}$.

In the zero-temperature, finite $\g$ limit, probabilistic  
methods can be used to show that  zero-energy ground states 
with finite entropy density exist for $\g<\g_{\rm c}$. However
the set of ground states gets splitted in an exponential number of clusters
with extensive Hamming distance (number of spins with different value)
separating them for 
$\g_{\rm d}<\g<\g_{\rm c}$~\cite{XOR_1,XOR_2}. 
A finite fraction of the spins
does not vary among the ground states of a particular cluster, while
they change when passing from a cluster to the other. 
At $\g_{\rm d}$ the fraction of frozen spins jumps from $0$
to a finite value $\phi_{\rm d}$. 
For instance $\g_{\rm d}\approx 0.818469$,
$\g_{\rm c}\approx 0.917935$, and $\phi_{\rm d}\approx 0.715332$ 
when $p=3$. 
This phase transition can be characterized in a purely geometrical
way. It corresponds to the appearence of a $2$-core,
i.e. a subgraph of $\cG$ such that any
of its vertices has degree at least $2$ in the subgraph.
The largest $2$-core in $\cG$, denoted as $\cG_{\rm core}$, includes 
$|\cG_{\rm core}| = O(N)$ vertices above $\g_{\rm d}$.
The subgraph of frozen spins will be referred to as ``backbone'',
$\cG_{\rm bone}$ and can be constructed recursively as follows.
Start from $\cG_{\rm bone}=\cG_{\rm core}$, add to it 
any hyperedge of $\cG$ having at most one vertex not yet in $\cG_{\rm bone}$,
and repeat until such an hyperedge exist.
In view of these remarks, it is not
surprising that several results of this paper can be generalized 
to kinetically constrained models (in this case the relevant geometrical
transition is the $k$-core percolation \cite{Pittel,RitortReview}). 

No exact result exists for the dynamics at any finite
value of $\g$. Standard generating function methods allow to 
formally derive a hierarchy of dynamical equations for multi-time
correlation and response functions~\cite{SeCuMo}.
Several approximation schemes have been 
proposed~\cite{SemerjianWeigt,CoolenEtAl} for this hierarchy. 
Such approximations have been developed mostly
within models with a continuous transition (typically, the model
(\ref{Hamiltonian}) with $p=2$). However, their generalization to
models undergoing a discontinuous transition is straightforward.
Unfortunately such approximations 
break down in the critical regime $T\downarrow T_{\rm d}(\g)$.

In the following we shall use repeatedly two useful properties 
of this model. First: as long as there exist at least one 
zero-energy ground-state, all such ground-states are equivalent. 
Suppose that $\sigma^{(*)}$ is such a ground state, and that
$\sigma$ is an equilibrium configuration for the Hamiltonian
(\ref{Hamiltonian}) at temperature $T$.
Define $\sigma_{i}'=\sigma^{(*)}_i\sigma_i$ (this is often called a 
``gauge" transformation). Then $\sigma'$ is distributed
according to the Boltzmann measure with an Hamiltonian of the form 
(\ref{Hamiltonian}), with the same graph $\cG$ and temperature
$T$ as the original model, but with $J_{i_1\dots i_p}=+1$ for all 
the interactions (ferromagnetic model).
Therefore any property of $\sigma^{(*)}$ 
which is invariant under this change of variable can be computed for the 
$\sigma^{(+)} = (+1,\dots,+1)$ ground state of the ferromagnetic model.
In particular, for $T=0$ this implies that the ground states of the 
$J=\pm 1$ model as observed from $\sigma^{(*)}$, ``look like"
the ground states of the ferromagnetic model as observed from 
$\sigma^{(+)}$.

The second property will be used uniquely in numerical simulations
and can be stated in two parts: $(i)$ As long as $\cG$ does not contain 
hyperloops (i.e. subgraphs such that each site has even degree)
the partition function of the model is {\em exactly} given
by the annealed computation, i.e. $Z_{N,\beta}(J) = 2^N(e^{-\beta}\cosh\beta)^M$.
$(ii)$ Under the same hypothesis, sampling the sign of the couplings 
$J \equiv\{J_{i_1\dots i_p}\}$ (for a given graph $\cG$) 
and then the configuration $\sigma$ 
according to the Boltzmann distribution is equivalent to sampling 
$\sigma$ from the uniform distribution and then the couplings 
as independent random variables with distribution
\begin{eqnarray}
\prob(J_{i_1\dots i_p}|\sigma) = \frac{1}{2\cosh \beta}\;
\, \exp\left\{\beta J_{i_1\dots i_p}\sigma_{i_1}\cdots \sigma_{i_p}\right\}\, .
\end{eqnarray}
The proof of this statement is deferred to Appendix~\ref{app:ProofProp}.
It is a consequence of the results of Refs.~\cite{XOR_1,XOR_2} that,
for $\g<\g_{\rm c}$ a random graph $\cG$ constructed as 
above does not contain any hyperloop with high probability.

Sampling first $\sigma$ and then $J$ is much simpler than 
the opposite. We shall adopt this procedure in order to generate 
thermalized configurations for simulating the equilibrium dynamics.
It is an open (and interesting) problem whether the same procedure 
can be useful in the region $\g>\g_{\rm c}$, $T>T_{\rm c}(\g)$
(at least for large enough systems).
%
%
\subsection{Averages, factor graphs and messages}

In this paper we shall denote by $\<\,\cdot\,\>$ expectation with 
respect to the Boltzmann measure at inverse temperature $\beta$,
and by $\E(\,\cdot\,)$ expectation with respect to the quenched disorder
(i.e. the graph $\cG$ and the couplings $J_{i_{1}\dots i_p}\in\{+1,-1\}$).
Connected correlation functions (see, for instance, \cite{ZinnJustin}
for a definition) are denoted 
by $\<\,\cdot\,\>_{\rm c}$.

\begin{figure}
\begin{center}
\includegraphics[width=9cm]{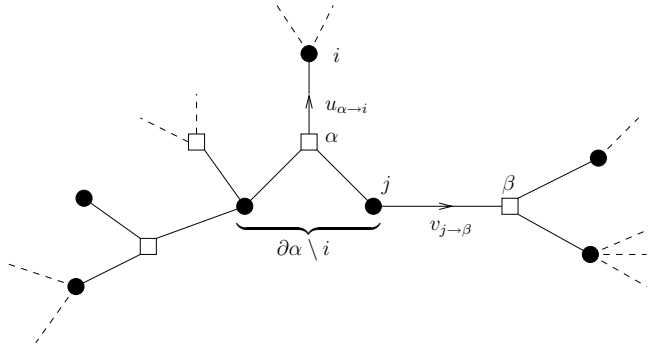}
\end{center}
\caption{Example of factor graphs.}
\label{fig_factorgraph}
\end{figure}

Factor graphs~\cite{Factor}
provide an useful representation of  models like the one 
studied in this paper. These graphs have two types of nodes, 
one for interactions
(``function nodes", squares of Fig.~\ref{fig_factorgraph}) and one 
for spins (``variable nodes", circles).
Edges connect each interaction with the spins involved in it. The 
graph is obviously bipartite.

We shall use indices $\a,\b,\dots$ to denote function nodes, and 
$i,j,k,\dots$ for variables.  Directed edges in the graph
will support~\emph{messages} of two types, $u_{\a \to i}$, 
and $v_{i \to \a}$, with various meanings depending on the context.
We denote by $\partial \a$  the set of variables in the interaction $\a$,
and by $\partial i$ the set of interactions the variable $i$ belongs to. In
this context $\setminus$ will denote the substraction of an element from a set.

Let $\cF$ be the factor graph associated to the model (\ref{Hamiltonian}).
The degree of function nodes in $\cF$ is fixed to $p$, whereas
the one of a generic  variable node converges to Poisson random variable
of parameter $p \g$.
A path in $\cF$ is a sequence of nodes, such that any two successive 
nodes are joined by an edge.
The distance between two variable nodes $i$ and $j$ is the
length (number of function nodes in it) of the shortest path graph 
joining them.  
Two variables are said to be ``adjacent" if their distance is equal to one,
and connected if there is a path joining them. A set of variables is connected
if its elements are pairwise connected.
An important property of  random graphs, on which we shall come back later,
is that, in the $N\to\infty$ limit, any finite neighborhood of a generic site
is, with high probability, a tree.

Finally, for asymptotic behaviors (here as $n\to\infty$) we write: 
$a_n\sim b_n$ 
if $(a_n/b_n)$ is bounded away from $0$ and $\infty$ for $n$ large enough; 
$a_n\simeq b_n$ if $(a_n/b_n)\to 1$. Further  $a\approx b$ if 
$a$ is numerically close to $b$.
%
%
\subsection{Glauber dynamics}

Most of our results can be applied to general 
single-spin flip Markov dynamics satisfying the detailed balance condition. 
Furthermore, the generalization to block (multi-spin flip) dynamics 
is in several cases straightforward as far as the block size is kept
finite as $N\to\infty$.

For the sake of definiteness, the reader may refer to Glauber dynamics
in continuous time. At each spin $\sigma_i$, a flip is proposed in a time 
interval $\de t$
with probability $\de t$ (i.e. according to a Poisson process of rate $1$).
Whenever a spin flip is proposed at site $i$, the new value of $\sigma_i$
is drawn from the conditional distribution, given the neighboring spins
\begin{eqnarray}
P_{\beta}(\sigma_i|\sigma_{[N]\backslash i}) \propto
\, \exp\Big\{\beta 
\sigma_i\sum_{\a\in\partial i}J_a\sum_{j\in\partial \a\backslash i}
\sigma_j\Big\}
\, ,\label{eq:GlauberRates}
\end{eqnarray}
(the constant hidden in the $\propto$ symbol is here independent of 
$\sigma_i$). Whenever precise non-asymptotic estimates on correlation
times will be given, they will refer to these transition rates.

The basic observables we shall use in order to probe this
stochastic dynamics are single spin correlation functions
\begin{eqnarray}
C_i(t) = \<\sigma_i(0)\sigma_i(t)\>\, .
\end{eqnarray}
Here expectation is taken with respect to the equilibrium dynamics, i.e.
$\sigma(0)$ is an equilibrated configuration, and $\sigma(t)$ is 
the configuration obtained by applying the above dynamics for a time 
interval $t$. One can also consider the global correlation 
function $C(t)$, defined by averaging $C_i(t)$ uniformly over the 
spin position.
%
\section{Origin of the dynamical transition}
\label{OriginSection}

Since purportedly,  the dynamics gets slower and slower upon approaching 
$T_{\rm d}(\g)$, it is quite natural to think that
spatial correlations in the equilibrium measure must become stronger in 
the same regime.
In this Section we first show that $n$-point correlations remain short
ranged at the dynamical transition. On the contrary, point-to-set correlations
diverge at $T_{\rm d}(\g)$. We show that this indeed implies 
a divergence in properly defined relaxation times.
%
%
\subsection{$n$-points correlation functions}

A simple way for characterizing correlations in the system
consists in considering $n$-points spin-glass susceptibilities
\begin{eqnarray}
\chi^{\rm SG}_{n,N} \equiv \frac{1}{N} \sum_{i_1\dots i_n}\<\sigma_{i_1}\cdots
\sigma_{i_n}\>_{\rm c}^2\, .\label{eq:SuscDefinition}
\end{eqnarray}
We are eventually interested in the thermodynamic limit.
The rationale for considering the spin-glass susceptibility is that 
$\E\<\sigma_{i_1}\cdots\sigma_{i_n}\>_{\rm c}=0$ unless 
$i_1=\cdots =i_n$ and $n$ is even~\footnote{Whenever 
$\{i_1,\dots,i_n\}$ contain an index $i$ repeated an odd number of times,
a simple gauge transformation implies that $\E\<\sigma_{i_1}\cdots
\sigma_{i_n}\>_{\rm c} = -\E\<\sigma_{i_1}\cdots\sigma_{i_n}\>_{\rm c}$.}. 
We claim that for any fixed $n$, the susceptibilities 
(\ref{eq:SuscDefinition}) remain finite  across the dynamic transition 
$T_{\rm d}(\g)$. More precisely, for all $T>T_{\rm c}(\g)$,
with high probability
\begin{eqnarray}
\chi^{\rm SG}_{n,N}\le C_n(\beta,\g)\, ,
\end{eqnarray}
for some bounded functions $C_n(\beta,\g)$. In other words,
$n$-points correlation functions give no hint\footnote{One may wonder
whether derivatives of $\chi^{\rm SG}_{n,N}$ with respect to $T$ 
(or $\g$) may give some useful information. It is not hard to 
generalize our arguments to show that this is not the case.} 
about the dynamic transition {\em as far as $n$ is finite}.
Here we will sketch a proof of this claim in the zero-temperature limit
(the approach is essentially rigorous in this case) for $n\le 2$. Technical
calculations, together with a cavity argument for $n>2$ or $T>0$ are 
presented in Appendix \ref{app:Correlations}.

Consider $\g<\g_{\rm c}$: the system will typically contain an 
unfrustrated ground state $\sigma^{(*)}$, so we apply a gauge transformation 
and reduce ourselves to the ferromagnetic model $J_{i_1\cdots i_p}=+1$.
Since $(\sigma_i^{(*)})^2=1$, the definition 
(\ref{eq:SuscDefinition}) remains unchanged in the new variables
(which we still denote as $\sigma$).
In the ferromagnetic system $\<\sigma_{i_1}\cdots\sigma_{i_n}\>_{\rm c}= 0$
or $+1$ for any $i_1,\dots,i_n$ and $n\le 3$. 
In fact $\<\sigma_{i_1}\cdots\sigma_{i_n}\> = 0$ or $1$
(for the not-connected correlation function). If
$\sigma_{i_1}\cdots\sigma_{i_n} = +1$ in all the ground states
then $\<\sigma_{i_1}\cdots\sigma_{i_n}\> = +1$ trivially.
If on the other hand there exist a ground state $\sigma^{(-)}$ such that
$\sigma_{i_1}^{(-)}\cdots\sigma^{(-)}_{i_n} = -1$, this is the case for 
exactly  one half of the ground states (for each ground state with
$\sigma_{i_1}\cdots\sigma_{i_n} = +1$ we can construct one of the other type
multiplying it by $\sigma_i^{(-)}$) and therefore
$\<\sigma_{i_1}\cdots\sigma_{i_n}\> = 0$. As for the connected correlation 
$\<\sigma_{i_1}\cdots\sigma_{i_n}\>_{\rm c}$, being sum of products
of ordinary correlations, it is an integer. Furthermore, it is non-negative 
because of Griffiths inequality for $n=1,2$, and by simple
enumeration of the possible values of 
$\<\sigma_{i_1}\sigma_{i_2}\sigma_{i_3}\>$, $\<\sigma_{i_1}\sigma_{i_2}\>$,
and so on, for $n=3$. Using the definition of $\<\cdots\>_{\rm c}$
it is finally easy to show that it cannot be larger than one.
We therefore proved the simplified expression
\begin{eqnarray}
\chi^{\rm SG}_{n,N} \equiv \frac{1}{N} \sum_{i_1\dots i_n}\<\sigma_{i_1}\cdots
\sigma_{i_n}\>_{\rm c}\, , \;\;\;\;\;
\mbox{for $T=0$, $n\le 3$, and $\g<\g_{\rm c}$}.\label{eq:SuscDefinitionZeroT}
\end{eqnarray}

Let $L = \{j_1,\dots,j_{l}\}$ with $j_k\in\{1,\dots,N\}$ a list of  
(not necessarily distinct) spin positions 
denote by $Z_{\cG}(L)$ the number of unfrustrated ground states 
such that $\sigma_i=+1$ for any  $i\in L$. We take $l$ to have Poisson
distribution with mean $N\lambda$ (with $\lambda\ge 0$) 
and $\{j_1,\dots,j_{l}\}$ i.i.d. uniformly distributed in 
$\{1,\dots,N\}$. Let 
\begin{eqnarray}
\psi_{\cG}(\lambda) \equiv \frac{1}{N}\, \E_{L}\log_2 Z_{\cG}(L)\, .
\end{eqnarray}
denote the ground state entropy of the system in which we forced 
$\sigma_i=+1$ for any $i\in L$. Simple calculus reveals 
that
\begin{eqnarray}
\left.\frac{\de \psi_{\cG}}{\de \lambda}\right|_0 = -\frac{1}{N}\sum_{i=1}^N
(1-\<\sigma_i\>)\, ,\;\;\;\;\;\;\;\;
\left.\frac{\de^2 \psi_{\cG}}{\de \lambda^2}\right|_0 = 
\frac{1}{N}\sum_{i,j}\<\sigma_i\sigma_j\>_{\rm c}\, .\label{eq:PotDerivatives}
\end{eqnarray}
Therefore, proving the thesis for $n\le 2$ is equivalent 
to proving that the first two derivatives of $\psi_{\cG}(\lambda)$ 
are bounded. 

By computing derivatives at $\lambda>0$, one can show that
$\psi_{\cG}(\lambda)$ is non-increasing and convex in 
$\lambda$.
Furthermore, since each constraint (either $\sigma_{i_1}\cdots\sigma_{i_p}=+1$ 
for $(i_1\cdots i_p)$ edge of $\cG$ or $\sigma_j=+1$ for $j\in L$)
at most halves the number of zero-energy ground states,
we have the bound $\psi_{\cG}(\lambda)\ge (1-\g-\lambda)$. 
Moreover, it is simple 
to derive an upper bound through an annealed computation
(we refer to Appendix \ref{app:Correlations} for the details).
Define
\begin{eqnarray}
\psi_{\rm ann}(\lambda) \equiv \sup_{\omega\in[0,1]}
\left\{ \entro(\omega)+\g\log_2\left[\frac{1+(1-2\omega)^p}{2}\right]
+\lambda\log_2(1-\omega)\right\}\, ,\label{eq:Annealed}
\end{eqnarray}
where $\entro(\omega) = -\omega\log_2 \omega-(1-\omega)\log_2(1-\omega)$ 
is the binary entropy function.
Then one can prove that,  given $\ve>0$, $\psi_{\cG}(\lambda)\le 
\psi_{\rm ann}(\lambda)+\ve\lambda^2$ for any $\lambda\ge 0$ 
with high probability.
It turns out that there exist $\g_{\rm ann}$, with $\g_{\rm d}\le\g_{\rm ann}
\le \g_{\rm c}$, such that, if $\g<\g_{\rm ann}$,
\begin{eqnarray}
\psi_{\rm ann}(\lambda) = 1-\g-\lambda+\frac{1}{2\log 2}\, \lambda^2+
O(\lambda^3)\, .\label{eq:AnnExp}
\end{eqnarray}
For instance we obtain $\g_{\rm ann} \approx 0.889493$, $0.967147$,
$0.989163$ for, respectively,
$p=3,4,5$.
These bounds imply in turn
\begin{eqnarray}
\left.\frac{\de \psi_{\cG}}{\de \lambda}\right|_0 = -1\, ,\;\;\;\;\;\;\;\;
\;\;\;
0\le \left.\frac{\de^2 \psi_{\cG}}{\de \lambda^2}\right|_0
\le \frac{1}{\log 2}\, .\label{eq:BoundsPot}
\end{eqnarray}
Thus proving that $\chi_{n,N}^{\rm SG}$, $n=1,2$, remains bounded for $T=0$
and $\g<\g_{\rm ann}$. In Appendix \ref{app:Correlations},
we argue that in fact the result remains true up to $\g_{\rm c}$.
%
%
\subsection{Point-to-set correlations}
\label{sec:PointToSet}

A much stronger criterion for correlations decay is obtained by 
considering observables depending on arbitrary number of spins
(i.e. not necessarily bounded by a constant independent
of $N$ and $\g$ as in the previous Section). 
Here we describe a concrete way for verifying such a criterion and
show that it allows to define a correlation length diverging at 
$T_{\rm d}(\g)$.

Fix a configuration $\sigma^{(0)}$ drawn from the equilibrium 
Boltzmann distribution at temperature $T$. Let $i\in \{1,\dots,N\}$ be 
a site in the system and $\ell$ a positive integer. Consider a configuration 
$\sigma$ distributed according to the Boltzmann measure, conditioned
to $\sigma_j=\sigma^{(0)}_j$ for all sites $j$ whose distance from $i$
is at least $\ell$. In other words, if $\cG_{\ell}(i)$ denotes
the subgraph including all the sites at distance smaller than $\ell$ from 
$i$ the distribution of $\sigma$ is
\begin{eqnarray}
P_{\beta,\ell}(\sigma) =\left\{\begin{array}{ll}
\exp\left\{-\beta H(\sigma)\right\}/Z_{\beta,\ell} &
\mbox{ if $\sigma_j = \sigma^{(0)}_j$ $\forall j\not\in\cG_{\ell}(i)$}\\
0 & \mbox{ otherwise}\, .
\end{array}\right.
\end{eqnarray}
If $\ell$ is small, $\sigma_i$ will be highly correlated with 
$\sigma^{(0)}_i$. However, we expect this correlation to decay as 
$\ell\to\infty$ if the temperature is high enough. 
We are therefore led to define a correlation length as follows
(a similar length scale has been discussed in
\cite{BiroliBouchaudMosaic})
\begin{eqnarray}
\ell_{i,*}(\ve) \equiv \min\left\{\,\ell :\, \E_{\sigma^{(0)}}
[\sigma^{(0)}_i\<\sigma_i\>_{\ell}]\le\ve \, , \right\}\, .
\label{eq:DefinitionPointToSet}
\end{eqnarray}
Here $\E_{\sigma^{(0)}}[\, \cdot\,]$ denotes expectation with respect
to the reference configuration $\sigma^{(0)}$ and $\<\,\cdot\,\>$
is the expectation with respect to the measure $P_{\beta,\ell}(\sigma)$.
Our definition depends upon the parameter $\ve\in (0,1)$.
However this dependency is irrelevant~\footnote{The single-site 
Edwards Anderson order parameter 
\cite{Ricci,Ricci2,ParisiLocal} must however be
larger than $\ve$ to get a non-trivial result.} and we can think of it as a 
fixed small number e.g. $\ve = 0.1$.

The main result of this Section is that $\ell_{i,*}(\ve)$ diverges as 
the dynamical transition line $T_{\rm d}(\g)$ is approached.
More precisely, with non-vanishing probability 
--with respect to the choice of the site $i$--,
we have (the thermodynamic limit being taken at the outset)
\begin{eqnarray}
\ell_{i,*}(\ve) \sim (T-T_{\rm d}(\g))^{-1/2}\, .
\end{eqnarray}
Equivalently, if the dynamical transition is approached by increasing
the number of constraints in the system:
 $\ell_{i,*}(\ve) \sim (\g_{\rm d}(T)-\g)^{-1/2}$.

Let us consider, for the sake of simplicity the case $T=0$ and 
$\g\uparrow\g_{\rm d}$. The finite temperature case will be 
discussed in App.~\ref{app:Correlations} 
(we also refer to \cite{Reconstr} for a more detailed discussion of this 
issue). The reference configuration $\sigma^{(0)}$ is, for $T=0$,
chosen uniformly at random among the exponentially numerous unfrustrated 
ground states. Without loss of generality we can assume that
the model is ferromagnetic and $\sigma^{(0)} =\sigma^{(+)}$ is the 
`all-plus' cofiguration. Repeating the type of arguments already
used in the previous Section, one can show that, at zero temperature, 
$\<\sigma_i\>_{\ell}$ is either $+1$ (if $\sigma_i$ is completely
determined by the boundary condition together with the requirement
of satisfying all the interactions) or $0$ (in the opposite case).
As a consequence $\ell_{i,*}(\ve)$ does not depend upon $\ve$: we have
$\<\sigma_i\>_{\ell} =1$ for $\ell<\ell_{i,*}$ and  
$\<\sigma_i\>_{\ell} =0$ for $\ell\ge \ell_{i,*}$. 

Let $\phi_{\ell}$ be the probability that $\<\sigma_i\>_{\ell} = 1$.
Obviously $\phi_0 = 1$. Furthermore, for any $\ell\ge  0$, the following
recursion relation holds
\begin{eqnarray}
\phi_{\ell+1} = \sum_{l=0}^{\infty} e^{-p\g}
\frac{(p\g)^l}{l!}[1-(1-\phi_{\ell}^{p-1})^l]
=1-\exp\big\{-p\g\phi_{\ell}^{p-1}\big\}\, .\label{eq:SphereRec}
\end{eqnarray}
This formula is proved by summing over the degree of site $i$ 
the probability that the degree is $l$ (factor $e^{-p\g}(p\g)^l/l!$)
times the probability that at least one of the interactions
has all the remaining spins determined by their $\ell$-th neighbors.

The dynamical transition point is given by
\begin{eqnarray}
\g_{\rm d} = \sup\Big\{\g\, :\, \phi>1-\exp\big\{-p\g\phi^{p-1}
\big\}
\mbox{ for } \phi\in(0,1]\Big\}\, .\label{eq:CriticalPoint}
\end{eqnarray}
Numerical values of $\g_{\rm d}$ are given in Table~\ref{TableSize}.
A simple asymptotic study of the recursion (\ref{eq:SphereRec})  shows that 
$\phi_{\ell}\to 0$ for $\g<\g_{\rm d}$, while 
$\phi_{\ell}\to \phi_{*}(\g)$ for $\g\ge\g_{\rm d}$,
$\phi_{*}(\g)$ being the largest solution of 
$\phi=1-\exp\big\{-p\g\phi^{p-1}\big\}$. Hereafter we shall use the
notation $\phi_{\rm d} = \phi_*(\g_{\rm d})$.
For any $p\ge 3$, $\phi_{\rm d}>0$ and 
$\phi_*(\g) = \phi_{\rm d} + {\rm cst}\sqrt{\g-\g_{\rm d}}+
O(\g-\g_{\rm d})$ for $\g\gtrsim \g_{\rm d}$.
As already mentioned, $\g_{\rm d}$ is the critical value for the 
appearence of a backbone (and a $2$-core) in the graph $\cG$ and 
is easy to realize that
its typical size is about $N\phi_*(\g)$ for $\g\ge\g_{\rm d}$.
In the present picture, the spins in the backbone are frozen because 
they are determined by spins arbitrarily far away. 

\begin{figure}
\begin{tabular}{cc}
\includegraphics[width=7.5cm]{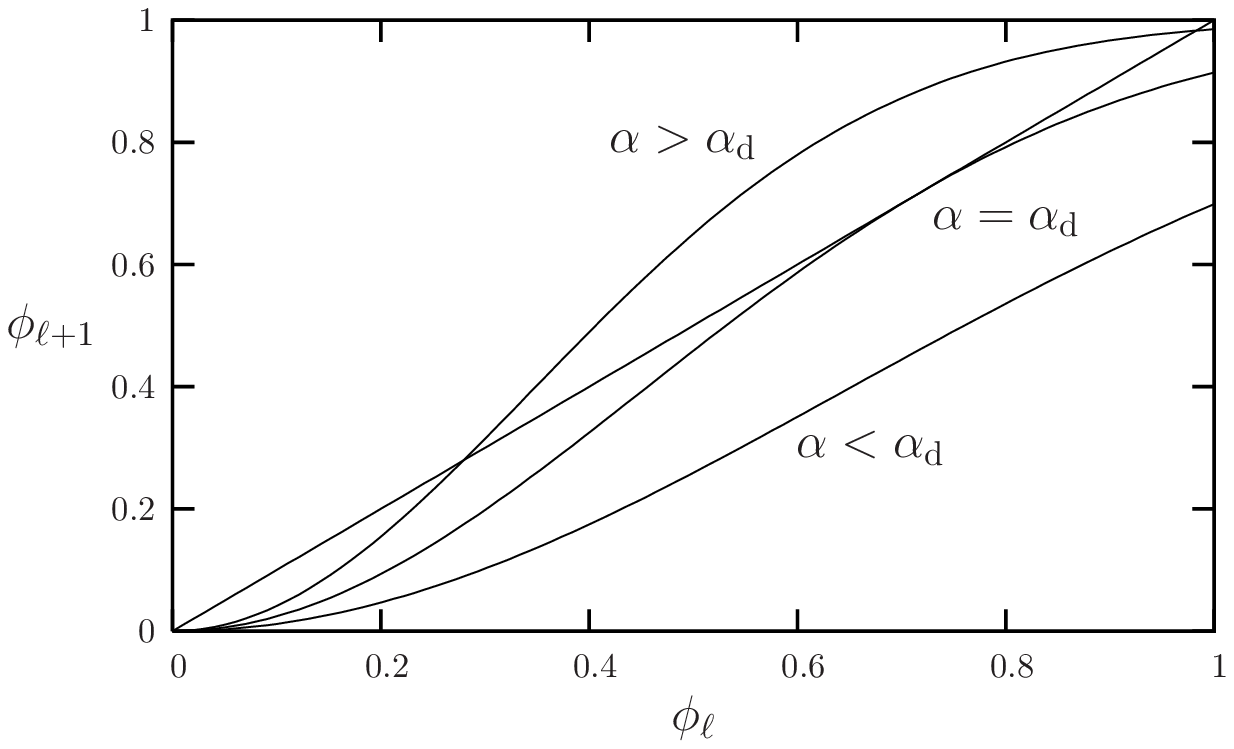}&
\includegraphics[width=7.5cm]{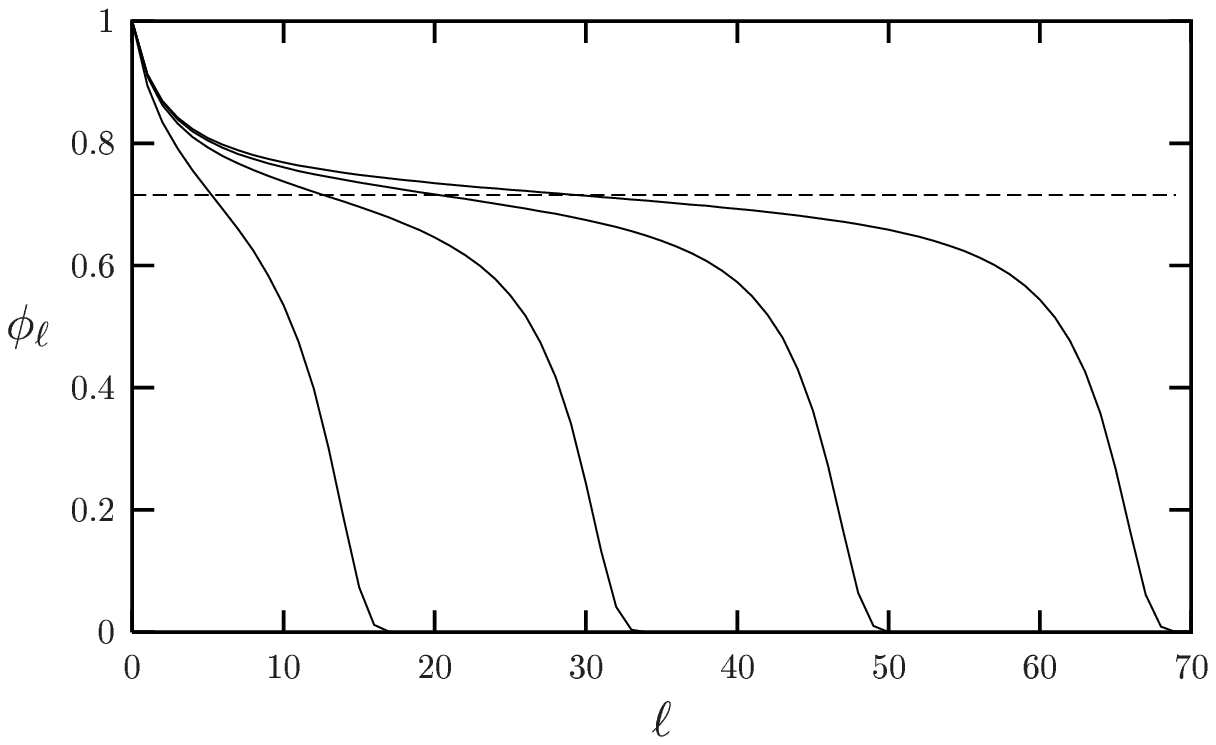}
\end{tabular}
\caption{Point-to-set correlation function for 
$p=3$. Left: graphical representation of the recursion
$\phi_{\ell+1} = 1-\exp\big\{-p\g\phi^{p-1}_{\ell}\big\}$.
Right: function $\phi_{\ell}$ for $\g = 0.75$, $0.8$, $0.81$ $0.814$.} 
\label{fig:Ball}
\end{figure}
As $\g_{\rm d}$ is approached from below, the convergence of
$\phi_{\ell}$ to $0$ becomes slower and slower. As can be seen in 
Fig.~\ref{fig:Ball}, $\phi_{\ell}$ first develops a plateau near $\phi_{\rm d}$,
whose length scales as $(\g_{\rm d}-\g)^{-1/2}$ and then 
decreases to $0$ right after this plateau (i.e. it decreases below any given
constant in a fixed number of iterations).
The behavior around the plateau is described by the scaling form
\begin{eqnarray}
\phi_{\ell} \simeq \phi_{\rm d} +  (\g_{\rm d}-\g)^{1/2}
{\cal K}\big[(\g_{\rm d}-\g)^{1/2}(\ell-\ell_0(\g))\big]\, ,
\end{eqnarray}
where
\begin{eqnarray}
\ell_0(\g) =  \frac{\pi}{2\sqrt{A_0 A_2}}\, (\g_{\rm d}-\g)^{-1/2}
\, ,\;\;\;\;\;\;\;
{\cal K}(x)  =  -\sqrt{\frac{A_0}{A_2}}\,\tan\big\{
\sqrt{A_0A_2}\, x\big\}\, ,
\end{eqnarray}
and
\begin{eqnarray}
A_0 = p \phi_{\rm d}^{p-1}\, e^{-p\g_{\rm d}\phi_{\rm d}^{p-1}}
\, ,\;\;\;\;\;A_2 = \frac{1}{2}p(p-1)\g_{\rm d}\phi_{\rm d}^{p-3}
e^{-p\g\phi^{p-1}}\big\{p(p-1)\g_{\rm d}\phi_{\rm d}^{p-1}-(p-2)\big\}\, .
\end{eqnarray}
Therefore, for any given sample, a fraction $\phi_{\rm d}$ of the sites $i$
is such that 
\begin{eqnarray}
\ell_{i,*}(\ve) =  \frac{\pi}{\sqrt{A_0A_2}}\, (\g_{\rm d}-\g)^{-1/2}
+O(1)\, ,
\end{eqnarray}
independently of $\ve\in (0,1)$.

\begin{figure}
\begin{tabular}{cc}
\includegraphics[width=8.cm]{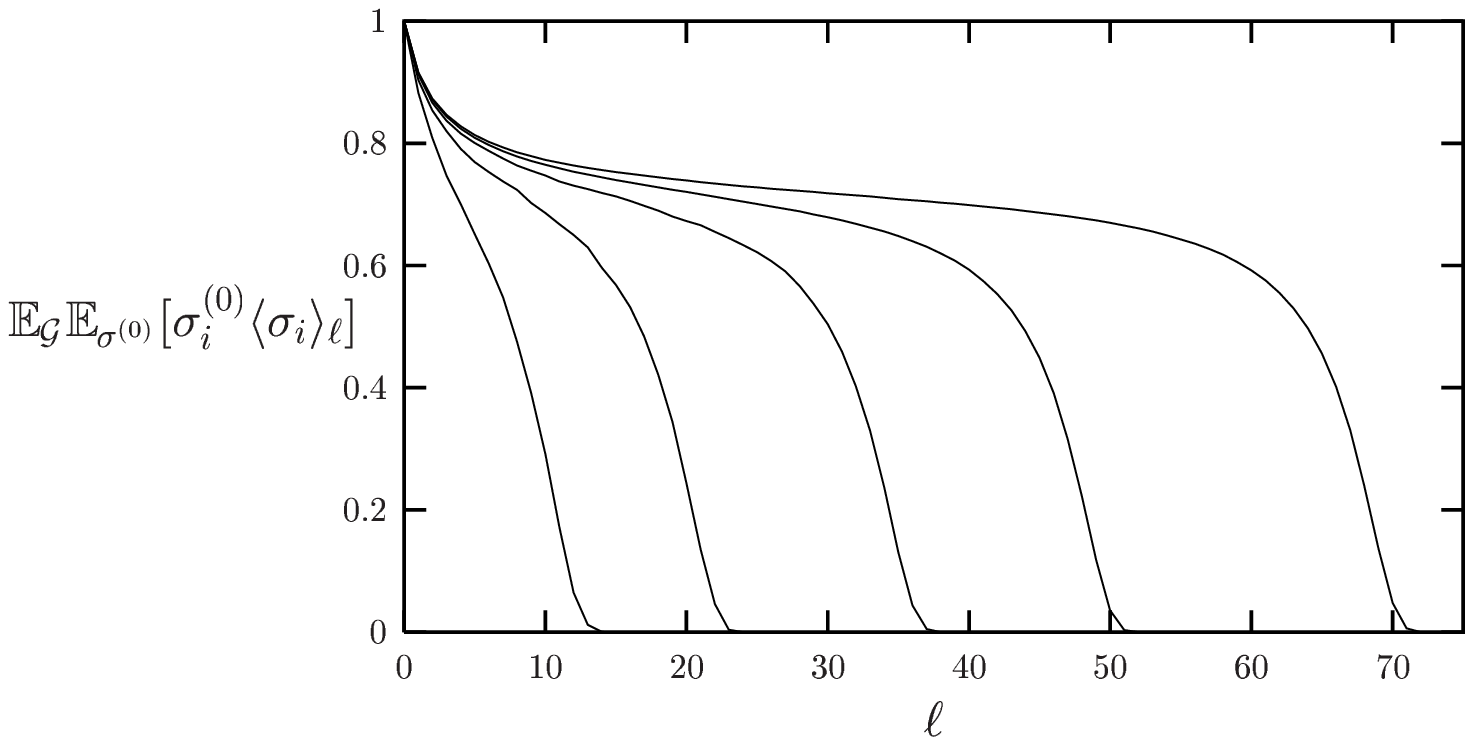}&\hspace{0.5cm}
\includegraphics[width=7.cm]{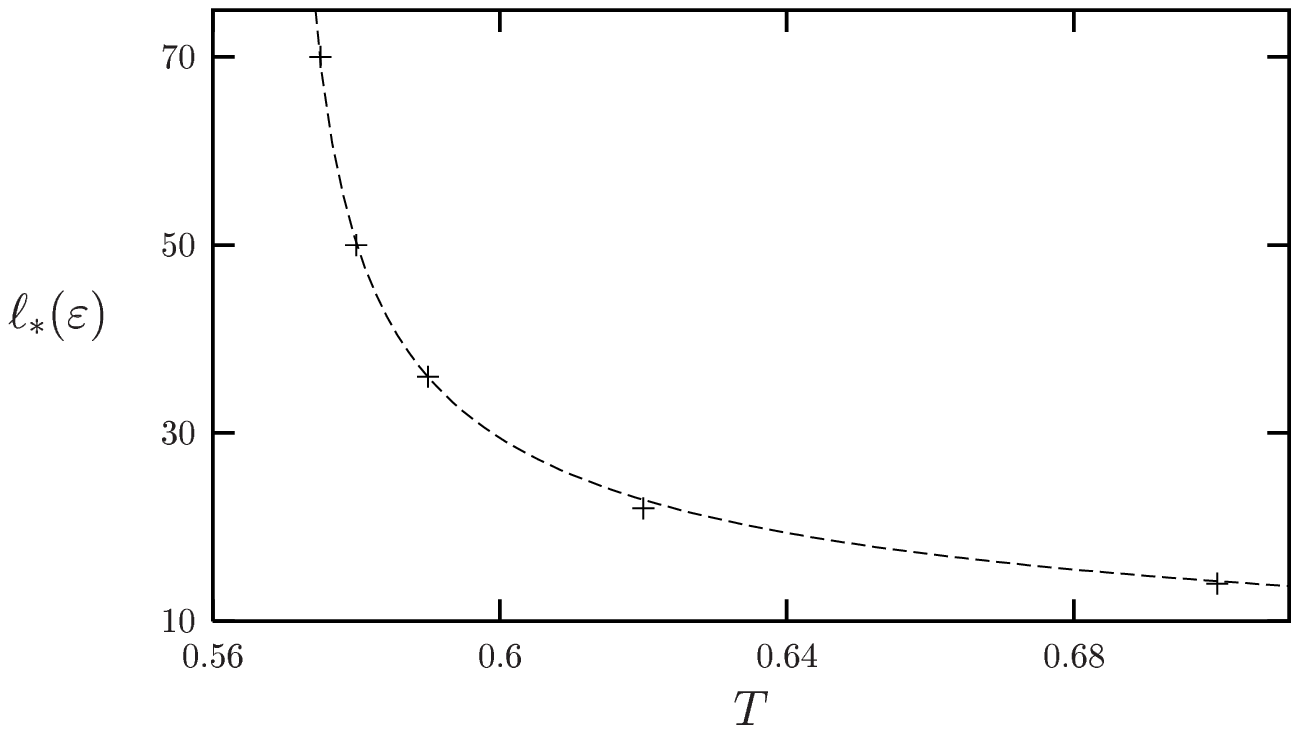}
\end{tabular}
\caption{Left: expected point to set correlation at $\g=1$ and,
from left to right, $T=0.7$, $0.62$, $0.59$, $0.58$ and $0.575$, as
computed using the recursive approach of Appendix~\ref{app:PointToSet}. Right:
point-to-set correlation length $\ell_*(\ve)$, as extracted from this 
data with $\ve=0.1$. The dashed line is a fit to this data according 
to the asymptotic expression $\ell_*(\ve) = C\, (T-T_{\rm d})^{-1/2}$,
with $T_{\rm d}=0.51695$.} 
\label{fig:BallFiniteT}
\end{figure}
To conclude this Section, in Fig.~\ref{fig:BallFiniteT} we
show the results of a recursive finite-$T$ calculation of  $\E_{\sigma^{(0)}}
[\sigma^{(0)}_i\<\sigma_i\>_{\ell}]$ as a function of $\ell$,
averaged over the graph realization. 
This allows to extract the typical point-to-set correlation length defined 
as (here $\E_{{\cal G}}$ denotes expectation over the graph realization)
\begin{eqnarray}
\ell_{*}(\ve) \equiv \min\left\{\,\ell :\, \E_{{\cal G}}\E_{\sigma^{(0)}}
[\sigma^{(0)}_i\<\sigma_i\>_{\ell}]\le\ve \,  \right\}\, .
\label{eq:DefinitionPointToSetAv}
\end{eqnarray}
Since $\E_{{\cal G}} \E_{\sigma^{(0)}}[\sigma^{(0)}_i\<\sigma_i\>_{\ell}]$ 
decreases rapidly to $0$ above $\ell_*(\ve)$, it is not hard to 
argue that $\ell_{i,*}(\ve)\le C_1 \ell_{*}(\ve)$ with high probability
and  $\ell_{i,*}(\ve)> C_0 \ell_{*}(\ve)$ 
for a finite fraction of sites/graphs ($C_0$, $C_1$ being two positive 
constants).
We also plot $\ell_*(\ve)$ as a function of the temperature for $\ve=0.1$.
As anticipated, the scaling $\ell_*(\ve)\sim (T-T_{\rm d}(\g))^{-1/2}$
is remarkably well verified allowing for a precise determination
of $T_{\rm d}(\g)$. The 
approach followed for this computation is described in Appendix 
\ref{app:PointToSet}.
%
%
\subsection{Implications for dynamics}
\label{sec:Implications}

As we have seen, at the dynamic transition line $T_{\rm d}(\g)$,
the point-to-set correlation length $\ell_*(\ve)$ diverges.
It is intuitively clear that this should induce a diverging
relaxation time, associated to
the (dynamical) glass transition. Imagine for instance to perturb the spin
$\sigma_i$. In order for a new equilibrated configuration 
(under the perturbed Hamiltonian) to be produced, the effect of the 
perturbation must propagate over a distance at least of the order of 
$\ell_{i,*}$, taking a time at least of order 
$\ell_{i,*}$. Therefore $\ell_{i,*}$ yields (up to a finite 
proportionality constant) a lower bound on the relaxation time 
for $\sigma_i$.

Remarkably, this handwaving argument can be turned into a mathematical 
derivation. For doing this, we need a precise definition for the 
relaxation time of $\sigma_i$. We consider the spin
autocorrelation function $C_i(t)$, and define
\begin{eqnarray}
\tau_i(\delta) = \min\{t>0 \, :\,  C_i(t)\le\delta\}\, .
\label{eq:TauDef}
\end{eqnarray}
Notice that $C_i(t)$ is a continuous and non increasing function of
$t$ (the last property follows from the spectral representation of
the Markov dynamics \cite{AldousFill}). Therefore $C_i(t)\le\delta$ 
for any $t\ge \tau_i(\delta)$. 
The precise value of $\delta$ is irrelevant, and it should be thought   
as a small fixed number.   Close to $T_{\rm d}(\g)$, we
expect $C_i(t)$ to develop a plateau. The definition 
(\ref{eq:TauDef}) correctly gives the slow time scale for the relaxation
of $\sigma_i$ as soon as $\delta$ is smaller than the plateau height, 
which is in turn equal to the local Edwards Anderson order 
parameter \cite{Ricci,Ricci2}.

We also need to introduce a cutoff radius
$\bar{\ell}_i$: this is the largest integer $\ell$ such that the subgraph
$\cG_\ell(i)$ is a tree.
\begin{propo}
Assume that $\bar{\ell}_i$,   $\ell_{i,*}(2\delta)>\log_2(4/\delta)$,
$\tau_i(\delta)<(\bar{\ell}_i-1)/(2\kappa_i e)$ and that the number 
of sites at distance $\ell$ from $i$ is at most $\kappa_i^{\ell}$
for  any $\ell >0$ and some $\kappa_i>0$.
Then  
\begin{eqnarray}
\tau_i(\delta)\ge \frac{1}{2\kappa_ie}\,\ell_{i,*}(2\delta)\, .
\label{eq:InequalityLength}
\end{eqnarray}
\end{propo}
\prooft
The proof is an application of some probabilistic arguments developed in 
\cite{HayesSinclair,DyerEtAl,BergerEtAl}. For greater convenience of the reader, 
we shall provide a self contained presentation of most of these ideas.

Let $\sigma^{(0)}$ be an equilibrium configuration for the system under 
consideration, $\ell\ge 0$ an integer and recall the
definition of $\cG_{\ell}(i)$. 
Consider two Markov trajectories 
$\{\sigma^{(1)}(t),\,t\ge 0\}$ and $\{\sigma^{(2)}(t),\,t\ge 0\}$ 
defined as follows. The initial condition is the same for both processes and 
is given by the reference configuration $\sigma^{(0)}$:
$\sigma^{(1)}(t=0) = \sigma^{(2)}(t=0) = \sigma^{(0)}$.
At times $t>0$, spin flips are proposed simultaneously in both systems 
according to the usual Glauber dynamics rule (each spin attempts a flip
according to a Poisson process of rate 1).
If the spin whose flip is proposed lies outside $\cG_{\ell}(i)$, 
the flip is never accepted for $\sigma^{(1)}$, and is accepted according 
to the usual transition probability for $\sigma^{(2)}$. If, on the other hand,
the spin which attempts a flip, let us say $j$, is within $\cG_{\ell}(i)$, 
the update is done in the two systems as 
follows. Let $\{p^{(1)}_+,p^{(1)}_-\}$ be the probabilities for
$\sigma^{(1)}_j$ to take values, respectively, $+1$ or $-1$ after the flip.
In the case of  Glauber dynamics, these are computed according to
Eq.~(\ref{eq:GlauberRates}). Denote by $\{p^{(2)}_+,p^{(2)}_-\}$ the same
transition probabilities for system $(2)$, and assume, for instance,
that $p^{(1)}_+\ge p^{(2)}_+$. Then, with probability
$p^{(2)}_+$, we set $\sigma^{(1)}_j = \sigma^{(2)}_j = +1$;
with probability $p^{(1)}_+-p^{(2)}_+$, we set $\sigma^{(1)}_j 
=+1$ and $ \sigma^{(2)}_j = -1$; finally, with probability
$1-p^{(1)}_+$,  $\sigma^{(1)}_j = \sigma^{(2)}_j = -1$.

Recall that the transition probabilities $p^{(\cdot)}_{+/-}$ 
only depend on the values of the spins on the neighbors of $j$. 
Therefore, the new values of $\sigma^{(1)}_j$, $\sigma^{(2)}_j$
will coincide whenever $\sigma^{(1)}_{j'} =  \sigma^{(2)}_{j'}$
for all the sites $j'$ at distance one from $j$. 
Furthermore, it is easy to see that when considered separately, the 
Markov processes $\{\sigma^{(1)}(t),\, t\ge 0\}$, 
$\{\sigma^{(2)}(t),\, t\ge 0\}$ have a very simple description.
The latter is the usual Glauber dynamics with equilibrated initial condition
$\sigma^{(0)}$. The former is Glauber dynamics for the subsystem
formed by spins in $\cG_{\ell}(i)$, with initial 
condition $\sigma^{(0)}$, and boundary condition 
(outside $\cG_{\ell}(i)$) also given by 
$\sigma^{(0)}$. The joint process is in fact a ``Markovian coupling" of these
two dynamics.

At $t=0$, $\sigma^{(1)}(t) = \sigma^{(2)}(t)$
by definition. Since the two dynamics are different outside $\cG_{\ell}(i)$
$\sigma^{(1)}(t)$ and $\sigma^{(2)}(t)$ will rapidly disagree there.
However, inside $\cG_{\ell}(i)$, disagreement can only propagate at a finite 
velocity, starting from sites at distance $\ell-1$ from $i$ and moving inward.
It is intuitively clear that, in order for the disagreement to reach 
$\sigma_i$, a time of order $\ell$ is needed. This intuition can be 
precised mathematically. Assume $\ell<\bar{\ell}_i$ and denote by
$N(\ell)$ the number of sites at distance $\ell$ from $i$. Then
\begin{eqnarray}
\prob\left\{\sigma_i^{(1)}(s) = \sigma_i^{(2)}(s)\; \forall\,
s\le t\right\}\ge 1-\left(\frac{et}{\ell}\right)^{\ell}\, N(\ell)\, .
\end{eqnarray}
This inequality is essentially adapted from \cite{HayesSinclair},
and we refer to this paper for a proof.

We now turn to the main steps of the proof.
Using the definition of $\{\sigma^{(1)}(t)\}$ and the fact that 
$\<\sigma^{(1)}_i(0)\sigma^{(1)}_i(t)\>$ is a non increasing
function of $t$, we have
\begin{eqnarray}
\E_{\sigma^{(0)}}[\sigma^{(0)}_i\<\sigma_i\>_{\ell}] =
\lim_{t\to\infty} \<\sigma^{(1)}_i(0)\sigma^{(1)}_i(t)\>\le
 \<\sigma^{(1)}_i(0)\sigma^{(1)}_i(t)\>\, .
\end{eqnarray}
Denote by
$\ind(t)$ the indicator function for the event that disagreement did not 
percolate to $i$ up to time $t$. More explicitly
\begin{eqnarray}
\ind(t) = \left\{\begin{array}{cc}
1 & \;\mbox{ if $\sigma_i^{(1)}(s) = \sigma_i^{(2)}(s)\; \forall\,s\le t$,}\\
0 & \;\mbox{ otherwise.}
\end{array}\right.
\end{eqnarray} 
Continuing from the above inequality we get, for $\ell<\bar{\ell}_i$,
\begin{multline}
\E_{\sigma^{(0)}}[\sigma^{(0)}_i\<\sigma_i\>_{\ell}] \le
 \<\sigma^{(1)}_i(0)\sigma^{(1)}_i(t)\> =
\<\sigma^{(2)}_i(0)\sigma^{(2)}_i(t)\, \ind(t)\> +
\<\sigma^{(1)}_i(0)\sigma^{(1)}_i(t)\, [1-\ind(t)]\> = \\
=  \<\sigma^{(2)}_i(0)\sigma^{(2)}_i(t)\>-
\<\sigma^{(2)}_i(0)\sigma^{(2)}_i(t)\, [1-\ind(t)]\>+
\<\sigma^{(1)}_i(0)\sigma^{(1)}_i(t)\, [1-\ind(t)]\>\le\\
\le  \<\sigma^{(2)}_i(0)\sigma^{(2)}_i(t)\> +
2 \<1-\ind(t)\>\le
C_i(t) + 2\left(\frac{et}{\ell}\right)^{\ell}N(\ell)
\, .
\end{multline}
Taking $t=\tau_i(\delta)$, we obtain 
\begin{eqnarray}
\E_{\sigma^{(0)}}[\sigma^{(0)}_i\<\sigma_i\>_{\ell}] \le
\delta+ 2 \left(\frac{e\tau_i(\delta)}{\ell}\right)^{\ell}N(\ell)
\, .
\end{eqnarray}
Now set $\ell = \max[\lceil 2\kappa_i e \tau_i(\delta)\rceil,
\lceil\log_2(2/\delta)\rceil ]$\footnote{Here and in the following,
$\lceil x \rceil$ denotes the smallest integer greater than or equal to $x$, and
$\lfloor x \rfloor$ the greatest integer smaller than or equal to $x$.}. 
Under the hypothesis of the proposition, $\ell<\bar{\ell}_i$ and 
$N(\ell)\le \kappa_i^{\ell}$. Therefore
\begin{eqnarray}
\E_{\sigma^{(0)}}[\sigma^{(0)}_i\<\sigma_i\>_{\ell}] \le
\delta+2\,2^{-\ell}\le 2\delta
\, .
\end{eqnarray}
By the definition of $\ell_{i,*}(\ve)$, the above equation implies 
$\ell\ge\ell_{i,*}(2\delta)$ thus proving our claim
(using the hypothesis $\ell_{i,*}(2\delta)>\log_2(4/\delta)$).
\endproof

Let us comment on the hypotheses of this Proposition.
For a random graph $\cG$, $\bar{\ell}_i$ is, with high probability,
close to $\log_{p(p-1)\g}N$ \cite{Bollobas}.
The hypothesis $\bar{\ell}_i>\log_2(4/\delta)$ is therefore satisfied
with high probability (although in any given sample there is
a vanishing fraction of sites $i$ for which this is not the case).

Furthermore, within any fixed distance $\ell$ from $i$, $\cF$ converges to
a (Galton-Watson) tree. The number of offspring is a Poisson variable 
of mean $p\g$ for variable nodes, while it is deterministic
and equal to $(p-1)$ for function nodes. Let $N_{\rm tree}(\ell)$
the number of $\ell$-th generation descendants in such a tree. It is 
easy to show that, for any $\kappa>p(p-1)\g$, 
$N_{\rm tree}(\ell)<\kappa^{\ell}$ with probability approaching one as
$\ell\to\infty$. 
On the random graph $\cG$, the number of neighbors $N(\ell)$  is in
fact smaller than on the tree. As a consequence, $\kappa_i$ is finite for 
most of the sites in the graph and, roughly speaking, of order 
$p(p-1)\g$ for a finite fraction of them. 
On the other hand, $\kappa_i$ can be much larger (diverging as $N\to\infty$)
for a vanishing fraction of sites. 
In particular, $\kappa_i$ must be at least as large as 
the degree of $i$, and the maximal degree in such a random graph 
scales as $\log N/\log\log N$.

Finally, the inequality (\ref{eq:InequalityLength}) is only interesting
in strong correlation regimes, such that $\ell_{i,*}(\ve)\gg 1$.  
Therefore, the hypothesis $\ell_{i,*}(2\delta)>\log_2(4/\delta)$ 
does not imply any loss of generality.

In the previous Section we argued that the typical point 
to set correlation length $\ell_*(\ve)$ diverges as  
$(T-T_{\rm d}(\g))^{-1/2}$ upon approaching the glass transition 
temperature. Therefore we obtained the following lower bound on the 
typical relaxation time:
\begin{eqnarray}
\tau(\delta)\gtrsim (T-T_{\rm d}(\g))^{-1/2}\, .\label{eq:Timescale}
\end{eqnarray}
As explained below, the usual MCT predictions agree with such a bound.
%
\section{From dynamics to rearrangements}
\label{RearrSection}

\subsection{On the relaxation time divergence}

The  inequality (\ref{eq:Timescale}) --along with its derivation-- 
explicitly shows how a divergent 
length scale induces a divergent time scale at the (dynamical) glass transition.
However, several reasons suggest that this lower bound is not tight. 
A first indication is provided by the usual predictions of MCT.
These imply 
\begin{eqnarray}
\tau(\delta)\sim (T-T_{\rm d})^{-\te_{\rm mct}}\, ,\;\;\;\;\;\;
\te_{\rm mct} = \frac{1}{2a_{\rm mct}}+\frac{1}{2b_{\rm mct}}\, ,
\label{eq:GeneralMCT1}
\end{eqnarray}
with $a_{\rm mct}$, $b_{\rm mct}>0$ determined by the equations
\begin{eqnarray}
\frac{\Gamma(1-a_{\rm mct})^2}{\Gamma(1-2a_{\rm mct})}
=\frac{\Gamma(1+b_{\rm mct})^2}{\Gamma(1+2b_{\rm mct})} = 
\lambda_{\rm mct}\, ,\label{eq:GeneralMCT2}
\end{eqnarray}
and $\lambda_{\rm mct}$ a positive, model-dependent constant between 0 and 1.
It is easy to show, using these equations (see also Sec.~\ref{sec_minsize}), 
that $a_{\rm mct}\le 1/2$, and therefore $\te_{\rm mct}\ge 1$.
Although MCT is not exact for the model studied here, we 
will argue below that several of its predictions are correct
(in particular, this can be shown in the fully connected limit 
$\g\to\infty$ \cite{KiTh}).
If this is accepted, it follows that Eq.~(\ref{eq:Timescale})
does not predict the correct critical exponent.

A more fundamental reason for (\ref{eq:Timescale}) not to be tight is the
following. 
This estimate was derived by showing that a region of linear size
$\ell_{i,*}(\ve)$ requires a time at least of order $\ell_{i,*}(\ve)$
to relax. This is true because a perturbation propagates at finite speed under
Glauber dynamics. However, the time required 
for equilibration on such a length scale can easily be much larger. 
Information can for instance propagate by diffusion (leading to 
$\tau\sim \ell^2$) or equilibration time may be
related to the volume rather than to the linear size of highly 
correlated regions.

In order to understand this point, we shall hereafter focus on the low 
temperature  limit at $\g<\g_{\rm d}$. In this regime, the system 
spends most of its time in quasi-ground states, and the leading
relaxation mechanisms will be ``jumps" from one such low-energy
configuration to a different one. Since the properties of these
jumps are likely to be continuous with respect to the energy of the starting 
and arrival points, we shall hereafter assume them to be exact ground states.

The time scale for relaxation of spin $\sigma_i$ can be estimated in this 
limit through a standard Arrhenius argument. We have
$\tau_i \sim \exp\{\beta \Delta_i\}$, where $\Delta_i$ is
the smallest energy to be overcome during a rearrangement of
the system which flips $\sigma_i$ (a jump). Due to the definition of our
Hamiltonian (\ref{Hamiltonian}), $\Delta_i$ is an even number. For
future convenience we shall write $\Delta_i = 2b_i$, and
refer to the Arrhenius law in the form
\begin{eqnarray}
\tau_i \sim \exp\{2\beta b_i\}\, .\label{eq:Arrhenius}
\end{eqnarray}
We will sometimes call $b_i$ the ``barrier" for spin $\sigma_i$ 
(although, strictly speaking, the energy barrier is $2b_i$).

A formal  definition of
rearrangements is given below. While it is clear that energy barriers 
are the relevant quantities determining time scales in the Arrhenius regime,
it is rather involved to evaluate them.
For pedagogical reasons we shall at first neglect barriers
and look for minimal size rearrangements.
%
%
\subsection{Rearrangements}
\label{sec_def_rearr}

A rearrangement $\R_i$ for the site $i$ is a set of sites such
that $i$ belongs to $\R_i$, and for every interaction $\a$, an even number
of the variables in $\partial \a$ belongs to $\R_i$. 
If one starts from a ground state and flips all the 
spins in $\R_i$, a new ground state is obtained with $\sigma_i$ flipped
(viceversa, any two ground states which differ at site $i$,
differ in a whole rearrangement $\R_i$).
A rearrangement $\R_i$ is ``simple" if it is   connected and
each interaction contains either zero or two sites of $\R_i$.
In the following we shall only consider simple rearrangements, 
and let this restriction understood.
It will be clear that, for our purposes (and in particular
within the $\g<\g_{\rm d}$ regime) there is no loss of generality 
in this restriction.
$\Re_i$ denotes the set of all simple rearrangements $\R_i$.

We shall consider observables defined on the set $\Re_i$, the simplest
one $n(\R_i)$ being the number of sites in the rearrangement. 
Let us discuss the general strategy on this example.  
The optimal
value of the observable $n$, and the set of optimal rearrangements with respect
to their size are, respectively,
\begin{equation}
n_i = \min_{\R_i \in \Re_i} n(\R_i) \ , \quad\quad \Re_i^{(n)} = 
\{\, \R_i \in \Re_i \,|\, n(\R_i) = n_i \,\} \ .
\end{equation}

Unless it diverges with the system size (which can be checked 
{\it a posteriori}), $n_i$, as well as the rearrangements in
$\Re_i^{(n)}$ can be constructed ``as if" the graph 
$\cG$ were a tree. A formal justification for this procedure
is obtained as follows. Consider the neighborhood $\cG_{\ell}(i)$ 
of radius $\ell$ around $i$. Let $\Re_{i,\ell}$ be the set of
rearrangements for the model whose factor graph is  $\cG_{\ell}(i)$. 
Denote by $n_{i,\ell}$ ($\Re_{i,\ell}^{(n)}$) 
the corresponding minimum rearrangement size (set of
optimal rearrangements). 
It is clear that, if $n_{i,\ell}<\ell$, the rearrangements in 
$\Re_{i,\ell}^{(n)}$ are also rearrangements for the whole graph 
$\cG$ and therefore $n_i\le n_{i,\ell}$. Furthermore, any rearrangement
belonging to $\Re_i$ but not to $\Re_{i,\ell}$ has size at least 
$\ell$. Therefore $n_i = n_{i,\ell}$ and $\Re_{i}^{(n)}
=\Re_{i,\ell}^{(n)}$. Consider now a sequence of random graphs $\cG$
of  increasing size: if $n_i$ remains finite in the thermodynamic limit, 
$\ell$ can be chosen large enough that $n_i< \ell$ and $\cG_{\ell}(i)$
is a tree with high probability.

A minimum-size rearrangement on a tree (and, by the argument above,
on a random graph as long as the its size is finite) can be constructed 
through the following recursive procedure. 
The root $i$ is by definition included in $\R_i$. In each of the interactions 
$\a \in \partial i$ of its
neighboorhood, exactly one variable, call it $j$, among the
$p-1$ distinct from $i$, has to be included in $\R_i$. 
Let $R_{j\to \a}$ be an optimal rearrangement for site 
$j$ in the subtree containing $j$ but not $\a$. Then
$\R_i$ is obtained by chosing, for each $\a$, the $j$
which minimizes $n(\R_{j\to\a})$.
For instance, each time the interaction $\a$  contains a node $j$
of degree $1$, an optimal choice is $\R_{j\to\a} = \{j\}$:
this branch of the rearrangement $\R_i$ contains just one node.

If $\cG$ is a random graph, any finite neighborhood of site $i$
converges to a Galton-Watson random 
tree~\footnote{More precisely, a {\em bipartite} tree with two
offspring distributions for the two types of nodes: 
for variable nodes, this is Poisson of mean $p\g$,
while function nodes have a fixed number $p-1$ of offsprings.} 
rooted at $i$, call it $\cT(i)$.
The distribution of $n_i$ is therefore asymptotically the same as the
one of minimal rearrangements for the root of such a tree. 
Although, for $p(p-1)\g>1$, $\cT(i)$ is infinite with non vanishing 
probability (corresponding to the percolation of $\cG$), 
the optimal rearrangement size turns out to be finite with probability
one up to $\g_{\rm d}$  ($>1/p(p-1)$), confirming the
validity of the procedure described above. 

\section{Rearrangements of minimal size}
\label{sec_minsize}

We consider here  rearrangements of minimal size, already defined in the 
previous Section. 
In computer science language, given a solution $\sigma$ 
of the XORSAT problem, we look for the solution $\sigma'$
which minimizes the Hamming distance from $\sigma$, under the condition
$\sigma_i'=-\sigma_i$. 

The minimal rearrangement for any given site $i$ is necessarily  connected.
The recursive construction of optimal rearrangements described above 
implies that, as long as $n_i$ is finite, optimal rearrangements are
simple.
%
%
\subsection{Recursion relations}

As shown in the previous Section, we can describe the computation of $n_i$
assuming that the underlying factor graph is a tree (we can picture it as
a large tree-like neighborhood of $i$).
We introduce two type of messages (`cavity fields') 
$\{ v_{j\to\a}\}$ and $\{u_{\a\to j}\}$ associated to 
directed  edges in the factor graph $\cF$. 
We define $v_{j\to\a}$ to be the minimum rearrangement size for site
$j$, relative to the subtree 
rooted at $j$ but {\em not including} 
$\a$ (in other words $v_{j\to\a}=\min n(R_{j\to\a})$. 
Analogously, $u_{\a\to i}+1$ is the size of the optimal 
rearrangement for site $i$, relative to the subtree rooted at $i$
and {\em including only} $\a$ among its neighbors (the $+1$ 
is introduced to simplify the equations below). The recursive 
construction described above implies the following relations
\begin{equation}
v_{i \to \a} = 1 + \sum_{\b \in \partial i \setminus \a} 
u_{\b \to i} \;,\;\;\;\;\;\;\;\;\;
u_{\a \to i} = \min_{j\in \partial \a \setminus i}[v_{j \to \a}] 
\ . \label{eq:MinSize_uv}
\end{equation}
Finally, the optimal rearrangement for $i$ in the original graph 
is obtained by combining optimal rearrangements for the subtrees 
rooted at $i$:
\begin{equation}
n_i = 1 + \sum_{\a \in \partial i} u_{\a \to i} \ .
\label{eq:MinSize}
\end{equation}
Given the factor graph $\cF$, the recursions (\ref{eq:MinSize_uv}),
(\ref{eq:MinSize}) can be solved by iteration (message-passing)
starting from the initial condition $v_{i \to \a}=u_{\a\to i}=0$.
We found that this iteration converges rapidly for typical random 
graphs at $\g<\g_{\rm d}$, leading to an algorithm of linear time 
complexity. 
%
%
\subsection{Probabilistic analysis}
\label{sec:ProbabilisticSize}

As already explained, the (asymptotic) probability distribution of $n_i$
can be computed by considering the random tree $\cT(i)$ rooted
at $i$. In this case, the messages $\{v_{j\to\a}\}$ and
 $\{u_{\a\to j}\}$ become random variables. 
Because of the construction of  $\cT(i)$, they are identically distributed.
We shall denote by $q_n$ the common distribution of the $v_{j\to\a}$'s
and by $\qh_n$ the distribution of the  $u_{\a\to j}$'s.  
The distribution of $n_i$ is also $q_n$ because of the memoryless 
property of Poisson random variables. Finally, as long as the corresponding
subtrees are disjoint, these random variables are independent.

The relations (\ref{eq:MinSize_uv}) can be interpreted as recursive 
distributional equations. Explicit expressions are easily written by using
the cumulative distributions
\begin{eqnarray}
Q_n = \sum_{n'\ge n} q_{n'} \, , 
\;\;\;\;\;\; \Qh_n = \sum_{n'\ge n} \qh_{n'} \, .
\end{eqnarray}
The recursive relations (\ref{eq:MinSize_uv}) imply:
\begin{eqnarray}
q_n & =  &
\sum_{l=0}^\infty p_l \sum_{n_1,\dots , n_l} \hat{q}_{n_1} \dots 
\qh_{n_l} \ \delta_{n,1+n_1+\dots+n_l}\, ,\label{eq:MinSizeDistr1} \\
\Qh_n & = & Q_n^{p-1} \ .
\label{eq:MinSizeDistr2}
\end{eqnarray}
Here $p_l = e^{-\g p} (\g p)^l / l!$, is the 
(Poisson) offsprings distribution at variable nodes.

Equations (\ref{eq:MinSizeDistr1}) and (\ref{eq:MinSizeDistr2}) 
uniquely determine $q_n$, $\qh_n$
and can be easily solved numerically: knowing the $q_n$ and $\hat{q}_n$
for $n \le m$ one can determine $q_{m+1}$, $\qh_{m+1}$.
The starting point is given by $q_1 = p_0$, $\hat{q}_1 = 1 - (1-p_0)^{p-1}$.

\begin{figure}
\includegraphics[width=8cm]{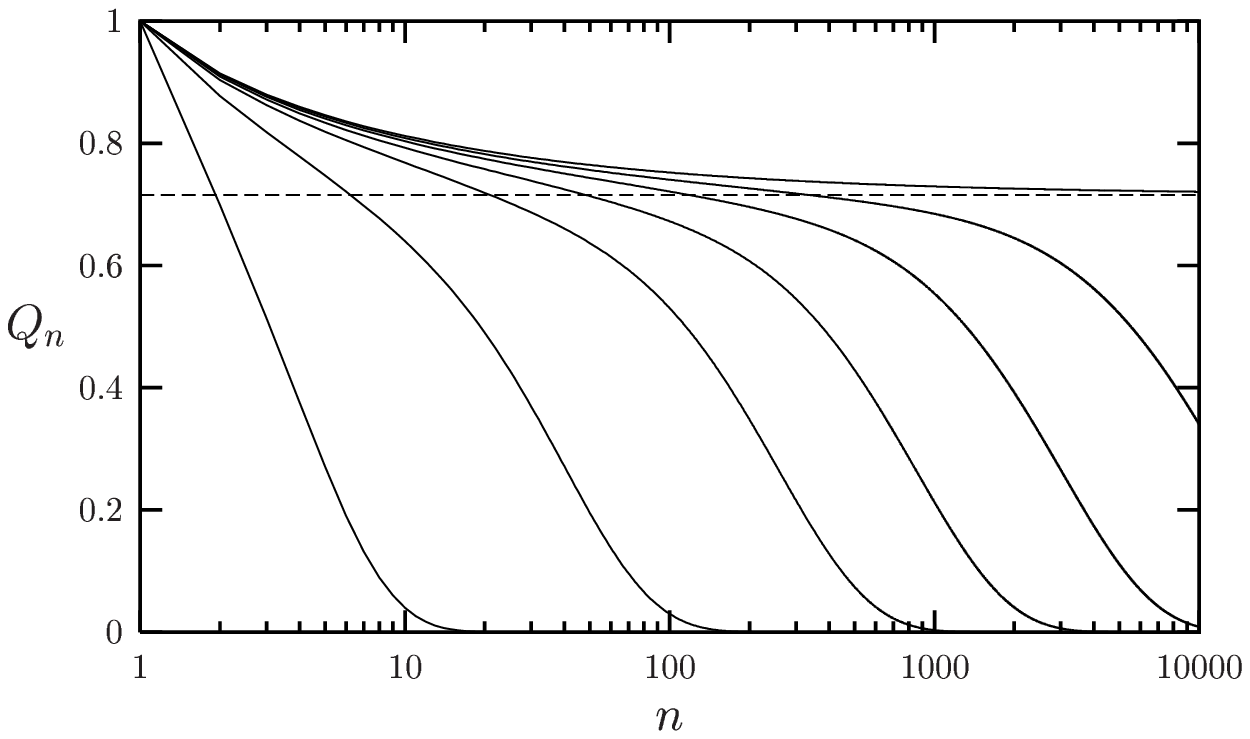}
\caption{The integrated distribution law of the minimal size of rearrangements.
$p=3$, from left to right 
$\g= 0.4, 0.7, 0.78, 0.8, 0.81, 0.815, \g_{\rm d}$.
The dashed horizontal line is the order parameter at the transition, 
$\phi_{\rm d} \approx 0.715332$.}
\label{fig_minsize_Qn}
\end{figure}

We present in Fig.~\ref{fig_minsize_Qn} 
the results of this computation. When the average 
degree $\g$ is raised towards its critical value $\g_{\rm d}$,
a plateau appears in the integrated probability law $Q_n$ (for $p \ge 3$), at
$Q_n\approx \phi_{\rm d}$. 
This means that a fraction $\phi_{\rm d}$ of the spins $\sigma_i$ have 
diverging minimum rearrangement size, consistently with the picture of 
freezing of some spins at the clustering transition $\g_{\rm d}$.
These are of course the same spins for which $\ell_i$ diverges upon 
approaching the transition, cf. Sec.~\ref{sec:PointToSet}. 
They will belong to the backbone after the addition of a small fraction of 
constraints.

%
%
\subsection{Asymptotic behaviour of the distribution}
\label{sec:AsymSizes}

The curves of Fig.~\ref{fig_minsize_Qn} are strongly reminiscent of the time
correlation functions obtained in the mode coupling theory of
supercooled liquids. Indeed, the equations (\ref{eq:MinSizeDistr1}) and
(\ref{eq:MinSizeDistr2}) have a structure similar to the one of schematic
MCT equations, cf. also App.~\ref{sec_app_minsize}. The analogy becomes even 
stronger if we consider the asymptotic behavior as $\g\to\g_{\rm d}$:
we present here a summary of results, while calculations are deferred to
App.~\ref{sec_app_minsize}. 

\begin{figure}
\includegraphics[width=7cm]{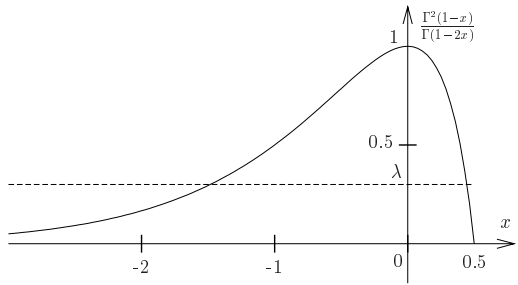}
\caption{The exponent $a$ (respectively $-b$) is the positive (respectively 
negative) root of the equation represented here, see 
Eqs.~(\ref{eq_MCT1_a}), (\ref{eq_MCT1_b}).} 
\label{fig_eq_MCT}
\end{figure}

For $\g=\g_{\rm d}$, the decay of $Q_n$ towards its plateau
value is algebraic, 
$Q_n \simeq \phi_{\rm d} + \mbox{cst}\, n^{-a}$ as $n\to\infty$
(here and below $\mbox{cst}$ denotes a generic positive constant).
The exponent $a$ is the positive solution of the transcendental equation
\begin{equation}
\frac{\Gamma^2(1-a)}{\Gamma(1-2a)}= \lambda \ ,
\label{eq_MCT1_a}
\end{equation}
where $\lambda= (p-2)/[\g_{\rm d} p (p-1)
\phi_{\rm d}^{p-1}]\in [0,1]$ is a non-universal parameter. 
A graphical representation of this equation is
provided in Fig.~\ref{fig_eq_MCT}.

In the critical region $\g \lesssim \g_{\rm d}$,
one can identify two distinct scaling regimes. The first one is 
$n\sim (\g_{\rm d}-\g)^{-1/2a}$, and corresponds to the behaviour of 
$Q_n$ around its plateau. We get
\begin{equation}
Q_n \simeq  \phi_{\rm d} +(\g_{\rm d}-\g)^{1/2}
\, \epsilon{\bf (}(\g_{\rm d}-\g)^{1/2a}n{\bf )}
\, ,
\end{equation}
where $\epsilon(\, \cdot\, )$ is a scaling function.
One can show that $\epsilon(t) \simeq \mbox{cst}\ t^{-a}$
as $t\to 0$ (thus matching the  behavior at $\g = \g_{\rm d}$ 
and $n$ fixed)
and $\epsilon(t) \sim - \mbox{cst} \ t^b$ as $t \to \infty$.
The positive exponent $b$ is fixed through
\begin{equation}
\frac{\Gamma^2(1+b)}{\Gamma(1+2b)}= \lambda \ ,
\label{eq_MCT1_b}
\end{equation}
see Fig.~\ref{fig_eq_MCT}. 

The second regime corresponds to the decay of $Q_n$ from the plateau
at $\phi_{\rm d}$ to $0$ and is defined by $n\sim (\g_{\rm d}
-\g)^{-\nu}$, where
\begin{equation}
\nu = \frac{1}{2a}+\frac{1}{2b} \ .
\end{equation}
We get the scaling form
\begin{equation}
Q_n \simeq Q_{\rm slow}{\bf (}(\g_{\rm d}-\g)^{\nu}n{\bf )}\, ,
\end{equation}
with $Q_{\rm slow}(t)$ a second scaling function. One finds that
$Q_{\rm slow}(t)\simeq \phi_{\rm d}-\mbox{cst}\ t^{b}$ as $t\to 0$,
while $Q_{\rm slow}(t)$ vanishes faster than exponentially
as $t\to\infty$.
Numerical values of the various exponents for some values of $p$ are collected
in Table~\ref{TableSize}.
\begin{table}
\begin{tabular}{| c || c | c | c | c | c | c |}
\hline
$p$ & $\g_{\rm d}$ & $\phi_{\rm d}$ & $\lambda$ & $a$ & $b$ & $\nu$ \\
\hline
\hline
3 & 0.818469 & 0.715332 & 0.397953 & 0.422096 & 1.221834 & 1.593787 \\
\hline
4 & 0.772280 & 0.851001 & 0.350174 & 0.433412 & 1.341647 & 1.526313 \\
\hline
5 & 0.701780 & 0.903350 & 0.320971 & 0.439997 & 1.421808 & 1.488035 \\
\hline
6 & 0.637081 & 0.930080 & 0.300707 & 0.444431 & 1.481191 & 1.462601 \\
\hline
7 & 0.581775 & 0.945975 & 0.285554 & 0.447677 & 1.527913 & 1.444121 \\
\hline
8 & 0.534997 & 0.956381 & 0.273649 & 0.450187 & 1.566174 & 1.429899 \\
\hline
9 & 0.495255 & 0.963661 & 0.263961 & 0.452205 & 1.598411 & 1.418505 \\
\hline
10 & 0.461197 & 0.969008 & 0.255868 & 0.453873 & 1.626162 & 1.409102 \\
\hline
\hline
\end{tabular}
\caption{Values of the critical parameters and exponents for the minimal
size rearrangements.}
\label{TableSize}
\end{table}

Notice that the scale $(\g_{\rm d}-\g)^{-\nu}$ has a very concrete 
interpretation.
For any $\ve>0$, one may ask what is the smallest size $n_\ve$, such that
all but a fraction $\ve$ of the sites have a minimal rearrangement 
whose size is at most $n_\ve$. If $\ve<\phi_{\rm d}$, one has
$n_{\ve}\sim (\g_{\rm d}-\g)^{-\nu}$.
%
\section{Rearrangements of minimal barrier}
\label{sec_minbarr}

In the low temperature limit, the time scale for the occurrence (flip) 
of a rearrangement $\R_i$, is given in terms of the corresponding 
energy barrier $b(\R_i)$ by the Arrhenius law, cf. Eq.~(\ref{eq:Arrhenius}).
In the previous Section we explained how to optimize $R_i$
with respect to a simple observable: its size $n(R_i)$. We want now to apply 
the same strategy to a more involved quantity, the barrier $b(R_i)$,
and use the result to estimate the relaxation time $\tau_i$.

Before dwelling into the actual computation, it is worth stressing
that energy barriers (and their distribution) are directly related
to auto-correlation functions. Consider a spin $i$ of
barrier $b_i$. If $b$ is a real number, fixed in the $\beta\to\infty$
limit, then
\begin{equation}
\lim_{\beta\to\infty} C_i(t=e^{2\beta b}) = 
\begin{cases} 1 \;\;\;\; \mbox{if}\; b<b_i\, , \\  0 \;\;\;\; \mbox{if}\; 
b>b_i\, ,
\end{cases} 
\end{equation}
Let $Q_{\lceil b \rceil}^{\rm site}$ (cf. also Sec.~\ref{sec:ProbabilisticBarrier}) 
be the fraction of sites $i$ such that
$b_i>b$. By averaging the above equation over the site, we obtain
\begin{equation}
\lim_{\beta\to\infty} C(t=e^{2\beta b}) = 
Q_{\lceil b \rceil}^{\rm site} \ .\label{eq:CorrDistr}
\end{equation}
%
%
\subsection{General considerations}
\label{sec:GeneralBarr}

Let us first expose the basic idea in an informal way.
Given a rearrangement $\R_i$, consider the paths in configuration 
space\footnote{These are sequences of configurations such that any
two successive configurations differ by one spin flip.}
which lead from a groundstate $\sigma$ to the one obtained
by flipping all the spins in $\R_i$.
The barrier $b(R_i)$ associated to this rearrangement will be the 
minimum among all paths of the maximal energy  along 
the path. If we let $b_i = \min_{R_i} b(R_i)$, the asymptotic 
relaxation time for spin $\sigma_i$ can be determined through 
Eq.~(\ref{eq:Arrhenius}).

Paths in
configuration space are definitely complex objects. Let us make the 
simplifying assumption that on the optimal ones, each variable of the system 
is flipped at most once. 
In the worst case, under this assumption we will get an upper bound on $b_i$.
We think that this upper bound is in fact tight: in Sec.~\ref{sec:LowerBarrier}
we shall provide a lower bound supporting this claim. Numerical experiments,
cf. Sec.~\ref{ScalingSection} also confirm this hypothesis.

Paths are thus defined by $(i)$ the set of
variables which are flipped, $\R_i$; $(ii)$ the order in which 
they are flipped, i.e. a permutation of the site indices in $\R_i$. 
Arguing as in Sec.~\ref{sec_def_rearr}, one 
can show that, as long as optimal rearrangements remain 
finite in the thermodynamic limit, we can assume that $R_i$ is simple.
Their typical size will be computed in Sec.~\ref{sec:SizeMinBarr}
validating this assumption for any $\g<\g_{\rm d}$.

Let us first consider the task of optimizing over $(ii)$, i.e. the
ordering of spins,   for a given choice of the rearrangement $\R_i$. 
Take the spins of the rearrangement to be the vertices of a graph, 
and put an edge between any two vertices corresponding to interacting
spins. The resulting graph will be a tree with high probability
as long as $\R_i$ is finite.
 
Imagine to put a ferromagnetic Ising model on this tree, with all spins 
initially up, and to flip them
in the order defining our paths in configuration space. The energy
of the original system along the path which flips $\R_i$ is 
equal to the one of this fictitious ferromagnetic Ising model. 
Therefore,  $b(\R_i)$ is just the energy barrier to be crossed
for reversing the spins of a ferromagnetic Ising model defined on a tree
associated to $\R_i$ (by flipping each spin exactly once).
In the following Section we shall study this problem and define one way
to solve it. Then we turn back to the optimal choice of the rearrangement
$R_i$, cf. point $(i)$.

%
%
\subsection{The minimal cutwidth problem and the disjoint strategy}

As a first step, we must compute the energy barrier for reversing
the spins of a ferromagnetic Ising model defined on a tree.
For the sake of simplicity, we shall define the energy of this fictitious 
model as the number of links between spins with different values 
(plus and minus). The energy in the original model can be recovered by 
multiplying by a factor $2$.

This problem was considered in Ref.~\cite{HenleyTree} which gave an
heuristic prediction of the behavior of $b_i$ for large trees, and 
in \cite{Angles} which considered the case of regular tree.

It is instructive to study two very simple instances of the problem. 
The first one is the case of an unidimensional graph of $n$ sites. The
optimal flipping order consists obviously in starting from one end of the 
chain and then proceed along it by flipping succesively neighboring variables.
The minimal energy barrier in this example is then 1. Consider as a second
simple case a star-like graph with one central site connected to $l$ neighbors.
After an instant of thought, one realizes that the minimal barrier is obtained
by flipping first half of the external spins, then the central one, and
finally the remaining half of the external sites. The energy barrier, achieved
just before or just after the flip of the central site, is 
$\lceil \frac{l}{2}\rceil$.

An useful graphical representation of the problem is the following. If one
draws the graph with its vertices aligned on an horizontal axis, ordered 
according to the temporal sequences in which the spins are flipped, the
energy at a given instant of time is simply the number of edges which are
drawn above this point of the temporal axis. The energy barrier is crossed
at the point where this number of edges is maximal. In fact the problem of
finding the ordering which minimizes the energy barrier is known in 
graph theory as the minimal cutwidth
problem, see~\cite{review_graph} for a review. On general
graphs minimal cutwidth is  NP-complete. However, it becomes polynomial
when restricted to particular families of graphs, such as 
trees~\cite{Yanna,Lengauer}.
We shall present now a simple strategy, known as {\em disjoint combination} 
\cite{Yanna}, which allows to construct ``quasi optimal" orderings.
The errors induced by this simple approach, together with the 
optimal strategy, are discussed in Sec.~\ref{sec:Yanna}.

\begin{figure}
\includegraphics[width=14cm]{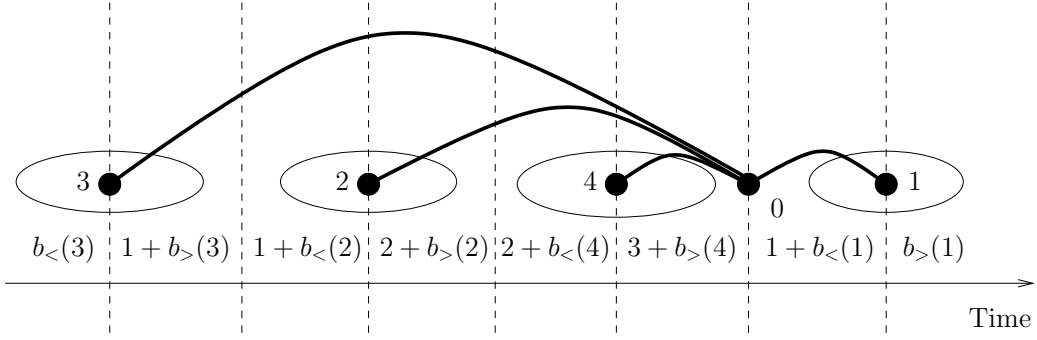}
\caption{The disjoint strategy. $l=4,j=3, \pi=(3,2,4,1)$, we used the shorthand
notations $\bb(i)=\bb(T_i,\o_i)$, $\ba(i)=\ba(T_i,\o_i)$. The numbers indicate
the maximal energy inside each time epoch, delimitated by dashed lines.}
\label{fig_disjoint}
\end{figure}

As often, a
tree-structured problem is efficiently tackled by a recursive approach: if
one knows a solution on a sub-tree, it is enough to combine the solution of
the different branches to obtain a solution on the original tree. More
precisely, let us consider a rooted tree $T$ composed of the root $0$ and $l$
sub-trees $T_1,\dots,T_l$, rooted at vertices $1,\dots,l$. 
We shall denote by $\o$ (respectively $\o_1,\dots,\o_l$) a permutation
of the vertices of $T$ (resp. $T_1,\dots,T_l$). Our aim
is to find the optimal $\o$, as defined above. 
Within the disjoint strategy, one restricts to 
trajectories $\o$ which forbid having two partially flipped sub-trees at 
the same time. In other words,
one chooses a permutation $\pi$ of $[1,l]$, an integer $j \in [0,l]$, and
flips first all variables of the sub-tree $T_{\pi(1)}$, then those of
$T_{\pi(2)}$, and so on until $T_{\pi(j)}$, then one flips the root $0$, and
proceed with the sub-trees $T_{\pi(j+1)},\dots,T_{\pi(l)}$. The permutation
$\o$ thus takes the form 
$(\o_{\pi(1)},\dots,\o_{\pi(j)},0,\o_{\pi(j+1)},\dots,\o_{\pi(l)})$. An example
is provided in Fig.~\ref{fig_disjoint}, where the rooted sub-trees are
represented as bubbles. The 
optimization with respects to $\o$ is now an optimization with respects to
$\pi,j$ and the sub-trajectories $\o_1,\dots,\o_l$. The crucial simplification
of the disjoint strategy is that the trajectories of the different
sub-trees are non-overlapping in time (this is clear on 
Fig.~\ref{fig_disjoint}), and can thus be performed independently
of each other. Indeed, let us define $b(T,\o)$ as the maximal energy 
encountered
during the flips of a rooted tree $T$ following the ordering $\o$, and
$\bb(T,\o)$ (resp. $\ba(T,\o)$) the maximal energy encountered before 
(resp. after) the root of $T$ has been flipped. From the graphical
representation of Fig.~\ref{fig_disjoint}, and 
considering the maximal energy  reached in each
period of time corresponding to a different sub-tree, it is easy to convince
oneself that
\begin{multline}
b(T,\o) = \max\Big\{\bb(T_{\pi(1)},\o_{\pi(1)}),1+\ba(T_{\pi(1)},\o_{\pi(1)}),
\dots,
j-1+\bb (T_{\pi(j)},\o_{\pi(j)}),j+\ba(T_{\pi(j)},\o_{\pi(j)}),\\
l-j+\bb(T_{\pi(j+1)},\o_{\pi(j+1)}),l-j-1+\ba(T_{\pi(j+1)},\o_{\pi(j+1)}),
\dots,\\
1+\bb(T_{\pi(l)},\o_{\pi(l)}),\ba(T_{\pi(l)},\o_{\pi(l)})\Big\} \ .
\end{multline}
The minimal barrier $b(T)$ within the disjoint strategy is then
obtained by minimizing over $\pi,j$ and the $\o_i$'s.
From the above expression it follows that the last 
optimization can be done independently, leading to
an expression of the form $b(T)=f_l{\bf (}\hat{b}(T_1),\dots,
\hat{b}(T_l){\bf )}$,
where
\begin{eqnarray}
f_l(\{\hat{b}_i\}) \equiv \min_{\pi,j} \max\big\{
\hat{b}_{\pi(1)},1+\hat{b}_{\pi(2)}, \dots, j-1+\hat{b}_{\pi(j)},
l-j-1+\hat{b}_{\pi(j+1)},\dots,1+\hat{b}_{\pi(l-1)},\hat{b}_{\pi(l)}\
\big\} \, ,\;\;\;
\label{eq_disj_b}
\end{eqnarray}
and $\hat{b}(T)$ is defined for an arbitrary rooted tree $T$ by
\begin{equation}
\hat{b}(T) = \min_\o \max[\bb(T,\o),1+\ba(T,\o)] \ .
\end{equation}
The last quantity is the minimal energy barrier for the tree $T$
supplemented by a link ({\em `anchor'}) between its
root and a fictitious spin which is never flipped, 
hence the shift $+1$ on the energy after the flip
of the root. Note that $\hat{b}(T)$ does not change if we intervert 
the role of $\bb$ and $\ba$ in the definition, 
a property which has been used in deriving 
Eq.~(\ref{eq_disj_b}).
In other words, the relevant information on the
sub-trees needed to choose the ordering $\o$ is not their optimal
barrier $b$ but the anchored barrier $\hat{b}$.
This happens because one has to take into 
account the energy of the links between the root $0$ and the roots of the
sub-trees.

The expression (\ref{eq_disj_b}) for $f_l$ can be simplified 
by noticing that for an optimal
permutation $\pi$, the minimum with respect to $j$ is reached for 
$j=\left\lceil \frac{l}{2}\right\rceil$. We get therefore 
\begin{equation}
f_l(\{\hat{b}_i\}) = \min_{\pi} \max_{i\in[1,l]}
\left[\hat{b}_{\pi(i)}+\left\lfloor\frac{i-1}{2}\right\rfloor \right] \, .
\end{equation}

A slightly different recursive equation holds (within the disjoint strategy) 
for the anchored barriers.
We have in fact  
$\hat{b}(T) =\hat{f}_l{\bf(}\{\hat{b}(T_1),\dots,\hat{b}(T_l)\}{\bf )}$,
where
\begin{eqnarray}
\hat{f}_l(\{\hat{b}_i\}) \equiv \min_{\pi,j} \max[
\hat{b}_{\pi(1)},1+\hat{b}_{\pi(2)}, \dots, j-1+\hat{b}_{\pi(j)},
l-j+\hat{b}_{\pi(j+1)},\dots,2+\hat{b}_{\pi(l-1)},1+\hat{b}_{\pi(l)}] 
\, .\;\;
\label{eq_disj_bhat}
\end{eqnarray}
Again the optimal choice of $j$ is $\left\lceil \frac{l}{2}\right\rceil$, and
the above expression simplifies to
\begin{equation}
\hat{f}_l(\hat{b}_1,\dots,\hat{b}_l) = \min_{\pi} \max_{i\in[1,l]}
\left[\hat{b}_{\pi(i)}+\left\lfloor\frac{i}{2}\right\rfloor \right] \ .
\end{equation}

For a tree $T$ containing only the root and no edge, one
defines $b(T)=0$, $\hat{b}(T)=1$. Using this initial condition
the above recursions can be efficiently 
applied to any tree.  In particular one can reproduce
the results of the two simple examples quoted before (linear and star-like
trees).
%
%
\subsection{Application to the $p$-spin model}
\label{sec:PspinBarr}

Let us now come back to the original problem of determining the minimal
energy barrier to flip one variable, say $\sigma_i$, in the 
$p$-spin model (\ref{Hamiltonian}), ending up in a ground state. With 
respect to the computation presented in the previous Section, we have one 
more degree of freedom to optimize on: the choice of the rearrangement
$\R_i$. This step can also be performed recursively.
For each of the function nodes $\a$ adjacent to $i$, we must choose 
a variable node $j\neq i$ in $\partial\a$, which belongs to $\R_i$.
The choice is of course dictated by the minimization of the anchored energy
barrier in the corresponding cavity graph. Using again the message notation,
we have
\begin{equation}
u_{\a \to i} = \min_{j\in \partial \a \setminus i}[v_{j \to \a}] 
\, ,\;\;\;\;\;\;\;\;
v_{i \to \a} = \hat{f}_{|\partial i|-1}
(\{u_{\b \to i}\}_{\b \in\partial i \setminus\a}) \ ,
\label{eq:MinBarr_uv}
\end{equation}
where by convention $\hat{f}_0=1$. Given the graph $\cG$, these 
equations can be efficiently solved through an iterative
(message-passing) procedure starting with the initial condition
$v_{i\to\a} = u_{\a \to i}=0$.
The minimal energy barrier to flip the variable $i$
is then computed using:
\begin{equation}
b_i = f_{|\partial i|}
(\{u_{\a \to i}\}_{\a \in\partial i}) \ ,
\label{eq_minbarr_b}
\end{equation}
with $f_0=0$: an isolated spin can be flipped without modifiying the energy
of the system.
%
%
\subsection{Probabilistic analysis}
\label{sec:ProbabilisticBarrier}

If we now consider a random graph $\cG$, the messages on the edges
of the factor graph are random variables, with laws $q_b$ for the messages
$v$, and $\hat{q}_b$ for the $u$'s. We use the same notation as in 
Sec.~\ref{sec_minsize}, there should not be confusion between the
two computations. Capital letters denote again cumulative distributions
$Q_b = \sum_{b'\ge b} q_{b'} \ , \Qh_b = \sum_{b'\ge b} \hat{q}_{b'}$.

Arguing as in Sec.~\ref{sec:ProbabilisticSize}, one can show that 
the recursions (\ref{eq:MinBarr_uv}) imply the following 
distributional equations
\begin{eqnarray}
q_b = \sum_{l=0}^\infty p_l \sum_{b_1,\dots,b_l} 
\hat{q}_{b_1} \dots \hat{q}_{b_l} \ \delta_{b,\hat{f}_l(b_1,\dots,b_l)} \ ,
\;\;\;\;\;\;
\Qh_b = Q_b^{p-1}\, ,
\label{eq_minbarr_distrib}
\end{eqnarray}
where $p_l = e^{-p\g}\, (p\g)^l/l!$ is the Poisson distribution 
of parameter $p\g$, and we set by convention $\fh_0=1$.
The distribution for the barriers~\footnote{The cumulative distribution 
of $b_i$ was simply denoted as $Q_b$ in Ref.~\cite{NostroLettera}.} 
$b_i$ is then obtained using Eq.~(\ref{eq_minbarr_b}):
\begin{equation}
q^{{\rm site}}_b = 
\sum_{l=0}^\infty p_l \sum_{b_1,\dots,b_l} 
\hat{q}_{b_1} \dots \hat{q}_{b_l} \ \delta_{b,f_l(b_1,\dots,b_l)} \ .
\label{eq_minbarr_distrib3}
\end{equation}
Here, by convention, $f_0=0$.

\begin{figure}
\includegraphics[width=10cm]{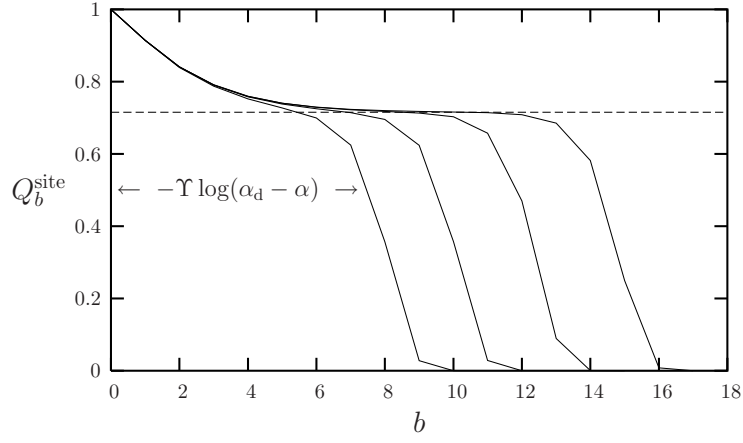}
\caption{Integrated law for the distribution of barriers, plotted
from left to right for $\g=0.816,0.818,0.8184,0.81846$.}
\label{fig_minbarr}
\end{figure}

The equations (\ref{eq_minbarr_distrib}) to 
(\ref{eq_minbarr_distrib3}) can be solved numerically, even if this task
is a bit more involved than for the distributions of minimal size
rearrangements. Some details can be found in App.~\ref{sec_app_minb_num}.
The results are plotted in Fig.~\ref{fig_minbarr}, for several values of 
$\g$ approaching $\g_{\rm d}$.
%
%
\subsection{Asymptotic behaviour of the distribution}
\label{sec_minbarr_as}

The critical behavior of~\footnote{For notational simplicity we refer to
$Q_b$ rather than to $Q_b^{\rm site}$: the two distributions have indeed
the same behavior.} $Q_b$, cf. Fig.~\ref{fig_minbarr},
presents some similarities with the distribution of  minimal 
size rearrangements. 
As $\g_{\rm d}$ is approached, a plateau develops at 
$Q_{b}\approx\phi_{\rm d}$. Eventually $Q_{b}$ decreases to $0$
at a scale $b$ diverging as $\g\to\g_{\rm d}$. 
The precise critical behaviour can be derived analytically revealing
both analogies and differences, cf. App.~\ref{sec_app_minb_ana}
for technical details.
Here is  an overview of the results.

\begin{figure}
\includegraphics[width=10cm]{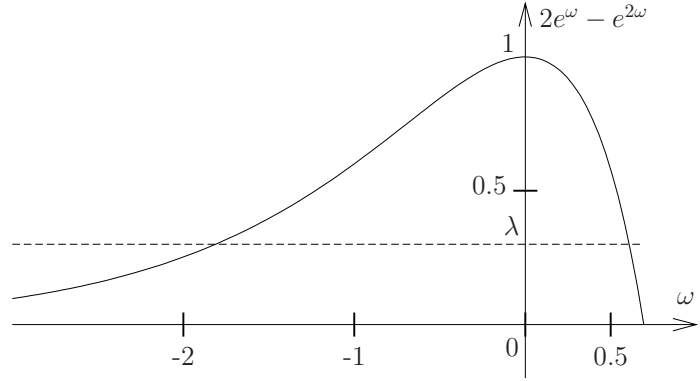}
\caption{The parameter $\omega_a$ (respectively $-\omega_b$) is 
the positive (resp. negative) root of the equation represented here,
cf. Eqs.~(\ref{eq_minbarr_oa}), (\ref{eq_minbarr_ob}).}
\label{fig_eq_omega}
\end{figure}

At the critical point $\g=\g_{\rm d}$, the plateau is approached
exponentially, 
$Q_b \simeq \phi_{\rm d} + \mbox{cst}\ e^{-\omega_a b}$
as $b\to\infty$.
The parameter $\omega_a$ is the unique positive root of the equation
\begin{equation}
2 e^{\omega_a} - e^{2 \omega_a} = \lambda \ ,
\label{eq_minbarr_oa}
\end{equation}
where $\lambda$ is defined as for minimal size rearrangements,
cf. Eq.~(\ref{eq_MCT1_a}). A graphical representation of this equation
is provided in Fig.~\ref{fig_eq_omega}, and numerical values of the solution
can be found in Table~\ref{TableBarrier}.

The scaling regime describing the plateau is defined by 
$\g\to \g_{\rm d}$ with $b-b_0(\g)$ fixed,
where $b_0(\g) \equiv -\frac{1}{2 \omega_a} \log (\g_{\rm d}-\g)$.
The form of the cumulative distribution in this limit is 
\begin{equation}
Q_b \simeq \phi_{\rm d}+(\g_{\rm d}-\g)^{1/2}\,\, 
\bar{\epsilon}_{b-b_0(\g)} \ .
\end{equation}
Here $\bar{\epsilon}_m$ is a scaling function defined on the integers. 
One can show that $\bar{\epsilon}_m\simeq \mbox{cst}\ e^{-\omega_a m} $
as $m\to -\infty$ and  $\bar{\epsilon}_m\simeq - \mbox{cst} \ e^{\omega_b m}$
as $m\to \infty$, where
\begin{equation}
2 e^{-\omega_b} - e^{-2 \omega_b} = \lambda \ .\label{eq_minbarr_ob}
\end{equation}
This implies that $Q_b$ decreases from the plateau  to $0$
on a scale  $b\simeq - \Upsilon \, \log(\g_{\rm d}-\g)$, where
\begin{equation}
 \Upsilon = \frac{1}{2 \omega_a} +\frac{1}{2 \omega_b} \ .
\label{eq_minbarr_bp0}
\end{equation}
More precisely, for any $Q_*\in (0,\phi_{\rm d})$, the smallest $b$ such that
$Q_b<Q_*$, say $b_*$, behaves as $b_* = - \Upsilon \, 
\log(\g_{\rm d}-\g)+O(1)$. 

Finally, as $b\to\infty$, the cumulative distribution
vanishes faster than exponentially: $Q_b\sim 
\exp\{-(2b)\log (2b)\}$. Barriers significantly larger than 
$- \Upsilon \, \log(\g_{\rm d}-\g)$ are therefore very rare
and related to sites of exceptionally large degree $l$ (leading,
as in the simple example of  star-like graphs, to
$b \approx \lfloor l/2\rfloor$). 
\begin{table}
\begin{tabular}{| c || c | c | c | c | c | c |}
\hline
$p$ & $\omega_a$ & $\omega_b$ & $\Upsilon$ & $\mu_a$ & $\mu_b$ & $\nu_{\rm barr}$ \\
\hline
\hline
3 & 0.574317 & 1.495740 & 1.204882 & 0.816504 & 3.903496 & 2.015719 \\
\hline
4 & 0.591180 & 1.640504 & 1.150551 & 0.798466 & 4.164498 & 1.944590 \\
\hline
5 & 0.601050 & 1.737456 & 1.119655 & 0.787987 & 4.341305 & 1.904838 \\
\hline
6 & 0.607719 & 1.809338 & 1.099093 & 0.780941 & 4.473350 & 1.878702 \\
\hline
7 & 0.612614 & 1.865937 & 1.084136 & 0.775785 & 4.577858 & 1.859868 \\
\hline
8 & 0.616408 & 1.912315 & 1.072615 & 0.771801 & 4.663830 & 1.845467 \\
\hline
9 & 0.619461 & 1.951413 & 1.063378 & 0.768599 & 4.736535 & 1.833994 \\
\hline
10 & 0.621990 & 1.985085 & 1.055750 & 0.765953 & 4.799310 & 1.824570\\
\hline
\hline
\end{tabular}
\caption{Exponents for the minimal barrier rearrangements.}
\label{TableBarrier}
\end{table}

%
%
\subsection{Yannakakis algorithm and single sample analysis}
\label{sec:Yanna}

The computation of energy barriers in the previous Sections relied on two
assumptions. First, we assumed that along an optimal trajectory in phase
space, each spin of the system is flipped at most once. We showed that the
problem is then reduced to the computation of the minimal cutwidth of
a tree (once the rearrangement $\R_i$ is given). 
The second hypothesis was the use of the disjoint combination
strategy to determine the minimal cutwidth. 
Here we begin by revisiting this second assumption, further analytical 
arguments will be presented in the next Section.

\begin{figure}
\includegraphics[width=14cm]{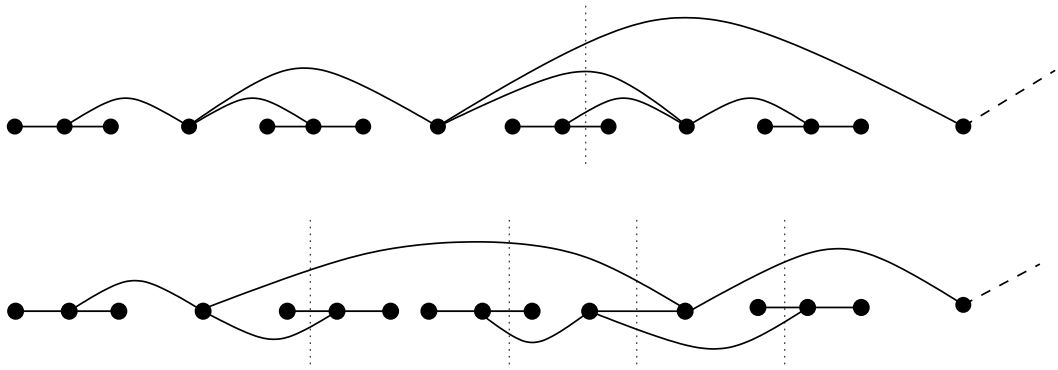}
\caption{Minimal cutwidth problem for a binary tree of 5 generations (only
one half of the tree is represented, the second follows by symmetry).
Top: ordering produced according to the disjoint combination strategy, with
a cutwidth of 4 reached on the dotted line. Bottom: an optimal ordering
of cutwidth 3.}
\label{fig_binary_counter}
\end{figure}

In general the disjoint combination strategy is not able to find the
minimal cutwidth of a tree, and provides only an upper bound on it.
The  simplest example is provided by a
binary tree (in which each vertex has two ``sons") of height (number
of generations) 5. As shown on the upper part of 
Fig.~\ref{fig_binary_counter}, the disjoint strategy predicts in this
case a minimal cutwidth equal to 4. However, the ordering
in the lower part of Fig.~\ref{fig_binary_counter} 
achieves a  cutwidth  equal to 3. This ordering is not
disjoint: there are points in the horizontal time axis such that
two height--3 subtrees are partially flipped. 

Yannakakis has presented in~\cite{Yanna} an algorithm which solves exactly
the minimal cutwidth problem for any tree in polynomial time. 
This procedure is quite involved. We shall recall here
a few selected features which are relevant for the application to
our case~\footnote{A C implementation
of the algorithm is available upon request to the authors.}. The
computation relies on the definition for any rooted tree $T$ 
of a cost function $c(T)$, which is a sequence of integers.
The length of $c(T)$ can, in principle, be as large as the number of 
vertices in $T$. The cost function
contains, along with the minimal cutwidth of $T$, additional informations
which were discarded in the disjoint strategy. 
Furthermore, it has two crucial properties:
\begin{itemize}
\item[$\bullet$] It can be computed recursively. If the root of $T$ has
$l$ sons, which are roots of the sub-trees $T_1,\dots,T_l$, there is
a function $f_l^{\rm Y}$ such that $c(T)=f_l^{\rm Y}(c(T_1),\dots,c(T_l))$.
\item[$\bullet$] The cost of two trees can be compared. More precisely, 
there exists an order relation among  costs, and the application
$f_l^{\rm Y}$ is monotonously increasing  with respects to this relation.
\end{itemize}

We want to compute the minimal barrier for a spin $i$
in a given $p$-spin sample, by optimizing over the rearrangement $\R_i$
and the flipping order of the spins in $\R_i$. 
Thanks to the two properties above, this task can be accomplished 
through an iterative message--passing procedure which uses 
Yannakakis strategy.
Messages $u_{\a \to i}$ and $v_{i \to \a}$ are now cost 
functions, obeying 
Eqs.~(\ref{eq:MinBarr_uv}), where
the minimum  is taken in the sense of the
order relation on cost functions, and $\hat{f}_l$ is replaced by 
$f_l^{\rm Y}$. When the fixed point of these equations has been reached,
the optimal cost function for any spin is given by 
$c_i=f_{|\partial i|}^{\rm Y}(\{ u_{\a \to i} \}_{\a \in \partial i})$.
The minimal energy barrier $b_i$ is then one piece of the information
contained in $c_i$.

\begin{figure}
\begin{tabular}{cc}
\includegraphics[width=8.cm]{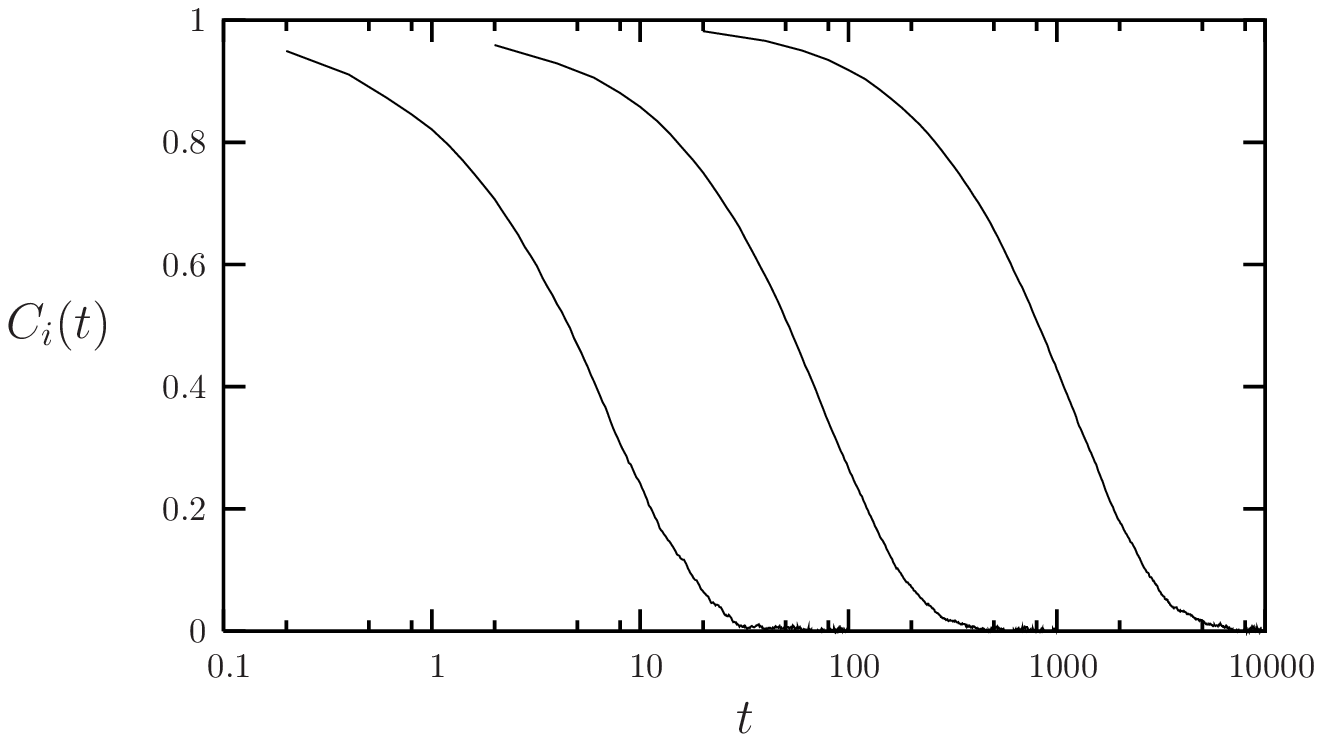} &\hspace{0.5cm}
\includegraphics[width=7.5cm]{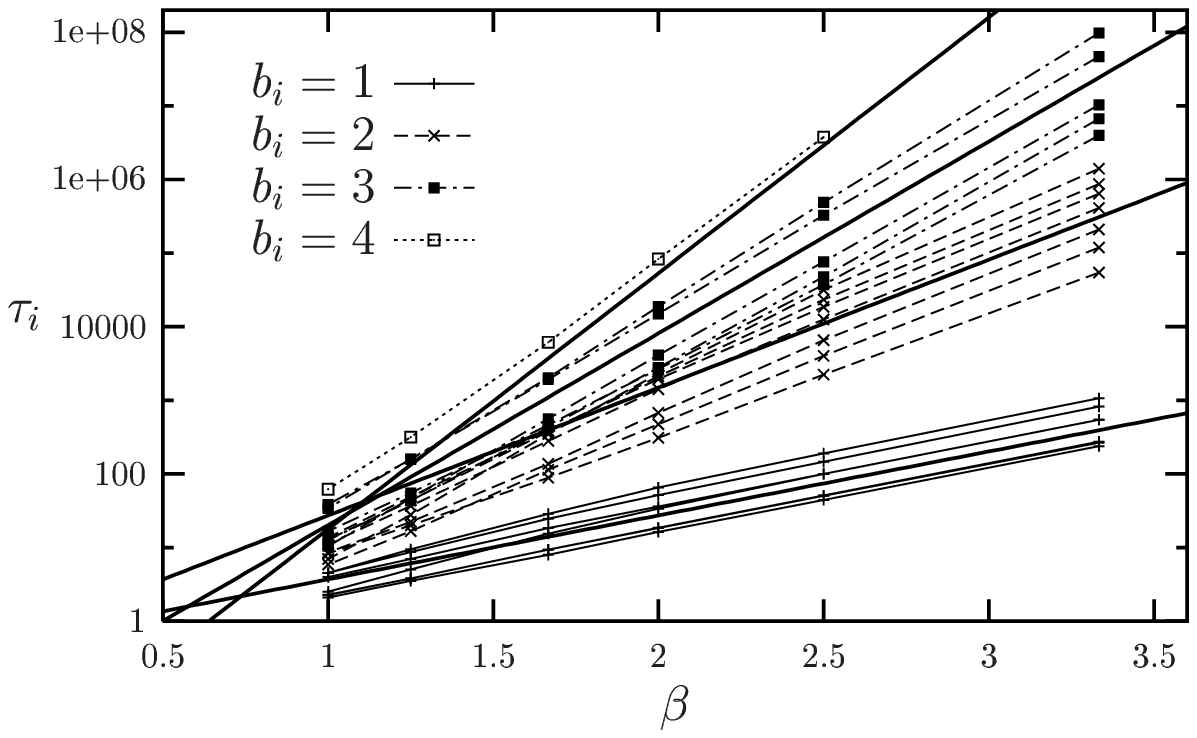}
\end{tabular}
\caption{Monte Carlo study of a sample of size $N=10^4$, $\g=0.6$.
Local correlation functions are measured by averaging over $2\cdot 10^3$ 
to $10^5$ runs depending on the temperature.
Left: local correlation function $C_i(t)$ for one spin,
from left to right $T=1$, $T=0.5$, $T=0.3$.
Right: local correlation times $\tau_i$ for 20 spins of the sample, defined as
$C_i(\tau_i)=0.5$. Bold solid lines
correspond to Arrhenius law (\ref{eq:Arrhenius}) with $b_i = 1,2,3,4$.
The symbols are chosen according to the computation
of $b_i$ with the algorithm of Sec.~\ref{sec:Yanna}: 
$7$ sites have $b_i=1$, 7 $b_i=2$, 
5 $b_i=3$ and 1 $b_i=4$.}
\label{fig_singlesample}
\end{figure}

As explained before, one expects that the local correlation time $\tau_i$
of spin $i$ is given, in the low temperature limit, by the Arrhenius
law (\ref{eq:Arrhenius}). We checked this fact by Monte Carlo
simulations.  Some results  are presented in Fig.~\ref{fig_singlesample}.
On a given sample $\cG$ of size $N=10^4$, and constraints density 
$\g=0.6$, we measured the correlation functions $C_i(t)$ for 
$20$ randomly chosen 
spins, at various temperatures. We generated the
equilibrium initial configurations efficiently using the property
proved in App.~\ref{app:ProofProp}.
In order to avoid the slowing down of usual
Metropolis-type algorithms in the low-temperature limit, we implemented an
$n$-fold Monte Carlo algorithm~\cite{nfold}. A few examples of correlation
functions are shown in the left panel of Fig.~\ref{fig_singlesample}.
We estimated the auto-correlation time $\tau_i$
from  $C_i(\tau_i)=1/2$. The behaviour of $\tau_i$ as a function
of the temperature is plotted in the right panel of the same figure
for $20$ different sites. 
As expected, $\log \tau_i(\beta)$ grows linearly
with $\beta$ (with integer slope) at small enough temperature.
The barriers $b_i$, computed with the algorithm of this 
Section, are in perfect agreement with the slopes. 
An interesting open problem would be 
the computation of the prefactor in the Arrhenius law, which might be related 
to the degeneracy of minimal barrier rearrangements.

The example of Fig.~\ref{fig_binary_counter} shows that the 
disjoint combination algorithm does not predict the correct
barrier for a general tree\footnote{On the other hand, it has been proved
\cite{Lengauer} that the disjoint estimate is always within a factor 
$2$ from the correct barrier.}.
However balanced trees are quite special. The relevant question is whether
for the random trees appearing in our model, the disjoint
strategy yields in most cases the correct prediction. In particular we
are interested in the  critical limit $\g \to \g_{\rm d}$.
Is the large rearrangements divergence 
$b\sim -\Upsilon \log(\g_{\rm d}-\g)$ modified when barriers
are computed adopting the exact Yannakakis strategy?

Unfortunately, the intricate character of the cost functions in 
Yannakakis algorithm forbids an analytical study of the corresponding 
barrier distribution along the lines of Secs.~\ref{sec:ProbabilisticBarrier} and
\ref{sec_minbarr_as}.  We thus resorted to a numerical computation of 
$Q_b^{\rm site,Y}$ by running the message passing
algorithm described above on large samples. 

\begin{figure}
\includegraphics[width=9cm]{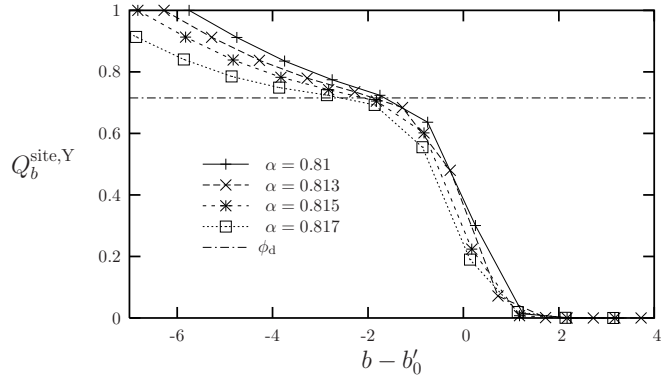}
\caption{The large scale part of the integrated distribution of barriers 
computed with Yannakakis algorithm.}
\label{fig_yanna}
\end{figure}

The distribution $Q_b^{\rm site,Y}$  obtained in this way has the same 
qualitative behaviour as $Q_b^{\rm site}$ (cf. Fig.~\ref{fig_minbarr}).
As $\g\to\g_{\rm d}$, a plateau 
develops at  $Q_b^{\rm site,Y}\approx \phi_{\rm d}$, whose length diverges at 
$\g_{\rm d}$.
The main consequence of passing from disjoint to Yannakakis
strategy is  a finite (as $\g\to\g_{\rm d}$) shift of the cumulative 
distributions towards smaller $b$'s. We plot in Fig.~\ref{fig_yanna}
$Q_b^{\rm site,Y}$ for several values of $\g$ as a function of
$b-b_0'(\g)$, where $b'_0 = - \Upsilon \log(\g_{\rm d}-\g)$ 
is the typical size of large barriers within the disjoint strategy,
cf Eq.~(\ref{eq_minbarr_bp0}). The 
good collapse of the part of the curve under the plateau 
supports the claim that the disjoint strategy is asymptotically 
optimal as $\g\to\g_{\rm d}$ and that the true barriers diverge indeed 
with the same law $- \Upsilon \log(\g_{\rm d}-\g)$.
Further numerical evidence of this claim will be discussed in 
Sec.~\ref{ScalingSection}.
%
%
\subsection{Lower bound on the minimal barriers}
\label{sec:LowerBarrier}

Starting from Sec.~\ref{sec:GeneralBarr}, we assumed so far that 
minimum barriers can be computed restricting to trajectories
such that each spin is flipped at most once along any of them
(`one-spin-flip' trajectories).
The numerical data presented in the previous Section give some credibility
to this hypothesis. If, on the other hand, the hypothesis were 
not correct, the calculations of the previous pages would only provide an 
upper bound on the barrier size.

In this Section we will show that, removing this simplifying
hypothesis, we can still construct a lower bound on the barrier
size. As a consequence, we can control the error introduced by
restricting to one-spin-flip trajectories. While the lower bound is not
tight, it gives further support to the claim that
the behavior $b\simeq -\Upsilon\log(\g_{\rm d}-\g)$ for the
typical barrier of large scale rearrangements is indeed exact.

We start by recalling that, as argued in Sec.~\ref{sec_def_rearr}, 
we can restrict
ourselves to the case of a tree graph $\cG$ (eventually by considering a
finite neighborhood of the site $i$). 
For simplicity of exposition,
we shall first assume that the rearrangement $\R_i$ is fixed (later on
we will have to optimize over $\R_i$).
As in Sec.~\ref{sec_minbarr}, we can therefore focus on the problem of
computing (a lower bound on) the energy barrier for flipping
all the spins of an Ising model on a tree $T$ rooted at $i$. Let 
$\sigma$ be a configuration of such a model and $E(\sigma)$ the
number of edges joining a plus and a minus spin in $T$ (one half of 
the usual energy). The barrier $b_i$ was originally defined as 
the minimum over all the trajectories $\{\sigma(t)\}$
joining the two ground states $\sigma^{(+)}$ and $\sigma^{(-)}$, 
of the maximum along the trajectory of
$E(\sigma(t))$. 

An equivalent (dual) definition can be obtained as follows.
Let $\kappa:\{\pm 1\}^T\to\{\pm 1\}$ be a function of the configuration 
$\sigma$ such that $\kappa(\sigma^{(+)})=+1$ and $\kappa(\sigma^{(-)}) = -1$.
Furthermore, let $\partial\kappa$ be the set of configurations $\sigma$
such that $\kappa(\sigma) = +1$ and $\kappa(\sigma^{(i)}) = -1$ for
some $i$. Here $\sigma^{(i)}$ is the configuration obtained from $\sigma$ by
flipping the $i$-th spin. It is easy to show that
\begin{eqnarray}
b_i = \max_{\kappa}\min_{\sigma\in\partial\kappa} E(\sigma)\, .
\label{eq:BarrierDualDef}
\end{eqnarray}

In order to construct a lower bound on $b_i$, we just need to 
exhibit a function $\kappa(\sigma)$. Computing the minimum over 
$\sigma\in\partial\kappa$ is of course quite difficult for a generic
function $\kappa(\sigma)$. Our plan is to define $\kappa(\sigma)$
recursively, exploiting the tree structure of the graph. This will allow
for a recursive evaluation of the minimum. More precisely,
for each site $j\in T$, we will introduce a function $\kappa_j(\sigma)$
depending on the values of $\sigma_j$ and its descendants. 
At any non-leaf sites $i$, $\kappa_i(\sigma)$ will be defined as a function
of $\{\kappa_{j}(\sigma)\}$, with $j$ running over the sons of $i$. 
For any leaf site $j$, we have $\kappa_j(\sigma)=\sigma_j$.
At the end we  set $\kappa(\sigma) = \kappa_i(\sigma)$. 
 
We are left with the task of choosing the recursion which determines
$\kappa_i$ in terms of the functions at its sons (to be denoted 
generically as $j$). A simple (but reasonably good choice) is
$\kappa_i(\sigma) = \Maj_{r_i,\Omega_i}(\{\kappa_j(\sigma)\})$, where
\begin{eqnarray}
\Maj_{r,\Omega}(\kappa_1,\dots,\kappa_l) = \left\{
\begin{array}{ll}
+1 & \mbox{if $\sum_{j\in\Omega}\kappa_j\ge l-2r$,}\\
-1 & \mbox{otherwise.}
\end{array}
\right.
\end{eqnarray}
Here $r$ is a non-negative integer and $\Omega$ a subset of $\{1,\dots,l\}$.
The  parameters $r_j$, and the sets $\Omega_j$ can be chosen in 
order to optimize the bound.
They will depend on the tree $T$, but not on the configuration $\sigma$.

The lower bound on the barrier implied by this choice of the function 
$\kappa$ can be computed recursively. Let $T_j$ be the subtree formed by
$j$ and its descendants. Denote by $b_j(+/-)$ the 
lower bound on the barrier for the subsystem $T_j$ under the condition that
$\sigma_j=+/-$ (in other words, the minimum in 
Eq.~(\ref{eq:BarrierDualDef}) is taken only over the configurations 
satisfying this condition). A moment of thought shows that
\begin{eqnarray}
b_i(+) & = & r_i+\min_{j\in\Omega_i}\big[\, b_j(+),\; b_j(-)+1\, \big]\, ,
\label{eq:Lower1}\\
b_i(-) & = & (n_i-r_i-1)+\min_{j\in\Omega_i}
\big[\,b_j(+)+1, \; b_j(-)\, \big]\, ,
\label{eq:Lower2}
\end{eqnarray}
where $n_i = |\Omega_i|$ is the number of descendants of $i$ which are 
involved in the definition of the function $\kappa_i(\sigma)$.

First notice that, by changing $r_j\leftrightarrow n_j-r_j-1$ for
all the sites $j\in T_i$, one  changes $b_i(+)\leftrightarrow b_i(-)$.
We can therefore assume, without loss of generality, that 
$b_i(+)\le b_i(-)$.
In fact it is not necessary to keep track of all the information
used in Eqs.~(\ref{eq:Lower1}), (\ref{eq:Lower2}).
We let $b_i = \min [ b_i(+),\; b_i(-)]$,
and define $\delta_i=0$ if $b_i(+) = b_i(-)$ and $=1$ otherwise.

The optimal choice of the parameters $r_i$'s is obtained as follows.
Suppose $\Omega_i$ is given
and let $b_{{\rm min},\Omega_i}= \min\{ b_j\, ; \, j\in \Omega_i\}$.
If $\delta_j = 1$ for all the $j\in\Omega_i$ such that 
$b_j = b_{{\rm min},\Omega_i}$
(in this case we will say that $\delta_{{\rm min},\Omega_i}=1$),
then we set $r_i = \lfloor\frac{n_i}{2}\rfloor$. In the opposite 
case (which we shall denote as $\delta_{{\rm min},\Omega_i} = 0$), 
the optimal choice is $r_i = \lfloor\frac{n_i-1}{2}\rfloor$.
Using these choices in Eqs.~(\ref{eq:Lower1}) and 
(\ref{eq:Lower2}), we get
\begin{eqnarray}
b_i  =  
b_{{\rm min},\Omega_i}+\left\lfloor\frac{n_i-1+\delta_{{\rm min},\Omega_i}}{2}
\right\rfloor\, ,
\;\;\;\;\;\;\;\;\;\;\;\;\;\;\;\;\;\;\;
\delta_i  =  \left\{
\begin{array}{ll}
1-\delta_{{\rm min},\Omega_i} & \mbox{for $n_i$ even,}\\
\delta_{{\rm min},\Omega_i} & \mbox{for $n_i$ odd.}
\end{array}
\right.\label{eq:LowerBoundFirst}
\end{eqnarray}
It is useful to notice that a natural order relation can be defined
on the couples $(b_i,\delta_i)$. We will say that
$(b,\delta)\preceq (b',\delta')$ if $b<b'$ or $b=b'$ and $\delta\le \delta'$.
It is simple to show that $(b_{{\rm min},\Omega_i},
\delta_{{\rm min},\Omega_i}) = \min\{ (b_j,\delta_j)\, ; \, j\in\Omega_i\}$ 
if the minimum is taken  with respect to this order relation. 
Furthermore, given $n_i$, the mapping (\ref{eq:LowerBoundFirst})
is monotonously increasing with respect to the same relation. 
We can therefore optimize over $\Omega_i$, at fixed $n_i$ as follows.
First order the decendants of $j$ by decreasing $(b_j,\delta_j)$.
Then form $\Omega_i$ by retaining
the first $n_i$ items of this list. If the list of ordered
couples is denoted as $\{(b_{[1]},\delta_{[1]});\, (b_{[2]},\delta_{[2]});\,
\dots\}$, then the optimized recursion is obtained by replacing 
$b_{[n_i]}$ to $b_{{\rm min},\Omega_i}$, and 
$\delta_{[n_i]}$ to $\delta_{{\rm min},\Omega_i}$ in 
Eq.~(\ref{eq:LowerBoundFirst}). We can finally optimize over $n_i$
to get
\begin{eqnarray}
(b_i,\delta_i) = \max_{1 \le n\le l_i}\left(
b_{[n]}+\left\lfloor\frac{n-1+\delta_{[n]}}{2}\right\rfloor
,\; (1-\delta_{[n]})\,\ind\{n\, \mbox{even}\}+\delta_{[n]}\ind\{n\,
\mbox{odd}\}\right)\, ,\label{eq:LowerBoundFinal}
\end{eqnarray}
where $l_i$ is the number of descendants of $i$, and the $\max$ is taken once
more with respect to the order relation $\preceq$.

As a simple application, one may consider the case of a $k$-generations
binary tree. We saw at the beginning of Sec.~\ref{sec:Yanna} that the disjoint
combination strategy overestimate the barrier in this case by a multiplicative
factor which approaches $2$ as $k\to\infty$. If 
we call $(b_k,\delta_k)$ the values of $b_i$ and $\delta_i$ at the 
root of such a tree Eq.~(\ref{eq:LowerBoundFinal}) implies the following 
recursion: $b_{k+1} = b_k+\delta_k$, $\delta_{k+1}= 1-\delta_k$.
This in turns yields $b_k = \lfloor k/2\rfloor$ which is essentially tight.

In applying the above method to the $p$-spin model (\ref{Hamiltonian})
we must consider the problem of optimizing over the rearrangement 
$\R_i$. Once again, we exploit the fact that the mapping
(\ref{eq:LowerBoundFinal}) is monotonously increasing with 
respect to the order relation $\preceq$. We can therefore
construct recursively $\R_i$
also in this case. For each site $j$ in $\R_i$ and interaction term $\a$, 
at least one more site $k$ adjacent to $\a$ must belong to $\R_i$. 
In fact, it can be argued that the minimal barrier is achieved by
simple rearrangements. We restrict to this case here: {\em exactly}
one more site $k$ (apart from $j$) belongs to $R_i$. This site
can be chosen by minimizing $(b_k,\delta_k)$ with respect to this order 
relation.

The whole procedure can of course be implemented in a message passing style.
Messages $u_{\a\to i}$ and $v_{i\to \a}$ are now 
couples $(b,\delta)$. They satisfy recursion relations of the form
(\ref{eq:MinBarr_uv}), with  $\hat{f}_{l}(\,\cdot\,)$ being 
replaced by the mapping (\ref{eq:LowerBoundFinal}). 
Also the probabilistic analysis can be carried out along 
the lines of Sec.~\ref{sec:ProbabilisticBarrier}. The only difference 
is that, due to the new degree
of freedom $\delta$, one has to keep track of joint distributions
$Q_{\delta,b}$, with $\delta=0,1$. The critical 
behavior of these distributions for $\g\to\g_{\rm d}$ is very close
to the one obtained in Sec.~\ref{sec_minbarr_as} within the 
disjoint strategy. The main difference is that the parameters 
$\omega_{a}$ and $\omega_b$ are replaced by $\omega'_a$ and 
$\omega'_b$ with $\omega'_{a/b}= 2\omega_{a,b}$.
As a consequence the lower bound on the typical barrier for large 
rearrangements is
$b_{\rm lb}\simeq -\frac{1}{2}\Upsilon
\log(\g_{\rm d}-\g)$. Together with the disjoint 
strategy analysis, this provides a rigorous proof that
$b = O{\bf (}- \log(\g_{\rm d}-\g){\bf )}$.

As stressed above, we do not think that the lower bound described in this 
Section is tight, even in the critical regime. It should be possible to
improve it at the price of a more intricate recursive definition of 
the functions $\kappa_i(\sigma)$ (in the definition used here, 
$\kappa_i(\sigma)$
depends only on the spins at the leaves of the rearrangement). 
On the other hand, it provides some support
for the conjecture that the behavior
$b\simeq -\Upsilon\log(\g_{\rm d}-\g)$ is asymptotically exact.
%
\section{Geometrical properties of optimal rearrangements}
\label{GeometricalSection}

In the previous Sections we studied two types of {\em optimal}
rearrangements: minimal size, and minimal barrier ones. Several
refined properties can be defined and studied. 
First of all, what is the trade off between the two criteria of
optimality? For instance, how larger are the minimal barrier 
rearrangements compared to the minimal size ones? 
Moreover, thanks to the underlying  Bethe lattice, some geometrical
properties of the rearrangements can be introduced.
For instance, one can study the distribution of the distances between 
the root of a rearrangement and the sites it contains. 

Finally, the diverging rearrangement size (and barrier) as 
$\g\to\g_{\rm d}$ implies that dynamics becomes more 
and more cooperative in this regime.
One possible characterization of correlations induced by the
increasing cooperativity is the following. 
Consider
a variable $i$ and the minimal size of its rearrangements, $n_i$. Imagine now
that a very strong external field is applied on another variable $j$,
preventing it from being flipped, and define $n_i^{(j)}$ as the minimal size
of the rearrangements for $i$ which exclude the frozen variable $j$. If
the two variables are sufficiently far apart, this additional constraint will
be irrelevant, and one will have $n_i^{(j)}=n_i$. However if $j$ gets closer
to $i$, it may happen that $j$ belonged to all optimal rearrangements for
$i$, which will imply $n_i^{(j)}>n_i$. How close the variables have to be
for this to happen allows to define a ``geometric correlation length".
We shall define a ``geometric response function" related to the difference
$n_i^{(j)}-n_i$, and a susceptibility by summing it over the
positions $i$ and $j$.
This turns out to be a close analog of the four-point correlation
function~\cite{Donati,Bennemann,FranzChi4A,FranzChi4B}, 
used in studies of the structural glass transition.

The above questions are considered in the rest of this Section.
Let us briefly list the main results. The size of minimal barrier 
rearrangements diverges as $(\g_{\rm d}-\g)^{-\nu_{\rm barr}}$
with  non-universal exponent $\nu_{\rm barr}>\nu$
(recall that $\nu$ is the analogous exponent for {\em minimal size} 
rearrangements).
The typical distance between the root
and a random node in a large rearrangement diverges as 
$(\g_{\rm d}-\g)^{-\zeta}$, with {\em universal} exponent 
$\zeta = 1/2$. Finally, the number of sites which are influential
for the optimal rearrangements of site $i$ (as measured through the above
susceptibility) diverges as $(\g_{\rm d}-\g)^{-\eta}$ with 
universal exponent $\eta = 1$.
%
%
\subsection{Average size of minimal barrier rearrangements}
\label{sec:SizeMinBarr}

\begin{figure}
\includegraphics[width=6cm]{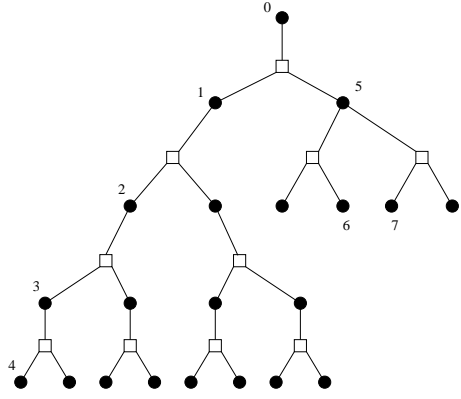}
\caption{An example where minimal barrier rearrangements and minimal sizes ones
are distinct.}
\label{fig_ex_contradiction}
\end{figure}

How large are minimal barrier rearrangements? Consider the simple example
of Fig.~\ref{fig_ex_contradiction}. The minimal barrier rearrangements
for the site $0$ have barrier 1 and size 5 (one of them is obtained  
by flipping the variables $0,1,2,3,4$ in this order). On the contrary the
minimal size rearrangements (for instance, the one formed by sites $0,5,6,7$) 
have barrier 2 and size 4.

This example shows that minimal barrier
rearrangements can be  strictly larger that minimal size ones.
It is however not clear whether the same remains true if
we focus on the critical behavior as $\g\to\g_{\rm d}$. 
To keep the level of difficulty of the computations at an acceptable level,
we compute barriers within the disjoint strategy, 
and assume it to be optimal in the critical regime. 

For a given site $i$, there can be several minimal barrier rearrangements 
of different sizes. We can resolve this ambiguity as follows. Distinct
rearrangements with equal barrier occur because the choice of
which spin to include at a given function node --i.e., the 
minimum in Eq.~(\ref{eq:MinBarr_uv})-- can be degenerate.
A random minimal barrier rearrangement~\footnote{Notice that this
procedure {\em is not} equivalent to picking a rearrangement 
uniformy at random among the ones of minimal size. However, it is 
more convenient for calculations. We expect the critical behavior
not to be affected by this choice.} can be sampled by removing such 
degeneracies, each time they occur, uniformly at random.
We call $m_i$ (to stress the difference with the minimal size
$n_i$) the average size of random minimal barrier rearrangements 
sampled in this manner.

The computation of $m_i$ can be performed recursively, exploiting the
local tree structure of the graph $\cG$. To this aim we supplement the messages
$u_{\a \to i}$, $v_{i \to \a}$ used to computed the barrier $b_i$,
cf. Eq.~(\ref{eq:MinBarr_uv}),  with messages 
$u'_{\a \to i}$, $v'_{i \to \a}$ defined as $m_i$ for
the corresponding cavity graphs. For instance, $v'_{i \to \a}$ is the
average size of the rearrangements of minimal barrier rooted at site $i$ in
the graph deprived of the interaction $\a$. The new messages
obey the following recursions:
\begin{eqnarray}
v'_{i \to \a} = 1+ \sum_{\b \in \partial i \setminus \a}
u'_{\b \to i} \, ,\;\;\;\;\;\;\;
u'_{\a \to i} &=& \frac{1}{{\cal N}_{\a \to i}} 
\underset{\underset{v_{j \to \a} = u_{\a \to i}}
{j \in \partial \a \setminus i} 
}{\sum}  v'_{j \to \a}
\, ,
\label{eq_joint_up_of_vp}
\end{eqnarray}
where ${\cal N}_{\a \to i}$ is the number of sites 
$j \in \partial \a \setminus i$ such that $v_{j \to \a} = 
u_{\a \to i}$ (in other words, the degeneracy of the minimum in 
Eq.~(\ref{eq:MinBarr_uv})).
The sum in the second equation corresponds to an uniform
average over the degeneracy at node $\a$, as discussed above. 
Finally, we have
\begin{equation}
m_i = 1+ \sum_{\a \in \partial i} u'_{\a \to i} \, .
\end{equation}

For random hypergraphs, the above recursions acquire
a distributional meaning. Note that the messages $u$ and $u'$ (or $v$ and
$v'$) on the same link are correlated. As messages $u'$ and $ v'$ can take 
non-integer values, we  will define
\begin{equation}
q_{b,m} \de m = \prob\left\{
v_{i \to \a}=b , v'_{i \to \a} \in [m,m+\de m]\right\}
\, ,
\end{equation}
Analogously $\qh_{b,m}$ is the joint distribution of
$u_{\a\to i}$ and $u'_{\a\to i}$. 
Marginalizing over $m$, one recovers the distributions
studied in Sec.~\ref{sec_minbarr}: 
$\int\!\de m\; q_{b,m}=q_b$.
The explicit distributional equations read:
\begin{eqnarray}
q_{b,m} &=& \sum_{l=0}^\infty p_l
\sum_{b_1,\dots,b_l} \int \prod_{j=1}^{l}
\de m_j \ \qh_{b_j,m_j}\;\; \delta_{b,\hat{f}_l(b_1,\dots,b_l)}
\delta(m-1-m_1-\dots-m_l) \ , \label{eq:SizeBarr0a}\\
\hat{q}_{b,m} &=& \sum_{k=1}^{p-1} {p-1 \choose k}
\int \prod_{j=1}^{k} \de m_j \ q_{b,m_j} \; Q_{b+1}^{p-1-i}\;\;
\delta \left(m - \frac{m_1+\dots+m_i}{i}\right)  \ .\label{eq:SizeBarr0b} 
\end{eqnarray}
Consider the partial averages $M_b = \int \de m \ q_{b,m} m$ and  
$\Mh_b = \int \de m \ \hat{q}_{b,m} m$.
Notice that $M_b/q_b$ is the average size of minimal barrier rearrangements,
conditioned on their barrier\footnote{To be more precise, the 
conditioning is on having an anchored barrier of $b$. This distinction does not
change  the critical behavior.}
 $b$. Equations (\ref{eq:SizeBarr0a}),
(\ref{eq:SizeBarr0b}) imply
\begin{eqnarray}
M_b &=& q_b + \sum_{l=1}^\infty p_l l \sum_{b_1,\dots,b_l} \Mh_{b_1}
\hat{q}_{b_2} \dots \hat{q}_{b_l} \delta_{b,\hat{f}_l(b_1,\dots,b_l)} \ , 
\label{eq_sizebarr_1} \\
\Mh_b &=& M_b \frac{\hat{q}_b}{q_b} \label{eq_sizebarr_2} \ .
\end{eqnarray}

\begin{figure}
\includegraphics[width=8cm]{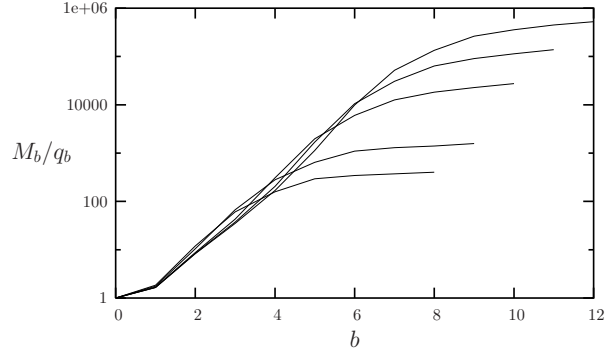}
\caption{The average size of minimal barrier rearrangements. From bottom to 
top, $\g=0.7,0.75,0.8,0.81,0.814$.}
\label{fig_sizebarr}
\end{figure}

These equations can be solved numerically by iteration, we defer the detailed
presentation of the method to App.~\ref{sec_app_sizeminbarr}. 
In Fig.~\ref{fig_sizebarr} we plot, for a few values of $\g$ approaching
its critical value, the quantity $M_b/q_b$ as a function of $b$. 

Two scaling regimes can be distinguished. For barriers of intermediate
size , $b\simeq -\frac{1}{2\omega_a}\log(\g_{\rm d}-\g)$, 
the average size $M_b/q_b$ grows exponentially with $b$. In the
large barrier regime $b\simeq -\Upsilon\log(\g_{\rm d}-\g)$,
the size saturate at $(M_b/q_b)\sim (\g_{\rm d}-\g)^{-\nu_{\rm barr}}$.
A careful asymptotic analysis reveals that
\begin{eqnarray}
\nu_{\rm barr} =\frac{\mu_a}{2\omega_a}+\frac{\mu_b}{2\omega_b}\, ,
\end{eqnarray}
where $\mu_{a/b}$ are positive solutions of the equations
\begin{eqnarray}
\frac{(e^{\mu+\omega}-1)(e^{-\omega}+e^{-\mu}-e^{-\omega-\mu})}
{(e^{\mu}-1)(e^{\omega}-1)} = \frac{\lambda}{2}\, ,
\end{eqnarray}
with $\omega = -\omega_a$ (for $\mu_a$) and $\omega=\omega_b$ (for $\mu_b$).
We obtain for instance $\nu_{\rm barr}\approx 2.0157$ for $p=3$ 
(cf. Table~\ref{TableBarrier} for the other values of $p$), 
which is strictly larger than the exponent for minimal size rearrangements, 
$\nu \approx 1.59379$.

Let us conclude by noticing that the ``inverse" computation,
namely computing the energy barrier for minimal size rearrangements,
is technically more challenging. However we expect the result to be 
analogous to the one derived here: the barrier for minimal size rearrangements
should diverge as $-\Upsilon_{\rm size}\log (\g_{\rm d}-\g)$ with
$\Upsilon_{\rm size}$ strictly larger than $\Upsilon$.
%
%
%
\subsection{Average depth of minimal size rearrangements}

We turn now to a study of the shape of optimal rearrangements. 
We define the depth of a rearrangement $\R_i$ as the sum of the 
distances of the sites $j\in\R_i$ from $i$. 

For a given graph $\cG$, consider a site $i$ with  minimal 
rearrangement size $n_i$, and call $t_i$ the 
average~\footnote{Here the average is taken with the same procedure described 
in Sec.~\ref{sec:SizeMinBarr}. Each time the minimum in 
Eq.~(\ref{eq:MinSize_uv}) is degenerate, one of the optimal branches 
is chosen uniformly at random.} depth of its optimal rearrangements. With this
definition, $t_i/n_i$  is the average distance between $i$ and  
sites in the optimal rearrangements for $i$. 
We shall call this quantity the ``radius" of 
optimal rearrangements.

\begin{figure}
\includegraphics[width=9cm]{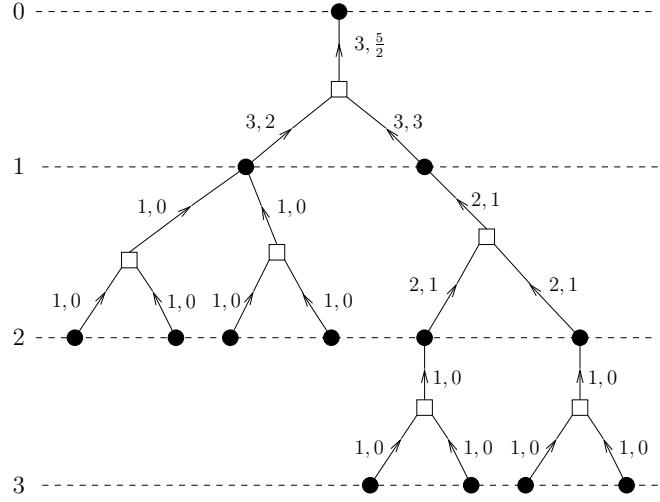}
\caption{An example of the computation of $t_i$. The numbers on the left
are the distances of the sites from the root. The value
of the messages $u,u'$ (resp. $v,v'$) are indicated next to the edges from
interactions to sites (resp. from sites to interactions). 
They verify the recursion equations 
(\ref{eq:MinSize_uv}) and (\ref{eq_minsize_up_of_vp}).}
\label{fig_ex_msg_Tn}
\end{figure}

Let us emphasize that
in general rearrangements of same size but different depths can coexist. A
simple example of this phenomenon is provided in Fig.~\ref{fig_ex_msg_Tn}.
Among the eight distinct optimal rearrangements (of size $n=4$) for the 
root, four are of depth 5 (they include the left branch in the first 
interaction) and four of depth 6 (right branch). All of them
are sampled with equal probability according to our procedure, therefore 
$t=11/2$, and the radius is $t/n=11/8$.

As sketched in Fig.~\ref{fig_ex_msg_Tn}, the quantity $t_i$ can
be computed in a recursive manner: 
along with the minimal size messages $u_{\a \to i}$
and $v_{i \to \a}$ which verify the recursion equations 
(\ref{eq:MinSize_uv}),  we introduce  
supplementary messages on every directed edge of the factor graph, 
$u'_{\a \to i}$
and $v'_{i \to \a}$. They are defined as $t_i$,
but refer to cavity graphs. For instance, 
$v'_{i \to \a}$ is the average depth of
the optimal rearrangements rooted in $i$ in the graph where the interaction 
$\a$ has been removed. They obey the  recursions:
\begin{eqnarray}
v'_{i \to \a} = \sum_{\b \in \partial i \setminus \a}
[ u_{\b \to i} + u'_{\b \to i}] \, ,
\;\;\;\;\;\;\;
u'_{\a \to i} &=& \frac{1}{{\cal N}_{\a \to i}} 
\underset{\underset{v_{j \to \a} = v_{\a \to i}}
{j \in \partial \a \setminus i} 
}{\sum}  v'_{j \to \a}
\ , 
\label{eq_minsize_up_of_vp}
\end{eqnarray}
where ${\cal N}_{\a \to i}$ is the number of sites 
$j \in \partial \a \setminus i$ such that $v_{j \to \a} = 
u_{\a \to i}$ (the degeneracy of the minimum in 
Eq.~(\ref{eq:MinSize_uv})).
In the first equation, the term $u_{\b \to i}$ arises because the
distance of a descendant of $i$ from the parent of $i$
is larger than its distance from $i$ by one.
Finally the site quantity can be computed in terms of the incoming messages:
\begin{equation}
t_i = \sum_{\a \in \partial i} [ u_{\a \to i} + u'_{\a \to i}] \ .
\end{equation}
The reader is invited to check the correctness of this procedure on the
example of Fig.~\ref{fig_ex_msg_Tn}.

Until now the computation has been performed for a site $i$ in a given sample $\cG$.
As in the previous Section, for a random graph $\cG$, the distribution 
of $t_i$ can be determined from the joint law
\begin{equation}
q_{n,t}\, \de t = \prob\left\{
v_{i \to \a}=n ,\, v'_{i \to \a}\in [t,t+dt]\right\} 
\ .
\end{equation}
Analogously, $\qh_{n,t}$ will denote here the joint distribution of
$u_{\a\to i}$ and $u'_{\a\to i}$.
Because of the memoryless property of Poisson random variables, $q_{n,t}$ is
also the joint distribution of  $n_i,t_i$. 
The recursion relations on the messages are easily translated into equations
for  $q_{n,t}$ and $\qh_{n,t}$:
\begin{eqnarray}
q_{n,t} &=&  \sum_{l=0}^\infty p_l
\sum_{n_1, \dots, n_l} \int\! \prod_{j=1}^{l}\de t_j\, \qh_{n_j,t_j} \;\;
\delta_{n,1+n_1+\dots+n_l}\; \delta(t-n_1-t_1-\dots-n_l-t_l)
\ , \label{eq_depth_qofqhat}\\
\hat{q}_{n,t} &=& \sum_{i=1}^{p-1} {p-1 \choose i}
\int 
\! \prod_{j=1}^{l}\de t_j\, q_{n_j,t_j} \;Q_{n+1}^{p-1-i}
\;\; \delta \left( t- \frac{t_1+\dots+t_i}{i}\right) 
\ . \label{eq_depth_qhatofq}
\end{eqnarray}

By definition
$\int \!\de t \ q_{n,t}=q_n$, is the distribution of
minimal sizes studied in 
Sec.~\ref{sec_minsize}. Hence $q_{n,t}/q_n$
is the distribution (with respect to the graph $\cG$, and the site $i$) of the 
average depth $t$ of optimal rearrangements, conditioned on their size $n$.
For large scale rerrangements of diverging size in the critical limit 
$\g \to \g_{\rm d}$, we expect this law to have a scaling form
\begin{eqnarray}
q_{n,t}/q_n \approx 
(\g_{\rm d}-\g)^{\zeta}\;{\cal S}\{
(\g_{\rm d}-\g)^{\zeta}t\,|\;(\g_{\rm d}-\g)^{\nu}n\}          
\end{eqnarray}

For  the sake of simplicity (and with the aim of determining the exponent
$\zeta$) we shall concentrate on the partial averages
$T_n = \int \de t \ t \, q_{n,t}$ and $\Th_n = \int \de t \ t \, 
\hat{q}_{n,t}$.
Hence $T_n/(n q_n)$ is the average radius of optimal rearrangements 
of size $n$. Equations (\ref{eq_depth_qofqhat}) and 
(\ref{eq_depth_qhatofq}) imply
\begin{eqnarray}
T_n &=& \sum_{l=1}^\infty p_l\, l \sum_{n_1, \dots, n_l} 
(\Th_{n_1}+n_1 \hat{q}_{n_1}) \hat{q}_{n_2} \dots \hat{q}_{n_l}\;
\delta_{n,1+n_1+\dots+n_l} \ , \\
\Th_n &=& T_n \frac{\qh_n}{q_n} \ . 
\end{eqnarray}
To lighten the notation, let us define 
$\Tt_n = \Th_n + n \hat{q}_n$. We finally obtain:
\begin{eqnarray}
T_n = \g p \sum_m q_{n-m} \Tt_m \ , \;\;\;\;\;\;\;
\Tt_n = T_n \frac{\qh_n}{q_n} + n \qh_n \ . \label{eq_depth}
\end{eqnarray}

\begin{figure}
\includegraphics[width=8cm]{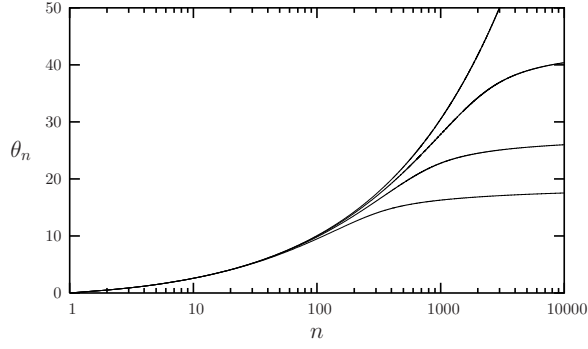}
\caption{The average radius $\theta_n$ in optimal rearrangements of
size $n$, from bottom to top $\g=0.8, 0.81, 0.815, \g_{\rm d}$.}
\label{fig_depth_res}
\end{figure}

As was the case for the distribution $q_n$, the sequence $T_n$
can be efficiently determined numerically by recursion on $n$, the
initial value being $T_1=0$. 
We present in Fig.~\ref{fig_depth_res} the output of
such a numerical computation. For clarity we plotted  
the average radius $\theta_n = T_n /(n q_n)$ as a function of the
size $n$.

Its critical behavior can be determined analytically. The details
of the computations are presented in App.~\ref{sec_app_depth}.
As for the distribution of minimal sizes $Q_n$, two scaling regimes
can be determined. For intermediate sizes, 
$n\sim (\g_{\rm d}-\g)^{-1/2a}$, the radius grows like a 
power law of $n$. In the large size rearrangements regime,
$n\sim (\g_{\rm d}-\g)^{-\nu}$, the radius saturate
at $\theta\sim (\g_{\rm d}-\g)^{-\zeta}$, 
where $\zeta =1/2$ is universal. This is consistent 
(as it should be) with the divergence of the point-to-set correlation length
defined in Sec.~\ref{sec:PointToSet}. 

It is not hard to realize that the same scaling 
$(\g_{\rm d}-\g)^{-1/2}$
must hold for the radius of large minimal barrier rearrangements. 
This is in fact the radius below which optimal rearrangements on site 
$i$ are insensitive to perturbations on other sites, see also next Section.
It  is also related to the number of iterations necessary for
iterations of the form (\ref{eq_minbarr_distrib})
to converge to their fixed point.

For large rearrangements, eliminating the dependency
upon $\g$ yields the scaling $\theta_n \sim n^{1/2\nu}$. This 
can be compared with the case of uniformly random trees, which can be 
interpreted as a mean field model for ordinary bond percolation.
In this case it is well known~\cite{AthreyaNey} that
$\theta_n\sim n^{1/2}$. Since $\nu>1$ always, large scale rearrangements 
are `denser' than percolation cluster.
%
%
\subsection{Geometrical susceptibility}
\label{sec:GeometricalSusceptibility}

\subsubsection{Minimal size rearrangements}

\begin{figure}
\includegraphics[width=10cm]{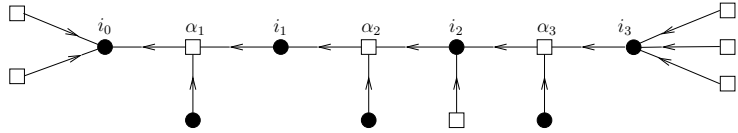}
\caption{A chain of $r=3$ interactions for the computation
of geometrical susceptibility.}
\label{fig_chi_chain}
\end{figure}

We turn now to the question of dynamical correlations induced
by optimal rearrangements, cf. the discussion at the beginning of this
Section. 
Consider the portion of a graph sketched in Fig.~\ref{fig_chi_chain},
where we isolated the root site $i_0$ and a chain of $r$ ($r=3$ in the
figure) sites $i_1,\dots,i_r$ linked by interactions 
$\a_1,\dots,\a_r$. We define  $n_{i_0}^{(i_r)}$ as the minimal size
of rearrangements for $i_0$ which do not include the variable $i_r$  
(one can imagine that a pinning external field has been applied on 
$\sigma_{i_r}$ thus preventing it from flipping). 
This is to be compared with 
$n_{i_0}$, the unconstrained minimal size of rearrangements for the root $i_0$.
The unconstrained size $n_{i_0}$ can be computed through the message passing
procedure described in Sec.~\ref{sec_minsize},  which makes use
of messages  $u_{\a \to i}$, $v_{i \to \a}$. 
In order to compute $n_{i_0}^{(i_r)}$, we introduce,
for each edge of the chain directed towards $i_0$, 
supplementary messages $u'_{\a_k \to i_{k-1}}$ 
and $v'_{i_k \to \a_k}$. 
The edges of the graph outside the chain bears the usual single message
(either $u_{\a \to i}$ or $v_{i \to \a}$).

The unconstrained (resp. constrained) size is given as follows in terms of 
messages:
\begin{equation}
n_{i_0} = 1 + u_{\a_1 \to i_0} + \sum_{\b \in \partial i_0 \setminus
\a_1} u_{\b \to i_0} \ , \;\;\;\;\;\;\;\;
n_{i_0}^{(i_r)} = 
1 + u'_{\a_1 \to i_0} + \sum_{\b \in \partial i_0 \setminus
\a_1} u_{\b \to i_0} \ .
\end{equation}
While the messages $u_{\a\to i}$, $v_{i\to\a}$ obey the usual 
relation (\ref{eq:MinSize_uv}) for minimal size rearrangements, for 
the new messages we have:
\begin{eqnarray}
v'_{i_k \to \a_k} &=& 1 + u'_{\a_{k+1} \to i_k} + 
\sum_{\b \in \partial i_k \setminus\{\a_k,\a_{k+1} \}} 
u_{\b \to i_k} \ , \\
u'_{\a_k \to i_{k-1}} &=& \min \left[ v'_{i_k \to \a_k} , 
\underset{j\in \partial \a_k \setminus \{ i_k, i_{k-1} \}}{\min} 
v_{j \to \a_k }   \right] \ .
\end{eqnarray}
Notice that these equations are formally identical to 
Eq.~(\ref{eq:MinSize_uv}).
The constraint of excluding $i_r$ from rearrangements whose minimal size is
$n^{(i_r)}_{i_0}$, is enforced through the boundary condition 
$v'_{i_r \to \a_r} =\infty$. In this way, the
variable $i_r$ is never included in the rearrangement. Equivalently
\begin{equation}
u'_{\a_r \to i_{r-1}} = 
\underset{j\in \partial \a_r \setminus \{ i_r, i_{r-1} \}}{\min} 
v_{j \to \a_r } \ .
\end{equation}

The two quantities $n_{i_0}$ and $n_{i_0}^{(i_r)}$ 
(as well as the messages $u_{\a\to i}$, $u'_{\a\to i}$ or
$v_{i\to\a}$, $v'_{i\to\a}$) are correlated, as both of them 
see the same ``environment".
We must therefore keep track of the probability distributions
\begin{equation}
q_{n,n'}^{(r-t)} = 
\prob\{v_{i_t\to\a_{t}} = n,\, v'_{i_t\to\a_t}=n'\}
\ ,
\end{equation}
and the analogous one (denoted as $\qh_{n,n'}^{(r-t)}$) for
messages $u_{\a_{t+1}\to i_t}$, $u'_{\a_{t+1}\to i_t}$.
Here probabilities are implicitly conditioned on the existence of a
chain of interactions of length $r$ between the root $i_0$ and the
constrained site $i_r$. 
Since the variable node degrees are Poisson random variables,
the couple $n_{i_0},n_{i_0}^{(i_r)}$ is distributed according to 
$q^{(r)}_{n,n'}$.
We also define the cumulative distributions
$Q_{n,n'}^{(r-t)}$ and $\Qh_{n,n'}^{(r-t)}$.  The first one is, for instance,
the probability that $v_{i_t\to\a_{t}} \ge n$ {\em and}
$v'_{i_t\to\a_t}\ge n'$. With these definitions, the distributional
equations read 
\begin{eqnarray}
\Qh_{n,n'}^{(r)} & = & Q_{n,n'}^{(r-1)} Q_{\max(n,n')}^{p-2} \, ,
\label{eq:SuscJoint1}\\
q_{n,n'}^{(r)} & = & \sum_{l=0}^{\infty}p_l
\sum_{n_1,n'_1}\sum_{m_1,\dots,m_l} \qh_{n_1,n'_1}^{(r)} \qh_{m_1}\dots 
\qh_{m_l} 
\delta_{n,n_1+m_1+\dots +m_l}\, \delta_{n',n'_1+m_1+\dots +m_l} \, , 
\label{eq:SuscJoint2}
\end{eqnarray}
where $q_n$, $\qh_n$ are the distributions of minimal size rearrangements,
i.e. the solutions of Eqs.~(\ref{eq:MinSizeDistr1}), (\ref{eq:MinSizeDistr2}).

Let $q_{n',{\rm con}}^{(r)}\equiv \sum_{n} q_{n,n'}^{(r)}$ 
(respectively $q_{n,{\rm unc}}^{(r)}\equiv \sum_{n'} q_{n,n'}^{(r)}$)
be the marginal distribution of the constrained 
(unconstrained\footnote{Note that the marginal distribution of unconstrained 
minimal sizes $q_{n,{\rm unc}}^{(r)}\equiv \sum_{n'} q_{n,n'}^{(r)}$
is distinct from $q_n$. In fact, in $q^{(r)}_{n,{\rm unc}}$, we condition
on the existence of a chain of length $r$ from the root.}) sizes. Define
\begin{equation}
s_{n}^{(r)} \equiv q_{n,{\rm con}}^{(r)}-q_{n,{\rm unc}}^{(r)}\, .
\end{equation}
The knowledge of $s_n^{(r)}$ allows to compute the expectation 
of any observable of the form $f(n_{i_0}^{(i_r)})-f(n_{i_0})$. For the sake of
simplicity, we shall hereafter focus on this function,
rather than solving Eqs.~(\ref{eq:SuscJoint1}), (\ref{eq:SuscJoint2})
for the joint distribution of $n^{(i_r)}_{i_0}$, $n_{i_0}$.
Further justification for considering on this type of functions
will be provided in the next Section.

Equations for $s_n^{(r)}$ are more easily written by introducing
the integrated version,
\begin{equation}
S_n^{(r)} = \sum_{n'\ge n} s_{n'}^{(r)} \ .
\end{equation}
as well as the corresponding quantities for $u$-messages:
$\sh^{(r)}_n$ and $\Sh^{(r)}_n$.
From Eqs.~(\ref{eq:SuscJoint1}), (\ref{eq:SuscJoint2}), and using the 
recursion equations (\ref{eq:MinSizeDistr1}), (\ref{eq:MinSizeDistr2}),
it is easy to obtain
\begin{eqnarray}
s_n^{(r)} = \sum_m \sh_{n-m}^{(r)} q_m \ , \;\;\;\;\;\;
\Sh_n^{(r)} = Q_n^{p-2} S_n^{(r-1)} \ .\label{eq:Snr}
\end{eqnarray}
Furthermore, the first of these equations can be rewritten in
terms of integrated sequences as $S_n^{(r)} = \sum_m \Sh_{n-m}^{(r)} q_m$,
cf.~App.~\ref{sec_app_geomsuc_sizes}.

\begin{figure}
\includegraphics[width=8cm]{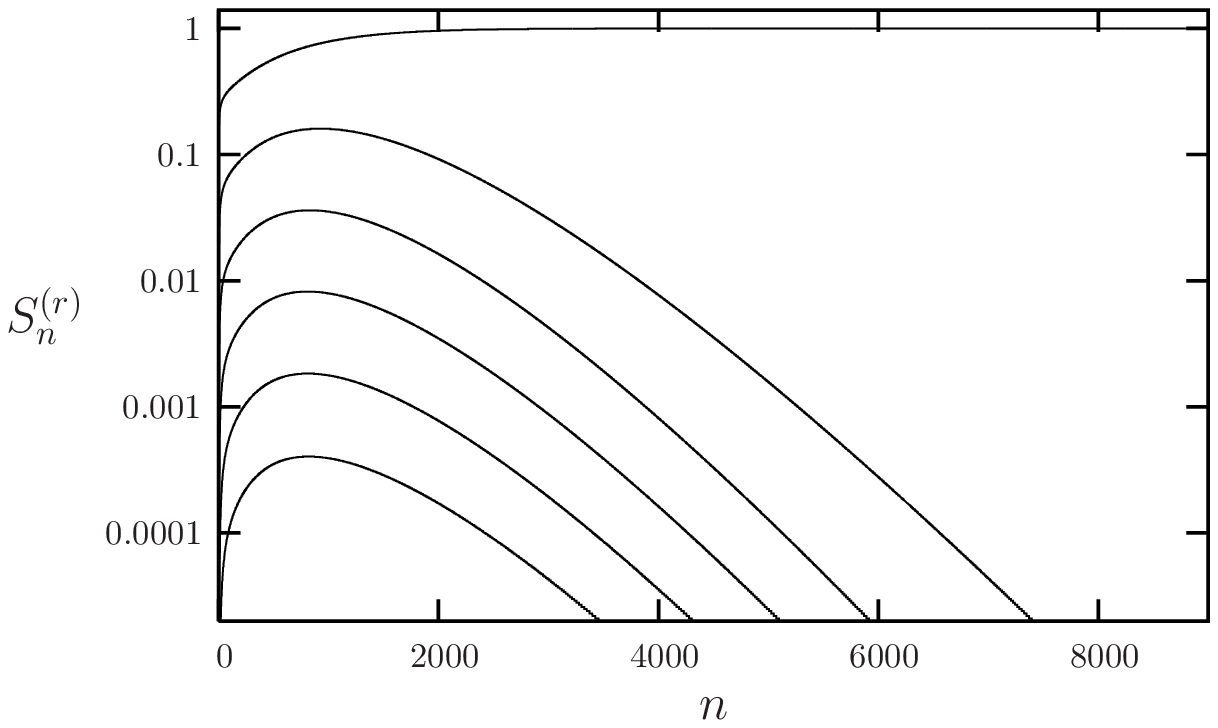}
\includegraphics[width=8cm]{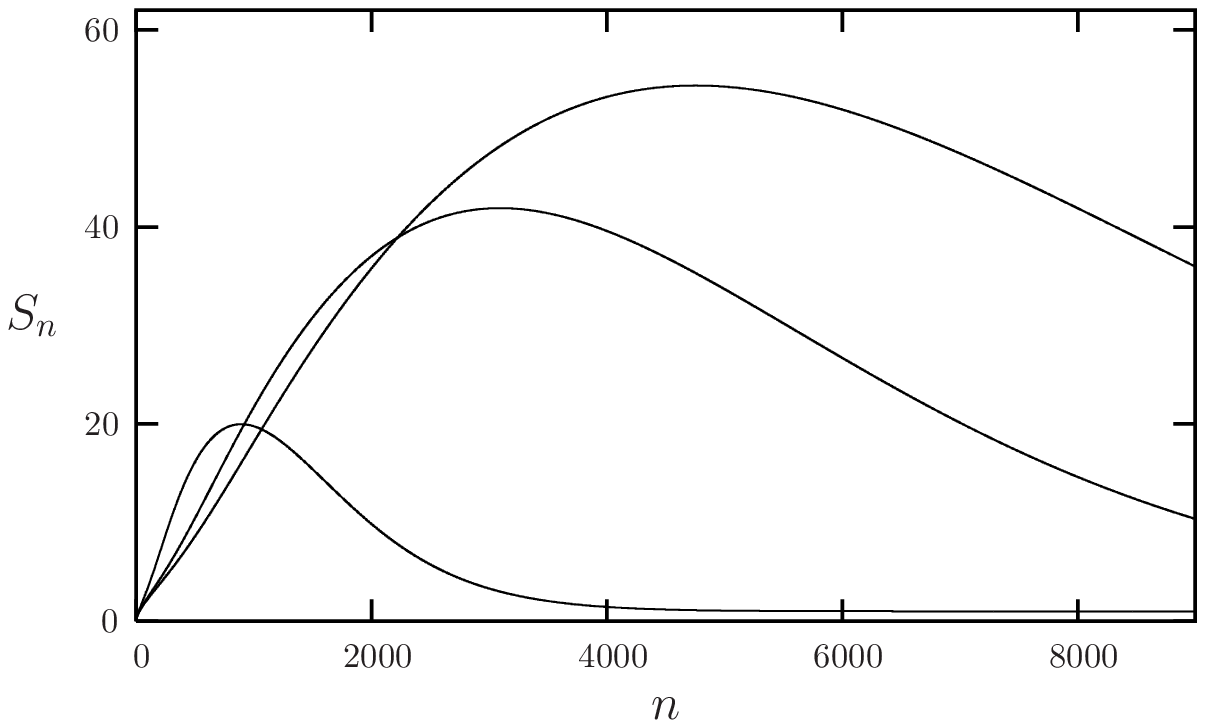}
\caption{Left: the response $S_n^{(r)}$ of the minimal rearrangement size
distribution  to a constraint on a site at distance $r$ of the root. 
We used $\g=0.8$ and, from top to bottom, $r=0,1,\dots,5$.
Right: the corresponding susceptibility $S_n$. From left to right, 
$\g=0.8,0.81,0.812$. }
\label{fig_Snr}
\end{figure}

For $r=0$ one has by definition $S_n^{(0)}=1-Q_n$.
Equations (\ref{eq:Snr}) can be used to compute $S_n^{(r)}$
efficiently by recursion (both on $n$ and $r$). 
We show the results in Fig.~\ref{fig_Snr}, left panel,
for $\g=0.8$ and a few increasing 
values of $r$. As $n\to\infty$, $S_n^{(r)}\to 1$  if $r=0$, 
(we are constraining the spin $\sigma_{i_0}$ itself not to be flipped:
there is no rearrangement of finite size satisfying such a condition), 
and $S_n^{(r)}\to 0$ if $r\ge 1$. 
Moreover one can clearly see $S_n^{(r)}\to 0$ as $r\to\infty$ at fixed $n$. 
Indeed the constraint of not
flipping $\sigma_{i_r}$ gets less and less relevant in this limit, 
and when $r$ is
much larger than the typical depth of the rearrangements 
$n_{i_0} \approx n_{i_0}^{(i_r)}$. From the studies of the depth of the
rearrangements, one expects that the values of $r$ where the constraint
becomes irrelevant scales like $(\g_{\rm d}-\g)^{-1/2}$ in the
critical limit.

Let us now define a susceptibility $S_n$ by summing the above ``response function"
$S_n^{(r)}$ over the position of the constrained site. 
As there are on average $(\g p(p-1))^r$ sites
at distance $r$ of the root, we have
\begin{equation}
S_n=\sum_{r=0}^\infty (\g p(p-1))^r S_n^{(r)} \, .
\end{equation}
We also define $\Sh_n=\sum_{r\ge 1} (\g p(p-1))^r \Sh_n^{(r)}$.
Intuitively, $S_n$ is a measure of the ``spine" of optimal
rearrangements of size $n$. By spine, we mean the subset of sites
which belong to {\em all} of the optimal rearrangements for $i$.

From Eq.~(\ref{eq:Snr}), we get
\begin{eqnarray}
S_n = 1-Q_n + \sum_m \Sh_{n-m} q_m \ , \;\;\;\;\;
\Sh_n = \g p (p-1) Q_n^{p-2} S_n \ .\label{eq:FinalSuscSize}
\end{eqnarray}
The numerical solution of these equations is shown in the right panel of
Fig.~\ref{fig_Snr}, for a few values of $\g$ approaching 
$\g_{\rm d}$. 
The behavior of these curves is strikingly similar to the one of the four 
point 
susceptibility close to the mode coupling transition in structural glasses
\cite{Donati,Bennemann,FranzChi4A,FranzChi4B}. 
Here $n$ plays the role of time in MCT, and $\g$ that of the temperature.
For a given value of $\g$, the spine size has a maximum 
$S_*(\g)$ which corresponds to rearrangements of total 
size $n_*(\g)$. When $\g\to\g_{\rm d}$ both $n_*(\g)$
and $S_*(\g)$ diverge.

An analytical study of the asymptotic behavior, 
cf.~App.~\ref{sec_app_geomsuc_sizes},
confirms this remark. The main result is that  
$n_*(\g) \sim (\g_{\rm d}-\g)^{-\nu}$ scales as the
typical size of large scale rearrangements, 
while $S_*(\g)\sim (\g_{\rm d}-\g)^{-\eta}$, 
with an universal value $\eta=1$. 
The same critical behavior was recently found for the
for the four point susceptibility of the fully-connected $p$-spin model, 
as well as in MCT~\cite{BiroliBouchaud}.
%
%
\subsubsection{Minimal barrier rearrangements}
\label{sec_geomsuc_barr}

A geometric susceptibility can be similarly defined in terms of 
minimal barrier rearrangements. In this context, $b_{i_0}^{(i_r)}$ is
the minimal barrier of a rearrangement for the variable $i_0$ which
excludes $i_r$, another variable at distance $r$ from $i_0$. To simplify
the analysis, we consider barriers obtained from the disjoint  
strategy. While exact barriers can only be computed through Yannakakis
approach, we expect this simplification not to alter the critical behavior.

Following the same steps as in the previous Section, one is lead 
to define a function $S_b^{(r)}$. This is the response of the minimum barrier
distribution for site $i_0$, to a perturbation applied on site
$i_r$. Using Arrhenius law, it is easy to establish the relationship 
between $S_b^{(r)}$ and physical observables. 
Assume that an external field $h$ has been
applied to site $i_r$ at distance $r$ from $i_0$.
Let $C^{(r)}(t;h)$ the expected (with respect
to the graph $\cG$) local correlation function for spin $\sigma_{i_0}$,
under the (equilibrium) modified Glauber dynamics. The response
$\chi^{(r)}(t) = C^{(r)}(t;\infty)-C^{(r)}(t;0)$ is given, in the low
temperature limit, by
\begin{eqnarray}
\lim_{\beta\to\infty} \chi^{(r)}(e^{2\beta b}) =  S^{(r)}_{\lceil b\rceil}\, ,
\label{eq:ArrheniusSusc}
\end{eqnarray}
which generalizes to susceptibilities the relation (\ref{eq:CorrDistr})
between barriers distributions and correlation functions.
Unlike usual response functions (in which the external perturbation
is infinitesimal), $\chi^{(r)}(t)$ is the response to an infinite, 
but localized perturbing field. Since in any case $\chi^{(r)}(t)\to 0$
as $r\to\infty$, this difference should be immaterial for the 
critical behavior.

Summing $S^{(r)}_b$ over the perturbation site $i_r$, one can define
a barrier susceptibility $S_b$.
Proceeding as in the previous Section, one can derive a set of coupled
equations which determine $S_b$: 
\begin{eqnarray}
S_b &=& 1-Q_b + \sum_{l=0}^\infty p_l \sum_{b_0, \dots , b_l} \sh_{b_0}
\qh_{b_1} \dots \hat{q}_{b_l} \;
\ind(b \le \hat{f}_{l+1}(b_0,\dots,b_l) ) \ , \label{eq_suscbarr1} \\
\Sh_b &=& \g p (p-1) Q_b^{p-2} S_b \ . \label{eq_suscbarr2}
\end{eqnarray}
These are the counterpart of Eqs.~(\ref{eq:FinalSuscSize})
for size susceptibilities. 

\begin{figure}
\includegraphics[width=8cm]{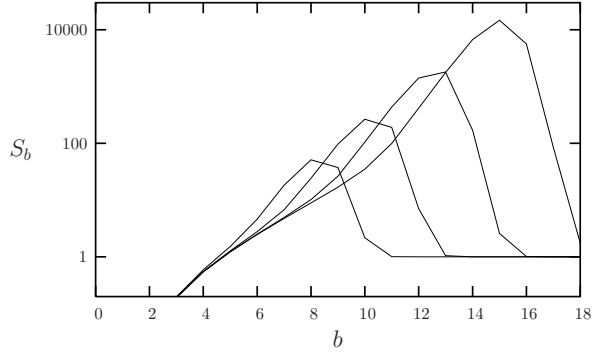}
\caption{Geometric susceptibility for minimal barrier rearrangements.
From left to right, $\g=0.816$, $0.818$, $0.8184$, $0.81846$.}
\label{fig_suscbarr}
\end{figure}

A numerical solution of these equations, 
cf.~App.~\ref{sec_app_geomsuc_barriers}, 
leads to the result shown in Fig.~\ref{fig_suscbarr}.
As for the size susceptibility $S_n$, the behavior of
its barrier analog $S_b$ is very close to the one of the four point
dynamical susceptibility in MCT.
The barrier height plays the role of time, and in fact they are related 
through Arrhenius law in the low temperature limit, 
cf.~Eq.~(\ref{eq:ArrheniusSusc}).
The maximum of $S_b$ corresponds to large scale barriers
$b_*(\g) \simeq - \Upsilon \log(\g_{\rm d} - \g)$. Its
height diverges at the dynamical transition $S_*(\g)\sim
(\g_{\rm d} - \g)^{- \eta}$, with the same universal value
$\eta = 1$. The details of this analytic study can be found in 
App.~\ref{sec_app_geomsuc_barriers}.
%
\section{Scaling analysis and numerical experiments}
\label{ScalingSection}

In the last Sections we focused on the zero temperature limit at
$\g<\g_{\rm d}$. Here we shall discuss the implications
of these results on the finite temperature behavior.

In order to simplify our discussion,
we take a step back and focus on the global relaxation 
time $\tau=\tau(\beta,\g)$. This is
defined from the global correlation function through the 
condition~\footnote{As for local relaxaton times, 
cf. Sec.~\ref{sec:Implications} the precise value of $\delta$ is irrelevant, as
long as it is smaller than the global Edwards Anderson parameter.
Note also that the relaxation time $\tau$ introduced here should not be confused with 
other global time scales, such as the inverse spectral gap, or the 
mixing time.} $C(\tau)=\delta$. Most of the arguments below
can be easily transposed to local times.

In the low temperature, $\g<\g_{\rm d}$ regime, Arrhenius law
yields $\tau\sim \exp(2\beta b_{\rm large})$, where $b_{\rm large}$
is the typical minimal barrier for large rearrangements.
Substituting the critical behavior of $b_{\rm large}$, we get
\begin{eqnarray}
\tau(\beta,\g)\sim (\g_{\rm d}-\g)^{-2\beta\Upsilon}\, .
\label{eq:LowTScaling}
\end{eqnarray}
This relation is valid if $\g\to\g_{\rm d}$ {\em after}
$\beta\to\infty$. On the other hand, at any $\beta<\infty$, 
the relaxation time does not diverge at $\g=\g_{\rm d}$, but 
at a somewhat higher value of $\g$ corresponding to the 
dynamical line, cf.~Fig.~\ref{phasediag}:
\begin{eqnarray}
\tau(\beta,\g)\sim {\bf (}
\g_{\rm d}(\beta)-\g{\bf )}^{-\te(\beta)}\, .
\label{eq:FiniteTScaling}
\end{eqnarray}
Equivalently, $\tau(\beta,\g)\sim {\bf (}T-T_{\rm d}(\g){\bf
  )}^{-\te(\g)}$. In Sec.~\ref{sec:Implications}, we derived a lower 
bound of this form implying $\te(\, \cdot\, )\ge 1/2$.

Let now $T=1/\beta$ be some fixed small temperature, and 
consider the behavior of $\tau(\beta,\g)$ as $\g$ is increased
towards $\g_{\rm d}$ and above. We expect that (\ref{eq:LowTScaling})
remains true as long as $\g$ is not too close to $\g_{\rm d}$.
Near $\g_{\rm d}$ the divergence is rounded off, and eventually
a crossover to the behavior (\ref{eq:FiniteTScaling}) takes place.
In other words, the dynamics crosses over from an ``activated regime'',
i.e. a regime for which Eq.~(\ref{eq:LowTScaling}) is accurate,
to a ``thermal'' regime in which  Eq.~(\ref{eq:FiniteTScaling}) holds.

At which values of $\g$ does the crossover takes place?
We will produce two distinct heuristic arguments that provide the same  answer.
The first one is very simple. The divergence (\ref{eq:LowTScaling})
is rounded off only if ${\bf (}\g_{\rm d}(\beta)-\g{\bf )}$
is significantly different from ${\bf (}\g_{\rm d}-\g{\bf )}$.
Equivalently, ${\bf (}\g_{\rm d}-\g{\bf )}\sim 
{\bf (}\g_{\rm d}(\beta)-\g_{\rm d}{\bf )}$. On the other hand
a cavity computation, cf. App.~\ref{app:LowTDline}, shows that,
as $\beta\to\infty$,
\begin{eqnarray}
\g_{\rm d}(\beta) = \g_{\rm d}+ x_{\rm d}\, e^{-2\beta}+
o(e^{-2\beta})\, ,\label{eq:LowTDynamical}
\end{eqnarray}
with $x_{\rm d}$ a finite, $p$-dependent constant. 
Therefore, the crossover regime corresponds
to $(\g_{\rm d}-\g)e^{2\beta} = O(1)$.

The second heuristic argument is more involved, but more instructive.
A finite temperature configuration can be characterized through
the locations of energy defects, i.e. frustrated interactions. 
In a low temperature equilibrium configuration, these defects are very
sparse, and their density is about $e^{-2\beta}$ (as long as
$\g<\g_{\rm d}$).  
Consider now a spin $\sigma_i$, and the 
associated minimal barrier rearrangement $\R_i$ of size $n_i$ and 
barrier $b_i$. 
Glauber dynamics will make use of energy defects
in order to accelerate the relaxation of $\sigma_i$, with respect
to the purely activated behavior $\exp(2\beta b_i)$.
A rough intuition of this phenomenon
is obtained by assuming that interactions containing an energy defect
can be removed from the system.
A first remark is that, as long as 
$n_ie^{-2\beta}\ll 1$, no energy defect is encountered
along the activated path (also allowing for slightly sub-optimal 
rearrangements).  Arrhenius behavior $\tau_i\sim\exp(2\beta b_i)$
is therefore not modified.
Focusing on large rearrangements, this 
implies that Eq.~(\ref{eq:LowTScaling}) remains correct as long as 
$(\g_{\rm d}-\g)^{-\nu_{\rm barr}}e^{-2\beta}\ll 1$
(i.e. $\g_{\rm d}-\g\gg e^{-2\beta/\nu_{\rm barr}}$).

This simple minded argument can be refined. We expect an energy
defect to be as effective in reducing the effective barrier
seen by spin $\sigma_i$, as a pinning field is in augmenting it.
In other words, the density of such defects must be compared
to the size of the large rearragements spine (i.e. to the peak of 
the geometric susceptibility), rather than to their overall size.
Using the results of Sec.~\ref{sec_geomsuc_barr}, we conclude once 
again that the crossover regime corresponds to 
$(\g_{\rm d}-\g)e^{2\beta} = O(1)$.

What is the behavior of $\tau(\beta,\g)$ in the crossover regime?
Inspired by the anomalous dynamical scaling for diluted ferromagnets
at the percolation transition \cite{Henley,Rammal}
we conjecture the scaling form\footnote{The
differences with respect to the form in Ref.~\cite{NostroLettera} 
are due to a different normalization of the temperature.}
\begin{eqnarray}
\tau(\beta,\g)\simeq e^{2\Upsilon\beta^2}\,{\cal L}
{\bf (}(\g-\g_{\rm d})e^{2\beta}{\bf )}\, .\label{eq:ScalingForm}
\end{eqnarray}
In order to match Eqs.~(\ref{eq:LowTScaling}) and
(\ref{eq:FiniteTScaling}), the scaling function ${\cal L}(\,\cdot\,)$
must have the following asymptotic behaviors
\begin{eqnarray}
{\cal L}(x) & \sim & \exp\left\{-\frac{1}{2}\Upsilon\, (\log |x|)^2
\right\}\;\;\;\;\;\;\;\;\; \mbox{as $x\to -\infty$,}\\
{\cal L}(x) & \sim & (x_{\rm d}-x)^{-\te_{\rm d}}
\;\;\;\;\;\;\;\;\;\;\;\;\;\;\;\;\;\;\;\;\;\;\;\;\;\; \mbox{as $x\to x_{\rm d}$,}
\end{eqnarray}
where $\te_{\rm d}\equiv\lim_{\beta\to\infty}\te(\beta)$.

\begin{figure}
\setlength{\unitlength}{1cm}
\includegraphics[width=8cm]{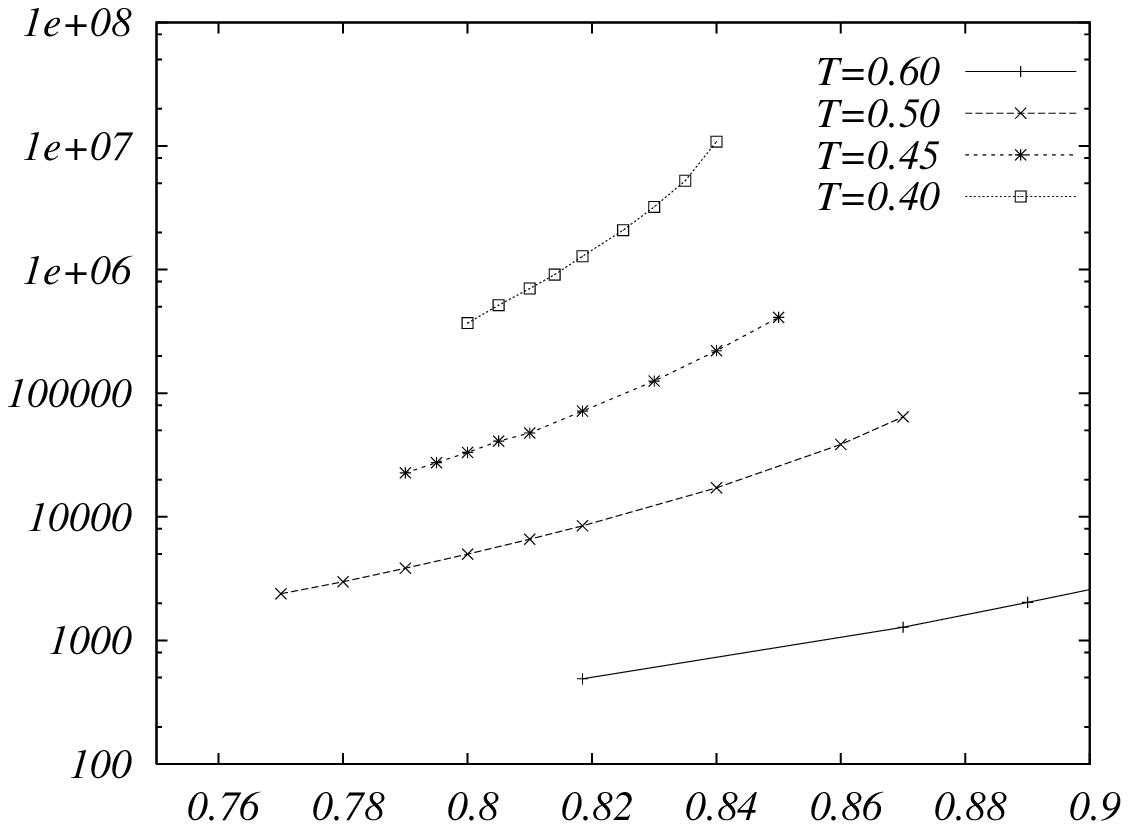}\hspace{1cm}
\includegraphics[width=8cm]{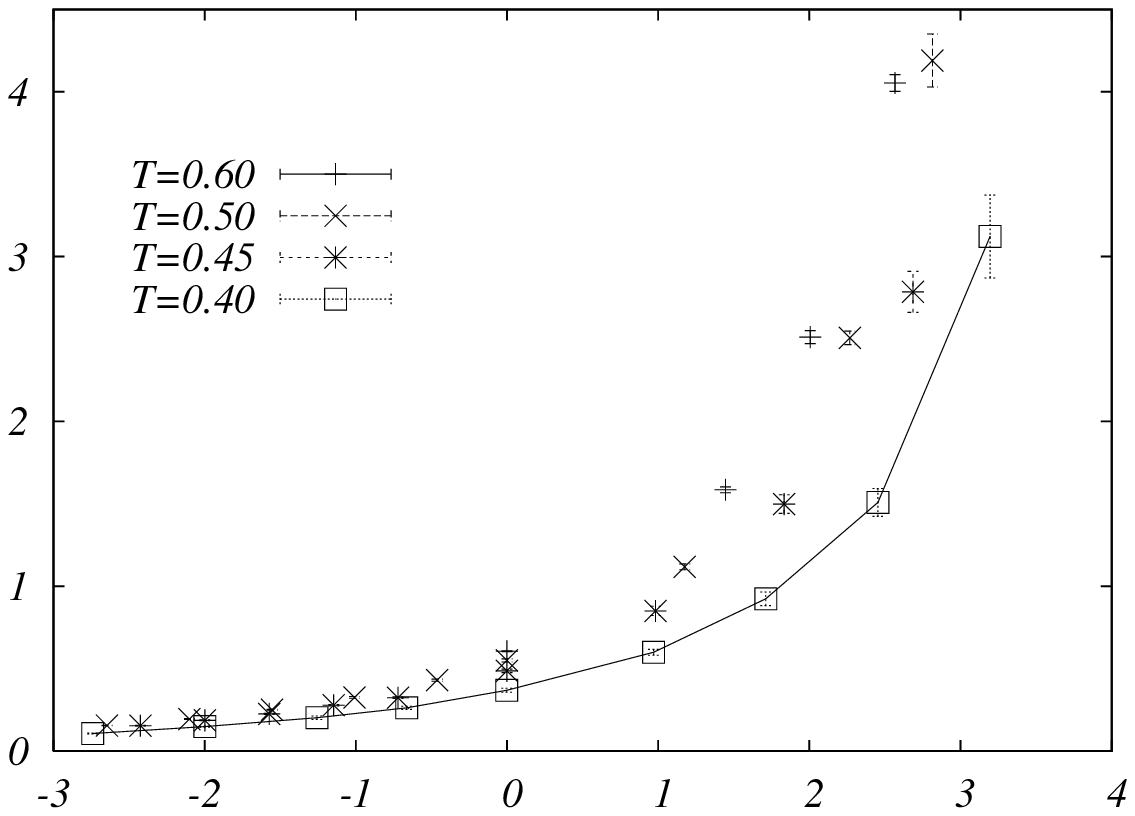}
\put(-13,-0.2){$\alpha$}
\put(-5,-0.2){$e^{2\beta}(\alpha-\alpha_{\rm d})$}
\put(-17,3){$\tau$}
\put(-8.8,3){$e^{-2\Upsilon\beta^2}\tau$}
\caption{Left: Global correlation times, as defined from the correlation 
function through the relation $C(\tau) = 1/2$.
Right: Scaling plot of the same data. Notice that no fitting parameter 
is used here.}
\label{fig_Times}
\end{figure}
\begin{figure}
\setlength{\unitlength}{1cm}
\begin{center}
\includegraphics[width=7.cm]{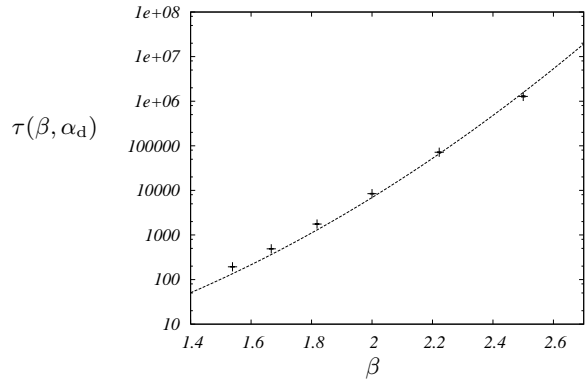}
\put(-8,3){$\tau(\beta,\g_{\rm d})$}
\put(-3.3,-0.2){$\beta$}
\end{center}
\caption{Super-Arrhenius behavior of the global correlation time at the
critical average degree $\alpha=\alpha_{\rm d}$, for $p=3$. 
The dashed line represent the theoretical prediction 
$\tau(\beta,\g_{\rm d}) = A\, e^{2\Upsilon\beta^2}$, with $A=0.45$ to fit 
the data.}
\label{fig_Super}
\end{figure}

We carried out extensive numerical simulations in order to check this
scaling hypothesis, along the lines exposed in Sec.~\ref{sec:Yanna}.
In Fig.~\ref{fig_Times}, left frame, we reproduce the resulting 
global correlation time for several values of $\beta$ as functions of
$\g$.
In Fig.~\ref{fig_Times}, right frame, we check Eq.~(\ref{eq:ScalingForm})
by  plotting $\tau(\beta,\g)$, in rescaled units,
for several values of $\beta$ and $\g$.
Despite the variation of the relaxation time
 over several orders of magnitude, an acceptable agreement is found.
Remarkably, Eq.~(\ref{eq:ScalingForm}) predicts a super-Arrhenius
divergence at this value of $\g$: $\tau(\beta,\g_{\rm d})\sim 
2\Upsilon\beta^2$. The prediction is well verified by
numerical data, cf. Fig.~\ref{fig_Super},
thus supporting the hypothesis made in
computing $\Upsilon$, cf. Sec.~\ref{sec_minbarr}.

While the numerical results do not prove unambiguously 
Eq.~(\ref{eq:ScalingForm}), they support it as a valid working hypothesis.
As in the case of diluted ferromagnets, Eq.~(\ref{eq:ScalingForm})
is consistent with temperature being  a relevant perturbation 
(in renormalization group sense) at the critical point
$(\g=\g_{\rm d},T=0)$. This in turns suggests that
the line $T=T_{\rm d}(\g)$ is ruled by a different universality
class with respect to this point.

\section{Conclusion}
\label{ConclusionSection}

In Section \ref{ModelSection} we stressed that the critical dynamics of 
Bethe lattice models was  understood only in 
the fully connected $\g\to\infty$ limit. In the previous pages we 
derived several results concerning the opposite limit $T\to 0$, 
$\g\to\g_{\rm d}$. We can now leverage on the knowledge 
concerning these two extremal points, in order to propose a complete
scenario for the whole dynamical line $(\g,T_{\rm d}(\g))$.

We found two universality classes\footnote{We use the term ``universality"
despite its content is not as clear in the present context as
it is for ordinary critical phenomena.} which correspond to distinct 
divergences of the slow time scale:
\begin{itemize}
\item Usual mode coupling theory (MCT) holds at the $\g = \infty$
point. The relaxation time divergence is as summarized in 
Eqs.~(\ref{eq:GeneralMCT1}), (\ref{eq:GeneralMCT2}). The parameter
$\lambda_{\rm mct}$ depends (continuously) on the model, and
the critical exponent $\te_{\rm mct}$ is, in general, irrational.
\item Activated mode coupling theory (aMCT) describes the $T\to 0$,
$\g\to\g_{\rm d}$ (with $(\g_{\rm d}-\g)e^{2\beta}\gg 1$)
behavior. The relaxation time diverges as in (\ref{eq:LowTScaling}),
with $\Upsilon$ determined by equations of the form
(\ref{eq_minbarr_oa}), (\ref{eq_minbarr_ob}), and (\ref{eq_minbarr_bp0}). 
Once again, $\lambda$ depends continuously on the model,
and $\Upsilon$ is, in general, irrational.
\end{itemize}
On the other hand, several important elements are common to 
the two universality classes. Among the others, the linear size
of cooperative regions scales as $\delta^{-\zeta}$, and the 
peak of the dynamical susceptibility as $\delta^{-\eta}$, 
with $\zeta=1/2$ and $\eta=1$
(here $\delta=T-T_{\rm d}$ for the first class, and $\delta=\g_{\rm d}-\g$
for the second)~\cite{BiroliBouchaud}.

In Sec.~\ref{ScalingSection} we argued that temperature is a relevant
perturbation at the $(\g=\g_{\rm d}, T=0)$ point. It is therefore 
natural to guess that the whole $\g>\g_{\rm d}$, $T=T_{\rm d}(\g)$
is controlled by the ordinary MCT class. Namely, there exist a 
continuous function $\lambda_{\rm mct}(\g)$, such that the
slow time scales behaves as in 
Eqs.~(\ref{eq:GeneralMCT1}), (\ref{eq:GeneralMCT2}).
The crossover away from the $(\g=\g_{\rm d}, T=0)$ point is described by 
Eq.~(\ref{eq:ScalingForm}). Finally the cooperativity
exponents $\zeta=1/2$ and $\eta=1$ remain constant on the whole line,
as well as in the crossover regime.

At any finite temperature, usual MCT behavior eventually overcomes
aMCT when the system is close enough to the dynamical line. 
On the other hand, activation can be important in a large
pre-asymptotic regime, depending on the way the dynamical line is approached.
Such a regime can be of experimental interest.

In the model  studied in this paper, the crossover is controlled 
by the scaling variable 
$x =(\g_{\rm d}-\g)e^{2\beta}$. This can be interpreted
as the ratio between two diverging sizes at the $(\g=\g_{\rm d}, T=0)$ point:
the spine of large scale rearrangements, which scales as 
$(\g_{\rm d}-\g)^{-1}$, and the inverse density of energy defects 
scaling as $e^{2\beta}$. In terms of observable quantities, we
could have written $x=T^2C_T/\chi_4^*$, where $C_T = \partial_T\<E(\sigma)\>$
is the system specific heat, and $\chi_4^*$ is the peak value of the 
dynamical four point susceptibility. 
In more general (non-mean field) systems, the same 
role is likely to be played by a ratio of the form $x=\xi_{\rm th}/\xi_4$.
Here $\xi_{\rm th}$ is a thermal correlation length, and $\xi_4$ 
is a dynamical correlation length extracted from four point
dynamical correlations. A similar scaling variable describes the crossover
in diluted ferromagnets at the percolation point~\cite{Stinchcombe}.

The methods developed in this paper can certainly be applied in a more 
general context. In particular, rearrangements can be defined for
kinetically constrained models (Kob-Andersen, Frederickson-Andersen 
for instance) on
Bethe lattices, and studied along the same lines~\cite{NostroKCM}. 
Interestingly, a recent numerical study~\cite{EPL_MauroCristinaGiulio} 
has investigated the relaxation time divergence in these models and
indicates a behavior of the type $\tau\sim(\rho_{\rm d}-\rho)^{-\te}$
($\rho$ being the particle density). 

As we mentioned in the Introduction, the problem studied here is known
in computer science as XORSAT, and is a simple example in the wider
family of random Constraint Satisfaction Problems \cite{CSP}. 
The dynamical glass transition, 
essentially studied in this context at zero temperature as a 
``clustering transition"
of the groundstates, is a feature shared by other examples of this family,
Satisfiability and Coloring in particular \cite{SAT,Coloring}. 
It would be of great 
interest to extend the refined decription of the changes in the 
landscape properties at the clustering transition we obtained in the 
XORSAT problem to these other models.

%
%
\section*{Acknowledgments}

This work has been partially supported by the integrated project EVERGROW
(n. 1935 in the complex systems initiative of the Future and 
Emerging Technologies directorate of the IST Priority, EU Sixth Framework),
and the Research Training Network STIPCO (HPRN-CT-2002-00319).
%
%
\appendix 

\section{Properties of the equilibrium distribution}
\label{app:ProofProp}

We prove here the statements $(i)$ and $(ii)$ made in Section
\ref{ModelSection}. In order to prove $(i)$, notice that the partition
function admits the usual high temperature expansion
\begin{eqnarray}
Z_{N,\beta}(J) = 2^N e^{-M\beta} (\cosh\beta)^M\left\{1+
\sum_{\omega\in {\rm hl} } \epsilon(\omega) (\tanh\beta)^{|\omega|}\right\}\, ,
\end{eqnarray}
where the (finite) sum over $\omega$ runs over all the hyperloops 
in  $\cG$. $|\omega|$ denotes the size of the hyperloop $\omega$ 
and $\epsilon(\omega)$ the product of the signs of the interactions in it.
If no hyperloop is present, only the first term survives.

Consider now $(ii)$. Let $\prob(J)$ be the distribution of the sign of the quenched
couplings, i.e. the uniform distribution over 
$\{\; J_{i_1\dots i_p}\in\{+1,-1\}\;\}$, and $\prob(\sigma|J)$ the Boltzmann
distribution for the configuration $\sigma$, given the couplings $J$.
Sampling $J$ from $\prob(J)$ and then $\sigma$ from the conditional
distribution $\prob(\sigma|J)$ is equivalent to sampling from the joint 
distribution
\begin{eqnarray}
\prob(J)\,\prob(\sigma|J) \equiv \prob(\sigma,J)\, .
\end{eqnarray}
Using the form of the Boltzmann distribution, we have
\begin{eqnarray}
\prob(\sigma,J) = \frac{e^{-M\beta}}{2^{M}Z_{N,\beta}(J)}\, \prod_{(i_1\dots i_p)\in\cG}
\, \exp\left\{\beta J_{i_1\dots i_p}\sigma_{i_1}\cdots \sigma_{i_p}\right\}\, .
\end{eqnarray}
This is in turn equivalent to sampling $\sigma$ form the marginal distribution 
$\prob(\sigma)$ and then $J$ from the conditional distribution 
$\prob(J|\sigma)$. We are left with the task of determining these 
distributions. Using the definition of marginal distribution, 
and the fact $(i)$ above, we get
\begin{eqnarray}
\prob(\sigma) \equiv \sum_{J} \prob(\sigma,J) = 
\frac{1}{2^{N}(2\cosh\beta)^M}\prod_{(i_1\dots i_p)\in\cG} 
\sum_{J_{i_1\dots i_p}}
e^{\beta J_{i_1\dots i_p}\sigma_{i_1}\cdots \sigma_{i_p}} = 
\frac{1}{2^{N}}\, ,
\end{eqnarray}
i.e. the marginal distribution of $\sigma$ is uniform. We can now 
apply the definition of Bayes theorem to get
\begin{eqnarray}
\prob(J|\sigma)\equiv\frac{\prob(\sigma,J)}{\prob(\sigma)} = 
\prod_{(i_1\dots i_p)\in\cG}\frac{1}{2\cosh\beta}\
\, \exp\left\{\beta J_{i_1\dots i_p}\sigma_{i_1}\cdots \sigma_{i_p}\right\}\, ,
\end{eqnarray}
which proves our claim.

\section{Correlations}
\label{app:Correlations}

This Appendix contains the technical details
of the calculation of equilibrium correlation functions presented in
Section \ref{OriginSection}.


\subsection{$n$-point correlation functions: zero-temperature potential}

We start by computing the annealed potential (\ref{eq:Annealed}).
By taking expectation with respect to the graph $\cG$, 
and  the constrained spins 
(but not on their number), it is elementary to obtain 
\begin{eqnarray}
\E_{L}\E_{\cG} Z_{\cG}(L) = \sum_{\omega}\binom{N}{N\omega}\,
\left[\frac{1+(1-2\omega)^p}{2}\right]^{N\g}\, (1-\omega)^{l}\, .
\label{eq:AnnealedFirst}
\end{eqnarray}
Here the sum runs over $\omega = 0, 1/N, 2/N,\dots, 1$.
Let us define the annealed potential at size $N$ as
$\psi_{\rm ann}(\lambda;N) \equiv N^{-1}\E_l\log_2 \E_{L}\E_{\cG} Z_{\cG}(L)]$. 
By a standard saddle point calculation, we get 
\begin{eqnarray}
\psi_{\rm ann}(\lambda;N) & =& \psi_{\rm ann}(\lambda)
+\frac{1}{N}\psi_{\rm ann}^{(1)}(\lambda) + O(N^{-2})
\, , \\
\psi_{\rm ann}^{(1)}(\lambda) & \equiv &-\frac{1}{2}\log_2[
|F_{\lambda}''(\omega_*)|\omega_*(1-\omega_*)]+
\frac{\lambda}{2(\log 2)^2(1-\omega_*)^2|F_{\lambda}''(\omega_*)|}\, .
\label{eq:AnnealedCorr}
\end{eqnarray}
Here $F_{\lambda}(\omega)$ 
is the function maximized in Eq.~(\ref{eq:Annealed}) 
and $\omega_*$ is the point realizing the maximum. 
As $\lambda\to 0$, we have 
$\psi_{\rm ann}^{(1)}(\lambda) = C_1\lambda+O(\lambda^{2})$.


As claimed in Sec.~\ref{OriginSection}, given $\ve>0$, $\psi_{\cG}(\lambda)\le 
\psi_{\rm ann}(\lambda)+\ve\lambda^2$ for any $\lambda\ge 0$ 
with high probability. Let us stress that this statement is stronger than
saying that, given $\ve,\lambda\ge 0$, $\psi_{\cG}(\lambda)\le 
\psi_{\rm ann}(\lambda)+\ve\lambda^2$ with high probability. 
On the other hand, it is required in order to estimate the derivatives of
$\psi_{\cG}(\lambda)$.
It can be proved by distinguishing  two regimes
(here $1/2<k<1$):
$\lambda \le CN^{-k}$ and 
$\lambda > CN^{-k}$. 
In the first regime, 
the term $(1-\omega)^l$ in Eq.~(\ref{eq:AnnealedFirst})
can be  replaced by $2^{-l}$. The relative error
is of order $(\omega-1/2)l \sim N^{-1/2}\cdot N^{1-k}$,
yielding $\E_{L}\E_{\cG} Z_{\cG}(L) = 2^{N(1-\g)-l}\, [1+
O(N^{1/2-k})]$. Since  $Z_{\cG}(L)$ is a power of $2$ by definition,
the probability that  $Z_{\cG}(L)>2^{N(1-\g)-l}$ is at most
$O(N^{1/2-k})$ (recall that $N\g$ is an integer). 
The desired result is obtained by taking expectation
with respect to $l$.

In the second regime, we apply Markov inequality
to the random variable  $Z_{\cG}(L)$ (here we condition on a typical value
of $\ell\approx N\lambda$):
\begin{eqnarray}
\prob\left\{\psi_{\cG}(\lambda)\ge \psi_{\rm ann}(\lambda)+\ve\lambda^2 
\right\}
\le 2^{\left\{-N\left[\psi_{\rm ann}(\lambda)+\ve\lambda^2-
N^{-1}\log_2 \E_{L}\E_{\cG} Z_{\cG}(L)\right]\right\}}\, .
\end{eqnarray}
The term $N^{-1}\E_{L}\E_{\cG} Z_{\cG}(L)$ can be safely estimated
by $\psi_{\rm ann}(\lambda;N)$ (deviations of $\ell/N$ from $\lambda$ 
being exponentially rare). The right hand side of the above
expression is therefore (neglecting small error terms)
 $2^{\{-N[\ve\lambda^2-N^{-1}\psi_{\rm ann}^{(1)}(\lambda) + 
O(N^{-2})]\}}$.
For any finite (as $N\to\infty$) $\lambda$, 
$N^{-1}\psi_{\rm ann}^{(1)}(\lambda)\ll \ve\lambda^2$.
If $\lambda\to 0$ (but $\lambda > CN^{-k}$) 
$N^{-1}\psi_{\rm ann}^{(1)}(\lambda)\sim N^{-1}\lambda\ll
\ve\lambda^2$. Therefore the probability that 
$\psi_{\cG}(\lambda)\ge \psi_{\rm ann}(\lambda)+\ve\lambda^2$ is
exponentially small.

\begin{figure}
\includegraphics[width=10.5cm]{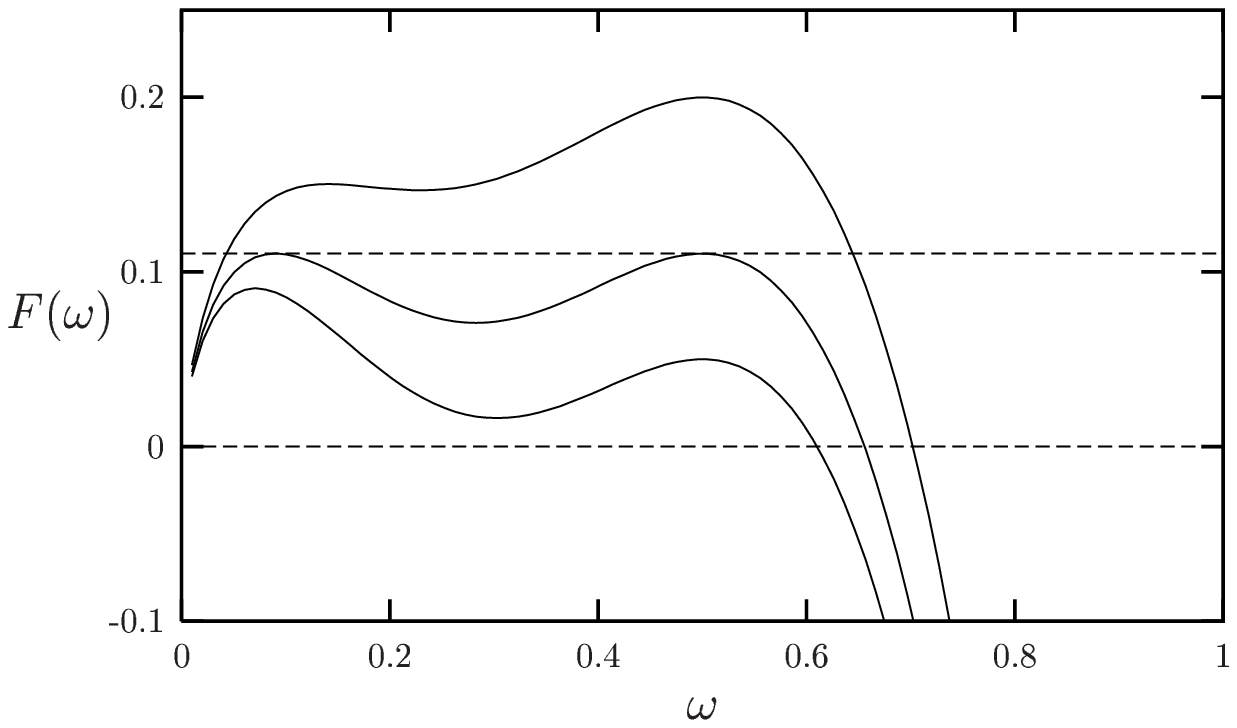}  
\caption{The function $F_\lambda(\omega)$ (cf. Eq.~(\ref{eq:Annealed})) 
for $\lambda=0$, and from bottom to top $\g=0.8$, $\g=\g_{\rm ann}\approx 0.889493$
and $\g= 0.95$.} 
\label{fig:AnnPot}
\end{figure}
In Fig.~\ref{fig:AnnPot} we plot the function $F_{\lambda}(\omega)$ for 
$p=3$, $\lambda=0$ and a few values 
of $\g$. For $\g<\g_{\rm ann}\approx  0.889493$,
$F_0(\omega)$ has a global maximum at $\omega= 1/2$ with 
$F_0(\omega) = 1-\g-\frac{1}{2}(2\omega-1)^2 +O((2\omega-1)^3)$
in its neighborhood. This implies the expansion (\ref{eq:AnnExp}).
The same is not true for $\g\ge\g_{\rm ann}$ because 
a second local maximum at $\omega<1/2$ overcomes the first.

The value of $\psi_{\cG}(\lambda)$ in the thermodynamic limit
can be computed through the cavity method. Alternatively,
one can use a leaf-removal argument in the 
spirit\footnote{We also refer
to \cite{Luby,Maxwell} for an analysis of this algorithm on graph ensembles
with general degree distributions. The results of \cite{Maxwell}
include Eq.~(\ref{eq:CavityPotential}).} of Refs.~\cite{XOR_1,XOR_2}. 
We get 
\begin{eqnarray}
\psi(\lambda) = \sup_{x\in[0,1]}\left\{-\g+p\g (1-x)x^{p-1}
+\g x^p + \exp[-\lambda-p\g x^{p-1}]\right\}\, .
\label{eq:CavityPotential}
\end{eqnarray}
For $\g<\g_{\rm c}$ and $\lambda = 0$, 
the $\sup$ is realized at $x=0$. 
A straigthforward calculation yields
\begin{eqnarray}
\left.\frac{\de \psi}{\de \lambda}\right|_0 = -1\, ,\;\;\;\;\;\;\;\;\;\;
\left.\frac{\de^2 \psi}{\de \lambda^2}\right|_0 = 1\, ,
\end{eqnarray}
in agrement with the rigorous bounds (\ref{eq:BoundsPot}). 
Higher derivatives can be computed as well and are finite for any
$\g<\g_{\rm d}$. In particular, the third derivative is easily 
seen to be related to the three points susceptibility $\chi^{\rm SG}_{3,N}$,
in analogy with Eq.~(\ref{eq:PotDerivatives}).
Under the assumption that the 
limits $N\to\infty$ and $\lambda\to 0$ can be interchanged, this 
implies the finiteness of $\chi^{\rm SG}_{n,N}$ for $T=0$, 
$\g<\g_{\rm d}$ and
$n\le 3$.
%
%
\subsection{$n$-point correlation functions: cavity argument} 

The argument of the previous Section could be generalized to 
finite temperature using the Franz-Parisi quenched 
potential~\cite{FranzPotential}. 
A more direct calculation consists in computing the correlation
functions $\<\sigma_{i_1}\cdots\sigma_{i_n}\>_{\rm c}$ using the cavity 
method.

\begin{figure}
\includegraphics[width=9cm]{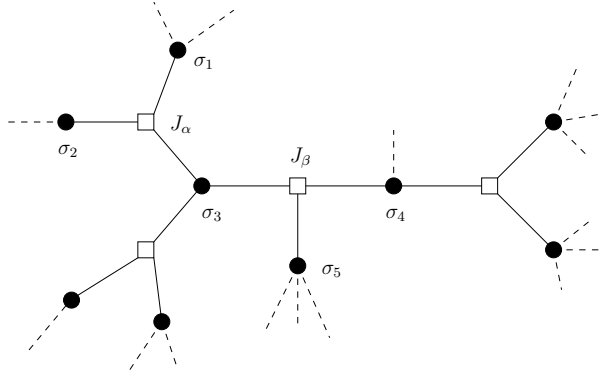}
\caption{Portion of the factor graph for the $p=3$ model. Dashed lines 
indicated other parts of the graph not represented here.} 
\label{fig:FactorConn}
\end{figure}
In order to simplify the exposition, consider the three-points 
function $\<\sigma_i\sigma_j\sigma_k\>$ for $p=3$. Generalizing
to the computation of {\em connected} correlation functions 
as well as to other values of $n$ and $p$ should be straightforward.
As a further simplification, consider the example in 
Fig.~\ref{fig:FactorConn} and $i=1$, $j =3$, $k=4$. 
The one-step replica symmetry breaking (1RSB) cavity expression 
for $\<\sigma_1\sigma_3\sigma_4\>$ reads
\begin{eqnarray}
\int \frac{1}{Z\{\rho_i\}}
\left\{\int z_{\rm cav}\{h_i\}^m
\<\sigma_1\sigma_3\sigma_4\>^{\{h_i\}}_{\rm cav}
\de\rho_1(h_1)\cdots \de\rho_5(h_5)\right\}\de \Qc[\rho_1]\cdots \de \Qc[\rho_5]\,.
\label{eq:CavityCorr}
\end{eqnarray}
Here $m$ is Parisi 1RSB parameter,
\begin{eqnarray}
\<\sigma_1\sigma_3\sigma_4\>^{\{h_i\}}_{\rm cav} &= &\frac{1}{ z_{\rm cav}\{h_i\}}
\sum_{\{\sigma_i\}}\sigma_1\sigma_3\sigma_4\;
\exp\{-\beta H_{\rm cav}(\{\sigma_i\};\{h_i\})\}\, ,\\
z_{\rm cav}\{h_i\} & = & \sum_{\{\sigma_i\}}\exp\{-\beta H_{\rm cav}
(\{\sigma_i\};\{h_i\})\} \, ,\\
 H_{\rm cav}(\{\sigma_i\};\{h_i\})  & =  & -J_{\a}
\sigma_{1}\sigma_{2}\sigma_{3}
-J_{\beta}\sigma_3\sigma_4\sigma_5-\sum_{i=1}^5 h_i\sigma_i\, ,
\end{eqnarray}
and $Z(\{\rho_i\}) = \int z_{\rm cav}\{h_i\}^m
\de\rho_1(h_1)\cdots \de\rho_5(h_5)$. Finally $\Qc[\rho]$ is the 1RSB
spin glass order parameter, i.e. a distribution over 
the space of symmetric distributions on $\mathbb{R}$: $\rho(h) =\rho(-h)$. 

If $m=1$ (which is the case for $T>T_{\rm c}(\g)$), these expressions 
simplify remarkably. One can in fact integrate explicitly over the $\{h_i\}$.
Since $\rho_i$ is symmetric, $\int\de\rho_i(h_i)\; e^{\beta h_i\sigma_i}$
does not depend on $\sigma_i$. The espression in curly brackets in
Eq.~(\ref{eq:CavityCorr}) is therefore given by
\begin{eqnarray}
\Big\{\cdots \Big\} = \sum_{\{\sigma_i\}}\sigma_1\sigma_3\sigma_4\,
e^{-\beta H_{\rm cav}(\{\sigma_i\};\{0\})}\;
\prod_{i=1}^5\int (2\cosh\beta h_i)\, \de\rho_i(h_i)\, .
\end{eqnarray} 
An analogous expression holds for $Z(\{\rho_i\})$. Taking the ratio
and integrating over $\Qc(\rho_i)$, we get the final result
\begin{eqnarray}
\<\sigma_1\sigma_3\sigma_4\> = \<\sigma_1\sigma_3\sigma_4\>^{\{0\}}_{\rm cav}\, .
\end{eqnarray}
Notice that the cavity fields on the right hand side are set to $0$.
In particular, for the case at hand $\<\sigma_1\sigma_3\sigma_4\> =0$.
More generally $\<\sigma_i\sigma_j\sigma_k\>$ is different from $0$
only if $i,j,k$ are connected by an interaction $-J\sigma_i\sigma_j\sigma_k$.
In this case we get $\<\sigma_i\sigma_j\sigma_k\> = \tanh\beta J$.
Proceding analogously, one can show that $\<\sigma_i\>=0$ 
and therefore $\<\sigma_i\sigma_j\sigma_k\>=
\<\sigma_i\sigma_j\sigma_k\>_{\rm c}$. Summing over all the triples we find 
that, in the thermodynamic limit,
\begin{eqnarray}
\chi_{3}^{\rm SG} = 6\g(\tanh\beta)^2\, .
\end{eqnarray}
for any $\g<\g_{\rm c}$. Proceeding analogously, one can compute 
higher order susceptibilities $\chi^{\rm SG}_n$ for any finite $n$.
%
%
\subsection{Point-to-set correlation functions at finite $T$}
\label{app:PointToSet}

In this Appendix we explain how to compute the point-to-set correlation length 
$\ell_*(\ve)$ and show that it diverges at $T_{\rm d}(\g)$.
Recall that the dynamical transition is usually defined using the 
following equations (one-step replica symmetry breaking with Parisi parameter 
$m$)
\begin{eqnarray}
\rho_i(h) & = & \frac{1}{{\cal Z}(\{\rh_{\a}\};m)}
\int\prod_{\a=1}^l\de\rh_{\a}(u_\a)\, 
\left(\frac{\cosh\beta h}{\prod_{\a=1}^l\cosh\beta u_\a}\right)^m\, 
\delta\left(h- u_1-\dots-u_l\right)\, ,\label{eq:1RSB1}\\
\rh_\a(u)  & = & \int\prod_{i=1}^{p-1}\de \rho_i(h_i)
\delta\left(u - u(h_1\dots h_{p-1})\right)\, ,\label{eq:1RSB2}
\end{eqnarray}
where $u(h_1\dots h_{p-1})\equiv\frac{1}{\beta}
\atanh[\tanh\beta\tanh\beta h_1\cdots \tanh\beta h_{p-1}]$.
The fields distributions $\rho_i(\,\cdot\,)$ and $\rh_\a(\,\cdot\,)$
are required to be symmetric.
The indices $i$ and $\a$ refer to nodes in the factor graph. 
When a random factor graph is considered,  $\{\rho_i(\,\cdot\,)\}$ and 
$\{\rh_\a(\,\cdot\,)\}$ become i.i.d. with common distributions
(respectively) $\Qc$ and $\Qch$, and $l$ becomes a 
Poisson random variable with mean $p\g$. 
The above equations acquire a distributional meaning. $T_{\rm d}(\g)$
is defined as the largest temperature such that a non-trivial
solution exists for these distributional equations with $m=1$.

Using the symmetry of $\rh_a(\, \cdot\,)$, it is easy to show that
${\cal Z}(m)=1$ for $m=1$. We can therefore
average over $\{\rho_i(\,\cdot\,)\}$ and 
$\{\rh_\a(\,\cdot\,)\}$. If we call, respectively  $Q(\,\cdot\,)$ and 
$\Qh(\,\cdot\,)$ the average distributions, we get
\begin{eqnarray}
Q(h) & = & \sum_{l=0}^{\infty}e^{-p\g}\frac{(p\g)^l}{l!}
\int\prod_{\a=1}^l\de \Qh(u_i)\, 
\left(\frac{\cosh\beta h}{\prod_{\a=1}^l\cosh\beta u_a}\right)\, 
\delta\left(h- u_1-\dots-u_l\right)\, ,\label{Average1RSB1}\\
\Qh(u)  & = & \int\prod_{i=1}^{p-1}\de Q(h_i)\;
\delta\left(u - u(h_1\dots h_{p-1})\right)\, .\label{Average1RSB2}
\end{eqnarray}

Consider now the calculation of the point-to-set correlation length
$\ell_{*}(\ve)$. First notice that, within the high temperature phase
the cavity fields vanish identically. This means that an 
equilibrium configuration within a neighborhood of a site $i$ can be 
constructed as follows. First fix $\sigma_i\in\{+1,-1\}$ uniformly at random.
Then for any of the interaction terms involving $\sigma_i$, let us
say $-J\sigma_i\sigma_{i_1}\cdots\sigma_{i_{p-1}}$, generate
the other spins from the distribution
\begin{eqnarray}
p_J(\sigma_{i_1},\dots,\sigma_{i_{p-1}}|\sigma_i) = \frac{1}{2^{p-1}\cosh\beta J}
\, \exp\big\{\beta J\sigma_i\sigma_{i_1}\cdots\sigma_{i_{p-1}}\big\}\, .
\end{eqnarray}
Then repeat the same operation with those neigbors of spins 
$i_1,\dots , i_{p-1}$ (as well as of the other neighbors of $i$), 
which have not yet been fixed. Repeat
recursively until all the spins within a distance $\ell$ from $i$
are determined. This procedure is well defined, 
because  the factor graph is a tree within any 
fixed distance from $i$.

Notice that, in order to generate the values of the spins adjacent
to $\sigma_i$, we did not need to know the graph structure at any distance 
from $i$ larger than $1$. As a consequence,
a random graph and a thermalized configuration for that graph  can 
be generated jointly   as follows (in the thermodynamic limit).
First generate the degree $l$ of site $i$ from a Poisson distribution 
with parameter $p\g$. Then generate the values of the spins which 
interact with $\sigma_i$, according with the rule given above. 
Repeat the same operation for the $(p-1)l$ neighbors of $i$, and 
continue recursively for any fixed distance $\ell$.

This gives a procedure for constructing the reference configuration 
$\sigma^{(0)}$ in our definition of $\ell_{*}(\ve)$, 
cf.~Eq.~(\ref{eq:DefinitionPointToSetAv}).
The recursive nature of this procedure allows for a recursive calculation of 
$\ell_*(\ve)$.
Condition now on the event that $\sigma^{(0)}_i=\sigma\in\{+ 1,-1\}$ and  
constrain a copy of the system to have spin $\sigma^{(0)}_j$ at site $j$,
for any $j$ whose distance from $i$ is 
at least $\ell$. Let $h$ be the effective field acting on site $i$
under this boundary conditions. This is of course a random quantity, because
the reference configuration as well as the underlying graph are random.
Denote by $P^{(\ell)}_{\sigma}(h)$ its distribution, and by 
$\Ph^{(\ell)}_{\sigma}(h)$ its distribution under the condition that 
the degree of $i$ is exactly $1$.
Then, it is easy to derive the recursions 
(here we assume, for the sake of simplicity, $J_{\a}=+1$, 
and use the shorthand
$p(\cdots|\cdot) = p_{+1}(\cdots|\cdot)$)
\begin{eqnarray}
P_{\sigma}^{(\ell)}(h) 
& = & \sum_{l=0}^{\infty}e^{-p\g}\frac{(p\g)^l}{l!}
\int\prod_{a=1}^l\de \Ph^{(\ell)}_{\sigma}(u_i)\, 
\delta\left(h- u_1-\dots-u_l\right)\, ,\label{Ball1}\\
\Ph^{(\ell)}_{\sigma}(u)  & = & 
\sum_{\sigma_1\dots\sigma_{p-1}}p(\sigma_1\cdots\sigma_{p-1}|\sigma)
\int\prod_{i=1}^{p-1}\de P^{(\ell-1)}_{\sigma_i}
(h_i)\;
\delta\left(u - u(h_1\dots h_{p-1})\right)\, ,\label{Ball2}
\end{eqnarray}
which hold for any $\ell>0$, together with the boundary condition
\begin{eqnarray}
P_{\sigma}^{(0)}(h) = \delta(h-\sigma\infty)\, ,\;\;\;\;\;\;\;\;\;\;
\Ph_{\sigma}^{(0)}(u) = \delta(u-\sigma\infty)\, .
\end{eqnarray}
These recursions can be easily approximated numerically using the population 
dynamics algorithm of \cite{MezardParisiBethe}. 
Notice that, by construction, $P_{+}(h) = P_{-}(-h)$ and
$\Ph_{+}(u) = \Ph_{-}(-u)$ and therefore a unique 
population need to be tracked. Furthermore, the recursion preserves the
following property
\begin{eqnarray}
P^{(\ell)}_+(-h) = e^{-2\beta h} P_+^{(\ell)}(h)\, ,\;\;\;\;\;\;\;\;
\Ph^{(\ell)}_+(-u) = e^{-2\beta u} \Ph_+^{(\ell)}(u)\, .\label{eq:Property}
\end{eqnarray}
The point-to-set 
length scale $\ell_*(\ve)$ can be estimated by computing the quantity
(here $\E_{{\cal G}}$ denotes expectation with respect to the sample 
realization)
\begin{eqnarray}
\E_{{\cal G}} \E_{\sigma^{(0)}}
[\sigma^{(0)}_i\<\sigma_i\>_{\ell}] = 
\int\!\de P^{(\ell)}_+(h)\;\tanh\beta h\, .
\end{eqnarray}
The data shown in Fig.~\ref{fig:BallFiniteT} have been obtained 
using this approach.

At high enough temperature, the iteration (\ref{Ball1}),
(\ref{Ball2}) converges to the trivial fixed point $P_+(h) = P_-(h) 
= \delta(h)$. As the temperature is lowered, the convergence becomes
slower and slower, and the plot of $\E_{{\cal G}} \E_{\sigma^{(0)}}
[\sigma^{(0)}_i\<\sigma_i\>_{\ell}]$ develops a plateau as a function of 
$\ell$. Eventually, the iteration converges to a non-trivial fixed point
of Eqs.~(\ref{Ball1}), (\ref{Ball2}), let us denote it by
$P^*_{\sigma}(h)$, $\Ph^*_{\sigma}(u)$. Exploiting the property
(\ref{eq:Property}) we can rewrite the fixed point distributions as
\begin{eqnarray}
P^*_{\sigma}(h) = \frac{e^{\beta h \sigma}}{\cosh\beta h}\, 
P(h)\,,\;\;\;\;\;\;\; \Ph^*_{\sigma}(u) = 
\frac{e^{\beta u \sigma}}{\cosh\beta u}\, \Ph(u)\,,
\end{eqnarray}
with $P(\,\cdot\,)$ and $\Ph(\, \cdot\,)$ two symmetric distributions.
Rewriting the fixed point condition, cf. Eq.~(\ref{Ball1}),
(\ref{Ball2})  in terms of $P(\,\cdot\,)$ and $\Ph(\, \cdot\,)$,
we recover the 1RSB equations (\ref{Average1RSB1}), (\ref{Average1RSB2}).
We thus proved that the existence of a non trivial fixed point for the 1RSB 
equations is equivalent to the existence of a non trivial fixed point 
for the recursion (\ref{Ball1}), (\ref{Ball2}). Under the assumption
that, as soon as such a fixed point appears $P^{(\ell)}_\sigma(\, \cdot\,)$ 
converges to it\footnote{A proof of this assumption in a similar problem,
due to J.~Martin and E.~Mossel, is reported in Ref.~\cite{Reconstr}.},
we thus proved that $\ell_*(\ve) = \infty$ for $T<T_{\rm d}(\g)$. 
\section{Asymptotia}

In  this  series of Appendices we shall work out the 
rearrangement critical behavior, i.e. both at $\g=\g_{\rm d}$ and as
$\g\uparrow \g_{\rm d}$. Throughout these Appendices $\delta\equiv 
\g_{\rm d}-\g$ will denote the distance to the critical point.

Before embarking in the calculations, it is interesting to
draw some parallels with the analytical treatment of the schematic MCT model,
and to collect technical properties we shall use in the following.

The time correlation function $C(\tau)$ of the schematic MCT 
(or of a fully-connected spherical spin-glass model at high temperature)
obeys an equation of the form
\begin{equation}
\dot{C}(\tau) = - C(\tau) + \int_0^\tau \!\de\tau' \; 
M(\tau-\tau') \dot{C}(\tau')\ ,
\end{equation}
where the memory kernel is a polynomial of the correlation function itself,
$M(\tau)=V{\bf (}C(\tau){\bf )}$. 
These two equations share some features with the ones
governing the distribution of minimal size rearrangements,
cf.~Eqs.~(\ref{eq:MinSizeDistr1}), (\ref{eq:MinSizeDistr2}): the first
one is a convolution on $\tau$ (resp. on $n$), the second a polynomial
relation between $M(\tau)$ and $C(\tau)$ (resp. between $\Qh_n$ and
$Q_n$). In both situations, one of the equations is simple in the ``direct 
space" of $\tau$ or $n$, the other being simpler in Laplace
transform (or generating function, see below).

As for MCT, the study of $Q_n$ (and similar quantities) in the critical regime,
can be carried out in three steps. First, one considers its decay 
exactly at the critical point,  with $n$ (or $\tau$) fixed. In a second 
step, one identifies an ``intermediate regime''
which describes  $Q_n$ around its plateau value. Matching these two regimes 
yields the scale  of $n$ in the intermediate regime. The 
``final regime'' describing the decay of $Q_n$
from its plateau to zero can also be described by a scaling function. Again,
the corresponding scale of $n$ is derived through a matching argument.

Given a function $a(t)$ defined for $t\ge 0$, the corresponding Laplace
transform will be denoted as
\begin{eqnarray}
\Lap[a](y) = \int_{0}^{\infty}a(t) e^{-yt}\;\de t\, .
\end{eqnarray}
The equivalent of Laplace transform for discrete variables is given by
generating functions (g.f.). We refer to \cite{Wilf,Odlyzko,Flaj_book}
for an extensive introduction to generating functions and
asymptotic combinatorics methods, and recall here a few basic definitions.
Given a sequence $a_n$ defined 
for $n=1,2,\dots$, the associated g.f. is the formal power series
\begin{equation}
\gf[a](x) = \sum_{n=1}^\infty a_n x^n \ .
\end{equation}
As for Laplace transform, convolutions of discrete sequences
translates into products of the corresponding generating functions.
We shall sometimes introduce (right) partial sums of a sequence, denoted with
the same upper-case letter:
\begin{equation}
A_n = \sum_{n'=n}^\infty a_{n'} \ .
\end{equation}
The corresponding g.f. $f = \gf[a]$ and $F = \gf[A]$ are then related by
\begin{equation}
f(x) = \frac{x-1}{x} F(x) + A_1 \ .
\end{equation}

The asymptotic behavior of a function shows up in the singularities
of its Laplace transform. For instance, if 
$a(t)\simeq A\, t^\mu$ when $t \to \infty$, then 
\begin{eqnarray}
\Lap[a](y)  = A\,\Gamma(1+\mu)
y^{-1-\mu}+\hat{a}_{\rm reg}(y)\, ,
\end{eqnarray}
where $\hat{a}_{\rm reg}(y)$ denotes a function having milder singularities
than $y^{-1-\mu}$ at $y\to 0$. A formally identical formula holds when
$a(t)\simeq A\, t^\mu$ when $t \to 0$ (in this case one has to look at
$y\to\infty$).

Similar relations holds for generating functions. 
If $a_n\simeq A\,n^{\mu}$ for large $n$, then
the associated generating function behaves as follows near $x=1$:
\begin{eqnarray}
\gf[a](x) = A\,\Gamma(1+\mu) (1-x)^{-1-\mu}+\hat{a}_{\rm reg}(x)\, .
\label{eq:GFBasic}
\end{eqnarray}
%
%
%
\subsection{Minimal size rearrangements}
\label{sec_app_minsize}

Throughout this Section, we shall denote by $f(x)$ (respectively $\fh(x)$)
the generating function of the distribution $q_n$ (resp. $\qh_n$)
of rearrangements minimal sizes. With this notation, 
Eq.~(\ref{eq:MinSizeDistr1}) can be written as
\begin{eqnarray}
f(x) = x\, \exp\big\{-p\g+p\g\fh(x)\big\}\, .\label{eq:MinSizeGF}
\end{eqnarray}
%
%
%

\subsubsection{Order parameter}

Let us first determine the order parameter of the transition, that is to
say the fraction of variables with diverging minimal rearrangement sizes. This
is defined as $\phi = \lim_{n \to \infty} Q_n=1-f(1)$. 
Similarly, let $\hat{\phi} = \lim_{n \to \infty} \hat{Q}_n
=1-\fh(1)$. From Eq.~(\ref{eq:MinSizeDistr2}) one obtains 
$\phh =\phi^{p-1}$, whereas Eq.~(\ref{eq:MinSizeGF}) evaluated at
$x=1$ leads  to
\begin{equation}
\phi = 1 - \exp[- p\g \phi^{p-1}] \ .
\label{eq_orderparam}
\end{equation}
This has a non vanishing solution only for $\g\ge\g_{\rm d}$,
where $\g_{\rm d}$ is the critical point defined in 
Eq.~(\ref{eq:CriticalPoint}).
The critical point and the value of the order parameter at this point can be 
determined by the two equations
\begin{eqnarray}
\phi_{\rm d} & = & 1 - \exp[- p\g_{\rm d}  \phi_{\rm d}^{p-1}] \, ,
\label{eq_orderparam_crit} \\
1 & = &\g_{\rm d} p (p-1) \phi_{\rm d}^{p-2} \exp[- p\g_{\rm d} 
\phi_{\rm d}^{p-1}] \, . \label{eq_orderparam_tangent}
\end{eqnarray}
The last of these equations expresses the fact that the two functions
of $\phi$ on the left and right hand sides of Eq.~(\ref{eq_orderparam})
become tangent at $\phi_{\rm d}$ when $\g =\g_{\rm d}$.
%
%
\subsubsection{Critical decay}

We shall first place ourselves right at the transition, $\g = \g_{\rm d}$,
and study the decay of $Q_n$ towards its asympotic value $\phi_{\rm d}$. 
Define
$\epsilon_n = Q_n - \phi_{\rm d}$ and $\eph_n = \Qh_n - \phh_d$.
The generating functions are given by
\begin{equation}
f(x)=1-\phi_{\rm d} - \frac{1-x}{x} \gf[\epsilon_n](x) \ ,
\end{equation}
and a similar expression for $\hat{f}(x)$. In terms of these 
quantities, Eq.~(\ref{eq:MinSizeDistr2}) becomes 
$\hat{\epsilon}_n = (\phi_{\rm d} + \epsilon_n)^{p-1} - \hat{\phi}_{\rm d}$.

Guided by the numerical solution of Eqs.~(\ref{eq:MinSizeDistr1}),
(\ref{eq:MinSizeDistr2}) we look for an asymptotic behavior of the form
$\epsilon_n \simeq  A \, n^{-a}$ as $n\to\infty$.
Here $a$ is a positive exponent and $A$ a
constant. Equivalently $q_n \simeq A  \, a \, n^{-a -1}$. 
Expanding the relation between $\epsilon_n$ and $\hat{\epsilon}_n$, we obtain
\begin{equation}
\hat{\epsilon}_n =  (p-1) \phi_{\rm d}^{p-2} \epsilon_n
+ \binom{p-1}{2} \phi_{\rm d}^{p-3} \epsilon_n^2 +O(\epsilon_n^3)\ .
\end{equation}
Our aim is now to determine the decay exponent $a$, by matching the 
singularities around $x=1$ in the equation on the generating functions.

Using Eq.~(\ref{eq:GFBasic}), we obtain
$\gf[\epsilon_n](1-s)\simeq A \,\Gamma(1-a)s^{-1+a}$ as $s\to 0$.
Assuming $0<a<1/2$, we apply the same formula to $\hat{f}$:
\begin{eqnarray}
f(1-s) & =& 1-\phi_{\rm d} - \frac{s}{1-s}\gf[\epsilon_n](1-s) \ , \\
\hat{f}(1-s) &\simeq& 1-\phh_d
- \frac{s}{1-s} (p-1)\phi_{\rm d}^{p-2} \gf[\epsilon_n](1-s) 
-  \frac{s}{1-s} \binom{p-1}{2}\phi_{\rm d}^{p-3} \gf[\epsilon_n^2](1-s) \ .
\end{eqnarray}
We can now plug these formulae in Eq.~(\ref{eq:MinSizeGF}), and match the 
two sides order by order as $s\to 0$. The terms of order $s^0$ are equal
thanks to the equation (\ref{eq_orderparam_crit}) on the order parameter.
The terms proportional to $s\gf[\epsilon_n](1-s)$ (of order
$s^a$) match because of Eq.~(\ref{eq_orderparam_tangent}).
Next, on the r.h.s. of  Eq.~(\ref{eq:MinSizeGF}) two terms
of order $s^{2a}$ appear (while none is present on the l.h.s.):
one is proportional to $s\gf[\epsilon_n^2](1-s)$, and the other
to $s^2\gf[\epsilon_n]^2(1-s)$. Requiring them to cancel yields
the relation  (\ref{eq_MCT1_a})
which fixes the exponent $a$.

The l.h.s. of Eq.~(\ref{eq_MCT1_a})
is shown in Fig.~\ref{fig_eq_MCT}. As long as $\lambda\in (0,1]$, 
the equation has a unique solution in $[0,1/2)$. The exponent
$a$ is given by this solution, thus verifying the hypothesis $a<1/2$.
It order to show that  $\lambda \le 1$ for any $p$, we use 
Eq.~(\ref{eq_orderparam_tangent}) to simplify the expression for $\lambda$:
\begin{equation}
\lambda = (p-2) \frac{1-\phi_{\rm d}}{\phi_{\rm d}} \ .
\end{equation}
Using the 
inequality $e^x \ge 1+x$ in Eq.~(\ref{eq_orderparam_crit}), we obtain
$\g_{\rm d} p \phi_{\rm d}^{p-2} \ge 1$.
Thanks to (\ref{eq_orderparam_tangent}) we get $(p-1)(1-\phi_{\rm d}) \le 1$,
which in turns implies the thesis.
%
%
\subsubsection{Intermediate regime}
\label{app:IntermediateSizes}

We shall now study the behaviour of the probability law of the minimal
rearrangements sizes slightly below the critical connectivity, i.e. 
as $\delta \to 0$.

Inspired by the distribution of zero-temperature point-to-set correlation 
lengths, studied in Sec.~\ref{sec:PointToSet}, we guess that an intermediate
scaling regime emerges when $Q_n = \phi_{\rm d}+O(\delta^{1/2})$.
Using the results of the previous Section, we expect this to happen
on a scale $n_0(\delta)\sim \delta^{-1/2a}$. For the time being, we 
only assume that $n_0(\delta)$ diverges at $\delta=0$ more rapidly than 
$\delta^{-1}$.

In order to blow up $Q_n$ around its plateau, we define, for any $t>0$
\begin{equation}
\epsilon(t) = \lim_{\delta \to 0} \delta^{-1/2} [Q_{n=t n_0}-\phi_{\rm d}] \ ,
\label{eq_def_epsilon_t}
\end{equation}
and similarly $\hat{\epsilon}(t)$.
Using Eq.~(\ref{eq:MinSizeDistr2}), we get
\begin{equation}
\hat{\epsilon}(t) = (p-1) \phi_{\rm d}^{p-2} \epsilon(t) + \binom{p-1}{2} 
\phi_{\rm d}^{p-3} \delta^{1/2} \epsilon(t)^2 +O(\delta)\, .
\end{equation}

In order to describe the plateau regime, we must properly rescale
the generating functions as $\delta \to 0$:
\begin{eqnarray}
f(1 - y/n_0) & = & 1-\phi_{\rm d} - \delta^{1/2} y 
\Lap[\epsilon](y) + o(\delta)\ , \\
\fh(1 - y/n_0) & = & 
1 - \phh_d - \delta^{1/2} (p-1) \phi_{\rm d}^{p-2}
y \Lap[\epsilon](y) 
 - \delta \binom{p-1}{2} \phi_{\rm d}^{p-3} 
y \Lap[\epsilon^2](y) + o(\delta)  .
\end{eqnarray}
In order to estimate the remainders, we assumed that $\epsilon(t)$
is smooth for $t>0$ and has a singularity at $t=0$ which is milder than
$t^{-1/2}$.

We can plug these formulae for $f(\,\cdot\,)$ and $\fh(\,\cdot\,)$,
into Eq.~(\ref{eq:MinSizeGF}) and compare the two sides order by order in
$\delta$.
The terms of order $\delta^0$ and $\delta^{1/2}$ match because of
Eqs.~(\ref{eq_orderparam_crit}) and (\ref{eq_orderparam_tangent}). On the
contrary the order $\delta$ determines the function $\epsilon(t)$ via
\begin{equation}
\lambda \Lap[\epsilon^2](y) =\frac{B}{y}\; +\;
y \left\{ \Lap[\epsilon](y) \right\}^2 \ ,\label{eq:SizePlateauEq}
\end{equation}
where  $B\equiv 2/(\g_{\rm d}^2 p (p-1)^2 \phi_{\rm d}^{p-3})$,
and $\lambda$ has been  defined above.
Inverse Laplace transforming the equation yields to
\begin{equation}
\lambda \ \epsilon(t)^2 = B\; +\;\frac{\de {\phantom y}}{\de t}\int_0^t  \!
\epsilon(u) \, \epsilon(t-u)\; \de u  \ .
\end{equation}
From this equation one can determine the asymptotics of $\epsilon(t)$, which
turn out to be:
\begin{equation}
\epsilon(t) \underset{t \to 0}{\sim}  \ t^{-a} \ , \qquad
\epsilon(t) \underset{t \to \infty}{\sim}  \ t^{b} \ ,
\end{equation}
where $a$ is the exponent already found in the previous Section,
and $b$ is the positive solution of Eq.~(\ref{eq_MCT1_b}).

This allows us to fix the scale $n_0$. Consider indeed
large (but independent of $\delta$) values of $n$, and the limit 
$\delta \to 0$. From the
study of the previous Section  at $\g=\g_{\rm d}$, we know that
$\epsilon_n \sim n^{-a}$. From the behavior of $\epsilon(t)$ we have 
$\epsilon_n \sim \delta^{1/2} \epsilon(n/n_0)\sim 
\delta^{1/2} (n/n_0)^{-a}$. Consistency in the limit $\delta \to 0$ implies
thus $n_0 \sim \delta^{-1/2a}$, confirming the qualitative argument given
in the beginning of this Section.
\begin{figure}
\includegraphics[width=7cm]{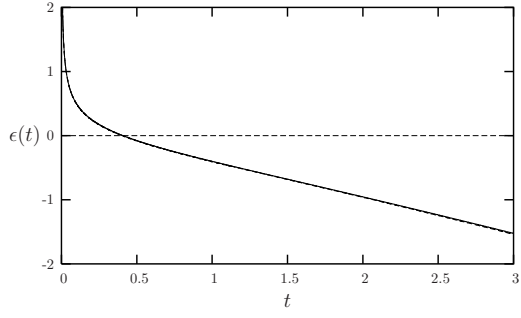}
\caption{The scaling function of the intermediate regime, $\epsilon(t)$, as
defined in Eq.~(\ref{eq_def_epsilon_t}). The datas used are at 
$\g = 0.816 , 0.817, 0.818$, and are almost superimposed.}
\label{fig_epsilon_t}
\end{figure}

In Fig.~\ref{fig_epsilon_t} we use the numerically determined  
distribution $Q_n$ in order to check the scaling form used in this Section.
%
%
\subsubsection{Final regime}
\label{sec_app_minsize_slow}

At what size scale $n_0'(\delta)$ does $Q_n$ exit the plateau regime
and decreases by a finite amount below $\phi_{\rm d}$?
Using the intermediate regime scaling, we have
$Q_{n_0'}-\phi_{\rm d} \simeq \delta^{1/2}\epsilon(n'_0 / n_0) = O(1)$.
From the behaviour of $\epsilon(t)$ at large $t$,
we get $n'_0(\delta) \sim \delta^{- \nu}$, where 
\begin{equation}
\nu = \frac{1}{2a}+\frac{1}{2b} \ .
\end{equation}

In order to describe this regime, we define the scaling limits
\begin{equation}
Q(t) = \lim_{\delta \to 0} Q_{n=t n'_0} \, , \;\;\;\;\;\;\;\;
\psi(y) = \lim_{\delta \to 0} f(1 - y/n'_0) \ ,
\label{eq_def_Qt}
\end{equation}
and analogous ones for $\Qh_n$ and $\fh(x)$ (with scaling functions
$\Qh(t)$ and $\hat{\psi}(y)$).
As usual, the scaling limits of generating functions
are related to Laplace transforms: $\psi(y) = 1- y \Lap[Q](y)$,
$\hat{\psi}(y) = 1- y \Lap[\Qh](y)$. Equations
(\ref{eq:MinSizeDistr2}) and (\ref{eq:MinSizeGF}) imply
\begin{equation}
 \Qh(t) = Q(t)^{p-1} \ , \ 
\psi(y) = \exp[-p\g_{\rm d}  + p\g_{\rm d} \, \hat{\psi}(y) ] \, .
\label{eq:FinalScaling}
\end{equation}

This set of equations determines $Q(t)$, the scaling function
of the decay of $Q_n$ between $\phi_{\rm d}$ and $0$
(this function was 
denoted as $Q_{\rm slow}(t)$ in Sec.~\ref{sec:AsymSizes}).
In particular they imply $Q(t) \simeq_{t \to 0} \phi_{\rm d} - {\rm cst} \ t^b$.
This behavior can be matched with the intermediate regime, cf. previous
Section, and can be used to confirm that $n'_0(\delta)\sim\delta^{-\nu}$.
\begin{figure}
\includegraphics[width=7cm]{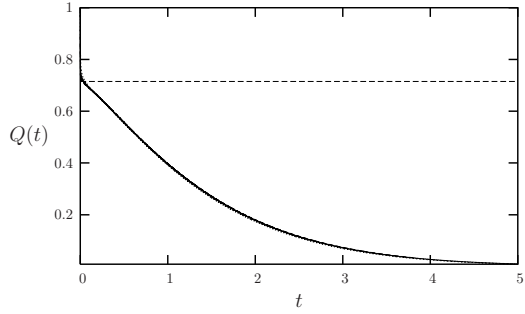}
\caption{The scaling function of the final regime, $Q(t)$, as
defined in Eq.~(\ref{eq_def_Qt}). The datas used are at 
$\g = 0.812, 0.816 , 0.817, 0.818$.}
\label{fig_Qt}
\end{figure}

In Fig.~\ref{fig_Qt} we check the scaling form used in this Section
against the numerical solution of Eqs.~(\ref{eq:MinSizeDistr1}),
(\ref{eq:MinSizeDistr2}).
%
%
\subsubsection{Extreme decay}

What is the probability of an ``anomalous" minimal size rearrangement,
i.e. a site $i$ such that  $n_i$ is much larger than for most of the other
sites? This question can be answered by studying the $n\to\infty$ 
asymptotics of the sizes distribution $Q_n$.

It is convenient to consider  first a slightly modified model, in 
which the maximum allowed variable degree $l_{\rm max}$ is a finite number
as $N\to\infty$. We refer to \cite{Bollobas} for graph ensembles of this type.
It is easy to generalize our treatment to such graphs: 
Eq.~(\ref{eq:MinSizeDistr2}) remains unchanged, while Eq.~(\ref{eq:MinSizeGF})
is modified to
\begin{eqnarray}
f(x) = x\sum_{l=1}^{l_{\rm max}} p_{l}\;\fh(x)^{l-1}\, .\label{eq:MinSizeGFGen}
\end{eqnarray}
Here $p_{l}$ is the probability that an edge chosen uniformly at random
in the factor graph $\cF$ is adjacent to a variable node of 
degree $l$. 

A numerical investigation of Eqs.~(\ref{eq:MinSizeDistr2}) and
(\ref{eq:MinSizeGFGen}) suggests the Ansatz
\begin{eqnarray}
Q_n = \exp\left\{-\omega\, n^{1+z}+o(n^{1+z})\right\}\, ,
\end{eqnarray}
with $z>0$.
Equation (\ref{eq:MinSizeDistr2}) implies 
$\Qh_n = \exp\left\{-(p-1)\omega\, n^{1+z}+o(n^{1+z})\right\}$. The 
corresponding generating functions can be estimated through the
saddle point method. We get
\begin{eqnarray}
f(x) =\exp\left\{z\omega^{-1/z}\left(\frac{\log x}{z+1}\right)^{\frac{z+1}{z}}
+o\left((\log x)^{\frac{z+1}{z}}\right)\right\}\, ,
\end{eqnarray}
as $x\to\infty$. The analogous expression for $\fh(x)$ is obtained
replacing $\omega$ by $(p-1)\omega$ in the last formula. 
Inserting into Eq.~(\ref{eq:MinSizeGFGen}) and matching the leading term as 
$x\to\infty$ determines the exponent $z$
\begin{eqnarray}
z = \frac{\log(p-1)}{\log(l_{\rm max}-1)}\, .\label{eq:Extreme}
\end{eqnarray}
In other words, the probability of very large rearrangements decreases
faster than exponentially. In a given graph $\cG$ of size $N$, the 
largest minimal size rearrangement $n_{\rm max}$ can be estimated by setting 
$NQ_{n_{\rm max}} = O(1)$, which implies
\begin{eqnarray}
n_{\rm max} \sim (\log N)^{\frac{1}{1+z}}\, .
\end{eqnarray}

Let us now reconsider the original model, with unbounded degree.
The expression (\ref{eq:Extreme}) yields $z\to 0$. A more careful study
shows that 
\begin{eqnarray}
Q_n = \exp\left\{ -n\, e^{k\, \varphi(n)\,[1+o(1)]}\right\}\, ,
\end{eqnarray}
where $\varphi(n)$ is a function diverging as $n\to\infty$,
but more slowly than any iterated 
logarithm\footnote{More precisely, it is a solution of the Abel 
equation~\cite{Kucma} $\varphi(n)=\varphi(\log n)+1$.} (i.e., for instance,
than $\log\log\log n$), and $k$ is a positive constant.
%
%
\subsection{Minimal barrier rearrangements}
\label{sec_app_minbarr}

We present in this Appendix  the resolution of 
Eqs.~(\ref{eq_minbarr_distrib}) on
the distribution of minimal barriers. 
We shall first discuss a more general class of
distributional equations. The representation developed in this context 
is then used to solve numerically Eqs.~(\ref{eq_minbarr_distrib}). 
Finally, we analyze the critical behavior of its solution.
%
%
\subsubsection{Transformation of the permutation functionals}

Equations (\ref{eq_minbarr_distrib}) 
involve functionals of the probability distribution $\qh$
which can be generalized to the following problem. Consider $l$ integers 
(fixed) $y_1,\dots,y_l$, that without loss of generality we assume 
to be ordered, $y_1 \le \dots \le y_l$, and define a function 
$f_l$ of $l$ other integers by
\begin{equation}
f_l(b_1,\dots,b_l) = \min_{\pi} \max_{i\in[1,l]}[b_{\pi(i)}+y_i] \ ,
\end{equation}
where the minimum is over the permutations $\pi$ of $[1,l]$.
Assume the $b_i$'s to be  i.i.d. random
variables with distribution $q$. What is the  distribution of
$f_l(b_1,\dots,b_l)$ ? Its cumulative distribution is
\begin{equation}
1-F_b^{(l)}=\sum_{b_1,\dots,b_l} q_{b_1}\dots q_{b_l} \,
\ind(f_l(b_1,\dots,b_l) \ge b) \ ,
\label{eq_perm_def}
\end{equation}
which is a functional of the law $q$.

A first simplification is that 
the minimum in the definition of $f_l$ is achieved
for a permutation $\pi_0$ which orders the $b_i$'s by their value
(and any such permutation is equivalent). 
In other words $x_i=b_{\pi_0(i)}$ with $x_1 \ge \dots \ge x_l$.
Now one has
\begin{equation} 
\ind(f_l(b_1,\dots,b_l) \ge b) =
\mathbb{I}\left(\max_{i\in[1,l]}[x_i+y_i] \ge b \right) =
1-\mathbb{I}\left(\max_{i\in[1,l]}[x_i+y_i] < b \right) =
1-\prod_{i=1}^l \mathbb{I}(x_i+y_i<b) \ .
\label{eq_perm1}
\end{equation}

It is useful to represent the $y_i$'s in a slightly different form. Let us
call $l'\in[1,l]$ the number of distinct elements in the set 
$\{y_1,\dots,y_l\}$, and $y'_1 < \dots < y'_{l'}$ their values. The 
degeneracies of the $y'$ are encoded in the coefficients $d_j$ defined for
$j\in[1,l']$ as the number of $y_i$ strictly smaller than $y'_j$. By definition
one has $0=d_1<\dots<d_{l'}<l$. As an  example, if for $l=6$,
$(y_1,\dots,y_6)=(3,3,5,6,6,8)$, one obtains $l'=4$, 
$(y'_1,\dots,y'_4)=(3,5,6,8)$ and $(d_1,\dots,d_4)=(0,2,3,5)$. These 
definitions can be used to remove the redundant constraints of 
Eq.~(\ref{eq_perm1}): if for instance $y_i=y_{i+1}$, 
$x_i+y_i<b \Rightarrow x_{i+1}+y_{i+1}<b$ as the $x_i$'s are decreasing. These
eliminations lead to
\begin{equation}
\ind(f_l(b_1,\dots,b_l) \ge b) = 
1-\prod_{j=1}^{l'} \mathbb{I}(x_{d_j+1}+y'_j<b) \ .
\label{eq_perm2}
\end{equation}
We now come back to the original variables $b$ and define $m_j$, 
for $j\in[1,l']$, as the number of $b_i$'s strictly smaller than $b-y'_j$.
This imply that 
\begin{equation}
x_1 \ge \dots \ge x_{l-m_j} \ge b-y'_j > x_{l-m_j+1} \ge \dots \ge x_l \ .
\end{equation}
We thus have $x_{d_j+1}+y'_j<b \Leftrightarrow m_j\ge l-d_j$. We can now
insert Eq.~(\ref{eq_perm2}) in the definition (\ref{eq_perm_def}) and trade
the sum on the $b_i$ against a sum on the $m_j$ respecting the above derived
constraints. 
\begin{equation}
F_b^{(l)} = \sum_{\{m_j\}}\hspace{-0.15cm}{}'\hspace{0.15cm} 
\frac{l!}{\prod_{j=1}^{l'-1}(m_j-m_{j+1})! m_{l'}!}
\prod_{j=1}^{l'-1}(Q_{b-y'_{j+1}}-Q_{b-y'_j})^{m_j-m_{j+1}} 
(1-Q_{b-y'_{l'}})^{m_{l'}} \ .
\label{eq_perm_sum}
\end{equation}
The combinatorial factor arises from the change of variables from
the $b$'s to the $m$'s, and the constrained sum is over
\begin{equation}
l=m_1 \ge \dots \ge m_{l'} \ , \quad m_j \ge l-d_j \ \forall j\in[1,l'] \ .
\label{eq_perm_constraint}
\end{equation}
In the following, we shall repeatedly  use the representation of $F_b^{(l)}$
provided by (\ref{eq_perm_sum}), (\ref{eq_perm_constraint}).

Before leaving this general preliminary, let us note that if the $b_i$'s are 
bounded from below by $b_{\rm min}$ (i.e. $q_b=0$ for $b<b_{\rm min}$), 
$f_l(b_1,\dots,b_l)$ is bounded from below by $b_{\rm min}+y_l$, hence 
$F_b^{(l)}=0$ for $b\le b_{\rm min}+y_l$.
%
%
\subsubsection{Application to the $p$-spin model and numerical resolution}
\label{sec_app_minb_num}

The equations (\ref{eq_minbarr_distrib}), determining the barrier distribution 
of the $p$-spin model (\ref{Hamiltonian}), can be rewritten as 
\begin{eqnarray}
Q_b = \sum_{l=0}^\infty p_l (1-F_b^{(l)}[\{\hat{q}\}]) \ , 
\;\;\;\;\;\;\;\;\;
\Qh_b = Q_b^{p-1} \ ,
\label{eq_Qb}
\end{eqnarray}
where $F_b^{(l)}$ is of the general form studied in the previous pages,
with the $b_i$'s drawn with the law $\qh$, and the $y_i$'s given
by $y_i = \left\lfloor \frac{i}{2} \right\rfloor$. Here we set by
convention $F^{(0)}_b =\ind(b\ge 2)$.

Following the notations of the previous Section, one has for a given value
of $l$: $l'=\left\lfloor \frac{l}{2} \right\rfloor+1$, $y'_i=i-1$, $d_1=0$ and
$d_j=2j-3$ for $2 \le j\le l'$. As the law $\hat{q}$ is supported on the 
positive integers, $F_b^{(l)}=0$ for $b\le l'$.
This allows to rewrite the equation (\ref{eq_Qb}) under the form:
\begin{eqnarray}
1-Q_b &=& p_0 \ \ind(b\ge2) + p_1 (1-\Qh_b) +
\ind(b\ge 3)\sum_{l=2}^{2b-3} p_l \ F_b^{(l)}[\{\hat{q}\}] \ ,
\label{eq_Qb1} \\
F_b^{(l)}[\{\hat{q}\}] &=&\!\! \sum_{\{m_j\}}\!\!{}'\;\frac{l!}
{\prod_{j=1}^{\left\lfloor l/2 \right\rfloor}(m_j-m_{j+1})! 
m_{\left\lfloor l/2 \right\rfloor+1}!}
\left(\prod_{j=1}^{\left\lfloor l/2 \right\rfloor}
\hat{q}_{b-j}^{m_j-m_{j+1}} \right)
(1-\hat{Q}_{b- \left\lfloor l/2 \right\rfloor})
^{m_{\left\lfloor l/2 \right\rfloor+1}}  ,
\label{eq_Qb2}
\end{eqnarray}
with the constraints on the sum in $F_b^{(l)}$:
\begin{equation}
l=m_1 \ge \dots \ge m_{\left\lfloor l/2 \right\rfloor+1} \ , \quad 
m_j \ge l+3-2j \ \ \forall j\in[2,\left\lfloor l/2 \right\rfloor+1] \ .
\end{equation}
These equations allow to determine the distribution of messages
$v_{i\to\a}$, $u_{\a\to i}$ defined in Sec.~\ref{sec:PspinBarr}
(anchored barriers).
The distribution of actual barriers to flip a spin in the system can
be expressed in a similar way,
\begin{eqnarray}
1-Q^{\rm site}_b&=& p_0 \ \ind(b \ge 1) + p_1 (1-\hat{Q}_b) + 
\ind(b \ge 2) \sum_{l=2}^{2b-2} p_l G_b^{(l)}[\{\hat{q}\}] \ ,\\
G_b^{(l)}[\{\hat{q}\}] &=& \!\!\sum_{\{m_j\}}\!\!{}'\; \frac{l!}
{\prod_{j=1}^{\lceil l/2 \rceil-1}(m_j-m_{j+1})! 
m_{\lceil l/2\rceil}!}
\prod_{j=1}^{\lceil l/2 \rceil-1}
\hat{q}_{b-j}^{m_j-m_{j+1}} 
(1-\hat{Q}_{b- \lceil l/2 \rceil+1})
^{m_{\lceil l/2 \rceil}}  ,
\end{eqnarray}
with the constraints on the sum in $G_b^{(l)}$:
\begin{equation}
l=m_1 \ge \dots \ge m_{\lceil l/2 \rceil} \ , \quad 
m_j \ge l+2-2j \ \ \forall j\in[2,\lceil l/2 \rceil] \ .
\end{equation}
By convention the products from $j=1$ to $j=0$ which appear in $G_b^{(2)}$ are
equal to 1.

Unlike for the minimal size distribution, Eqs.~(\ref{eq_Qb}) 
do not provide immediately an algorithm for computing $Q$ by recursion over 
$b$. The reason is that the 
anchored barrier of a rooted tree is at least as large, but not 
necessarily larger, than the anchored barriers of its sub-trees.
Consider now the process of merging $l$ sub-trees (this corresponds to
iterating (\ref{eq_Qb}) once). If sub-trees barrier distribution 
are known only up to barrier height $b-1$, the barrier distribution of the  
new, larger  tree is necessarily undetermined at height $b$.

We can propose two ways to solve this difficulty. The first one is 
to derive a well-behaved recursion on $b$. Assume that $Q$ and $\Qh$ 
(resp. $q$ and $\qh$) are known upto height $b-1$ (resp. $b-2$). 
The only unkwnown of the right hand side of Eq.~(\ref{eq_Qb1}) is 
$\qh_{b-1}$, which appears linearly because of the conditions 
$m_1=l, m_2\ge l-1$ in the sum (\ref{eq_Qb2}). We can thus write
\begin{equation}
1-Q_b = A_b\,\qh_{b-1}  + B_b \ ,
\end{equation}
where the coefficients $A_b$ and $B_b$ are computable from 
$\qh_1,\dots,\qh_{b-2}$ using Eq.~(\ref{eq_Qb2}). 
Using $\qh_{b-1}=\Qh_{b-1}-\Qh_b$, and $\Qh_b=Q_b^{p-1}$,
one ends up with
\begin{equation}
1- Q_b + A_b Q_b^{p-1} = A_b  \,\Qh_{b-1} + B_b \ .
\end{equation}
One can solve this equation for $Q_b$ (recall that $\Qh_{b-1}$ is assumed to
be known at this stage) and
therefore determine $q_{b-1}$, $\Qh_b = Q_b^{p-1}$ and $\qh_{b-1}$.
This allows to continue the recurrence at the next barrier height. 

The second method is iterative: make an initial guess for the law $Q$,
compute the law $\Qh_b=Q_b^{p-1}$, cf.~Eq.~(\ref{eq_Qb}), and inject it in 
(\ref{eq_Qb1}) to recompute $Q$. Any fixed point of this procedure 
corresponds to a solution of Eqs.~(\ref{eq_Qb}).  Since the 
solution is unique, whenever the procedure converges,
it provides a good numerical approximation of this solution.
It can be argued\footnote{A complete proof is beyond the scope of this 
Appendix. The basic idea is that the function $Q_b^{(\ell)}$ 
obtained after $\ell$ iterations, is the minimal barrier distribution 
for a subgraph of radius $\ell$ around a random node $i$.
Convergence follows from the remark that this barriers become
larger (and therefore  $Q_b^{(\ell)}$ worsen, in a sense that 
can be made precise) as $\ell$ increases.} 
that the iterative procedure indeed converges for
any $\g<\g_{\rm d}$, if the initial condition $Q^{(0)}_b  = \ind(b\le 1)$
is used. 
%
%
\subsubsection{Asymptotics}
\label{sec_app_minb_ana}

In this Appendix we work out the asymptotic behavior of minimal barrier 
distributions, cf.~Sec.~\ref{sec_minbarr_as}. 
Motivated by the results on minimal size rearrangements, 
cf.~Sec.~\ref{sec_app_minsize}, we look for a plateau at 
$Q_b\approx \phi_{\rm d}$, $\Qh_b\approx\phh_{\rm d}$. 
We thus define
\begin{equation}
Q_b=\phi_{\rm d}+\epsilon_b \ , \;\;\;\;\;\;\; 
\Qh_b=\phh_{\rm d}+\eph_b \ .
\end{equation}
With this definition one can
set up an expansion of the functionals $F_b^{(l)}$ in powers of 
$\hat{\epsilon}$. We shall need in the following this expansion upto terms
of order $\hat{\epsilon}^2$. 
Four types of terms appear at this order in Eq.~(\ref{eq_Qb2}) :
\begin{itemize}
\item[$\bullet$] $m_1 = \dots = m_{\left\lfloor l/2 \right\rfloor+1}=l$, which
gives a term of order 1.
\item[$\bullet$] $m_1 = \dots =m_s=l$, 
$m_{s+1}= \dots = m_{\left\lfloor l/2 \right\rfloor+1}=l-1$, 
with $1 \le s<\left\lfloor l/2 \right\rfloor+1$, of order $\hat{\epsilon}$.
\item[$\bullet$] $m_1 = \dots =m_s=l$, 
$m_{s+1}= \dots = m_{\left\lfloor l/2 \right\rfloor+1}=l-2$, 
with $2 \le s<\left\lfloor l/2 \right\rfloor+1$, of order $\hat{\epsilon}^2$.
\item[$\bullet$] $m_1 = \dots =m_s=l$,
$m_{s+1} = \dots =m_t=l-1$, 
$m_{t+1}= \dots = m_{\left\lfloor l/2 \right\rfloor+1}=l-2$,
with $1 \le s<t<\left\lfloor l/2 \right\rfloor+1$, of order $\hat{\epsilon}^2$.
\end{itemize}
Summing this terms and expanding the result upto
order $\hat{\epsilon}^2$, we obtain:
\begin{equation}
F_b^{(l)} = (1-\phh_{\rm d})^l - l (1-\phh_{\rm d})^{l-1} \eph_b + 
\binom{l}{2} (1-\phh_{\rm d})^{l-2} [2 \eph_b \eph_{b-1} 
- \eph_{b-1}^2 ] + O(\eph^3) \ .
\end{equation}
Inserting this expansion in Eq.~(\ref{eq_Qb1}), assuming $b\ge 3$ (we are
interested in the large $b$ behavior) leads to
\begin{multline}
1-\phi_{\rm d}-\epsilon_b = 
\left[ \sum_{l=0}^{2b-3} p_l (1-\hat{\phi}_{\rm d})^l \right]
- \left[ \sum_{l=0}^{2b-3} p_l l (1-\hat{\phi}_{\rm d})^{l-1} \right] 
\hat{\epsilon}_b+\\
+  \left[ \sum_{l=0}^{2b-3} p_l \binom{l}{2} (1-\hat{\phi}_{\rm d})^{l-2} 
\right]
[2 \hat{\epsilon}_b \hat{\epsilon}_{b-1} - \hat{\epsilon}_{b-1}^2 ]
+ O(\hat{\epsilon}^3) \ .
\end{multline}
Since we are interested in the large $b$ behavior, we can extend the sum on 
$l$ upto infinity with a negligible error\footnote{More precisely,
we shall eventually focus on $b=O{\bf (}\log(1/\delta){\bf )}$, and retain
terms in this equation up to $O(\delta)$. On the other hand,
extending the sums up to $l=\infty$ produces errors of
order $1/b!\doteq \exp(-b\log b)\doteq (\log(1/\delta))^{\log\delta}\ll \delta$
(to the leading exponential order).}. 
Moreover we can expand the first equation in (\ref{eq_Qb})
to express $\hat{\epsilon}_b$ in terms of $\epsilon_b$. Ordering
the different powers of $\epsilon$, we get
\begin{multline}
\left[1-\phi_{\rm d}-e^{-\g p \hat{\phi}_{\rm d}} \right]
+ \epsilon_b \left[\g p (p-1) \phi_{\rm d}^{p-2} e^{-\g p \hat{\phi}_{\rm d}}
- 1\right]+\\
+\frac{\g p (p-1)(p-2)}{2} \phi_{\rm d}^{p-3} e^{-\g p\hat{\phi}_{\rm d}} 
\epsilon_b^2
-\frac{(\g p (p-1))^2}{2} \phi_{\rm d}^{2p-4} e^{-\g p\hat{\phi}_{\rm d}}
(2 \epsilon_b \epsilon_{b-1} - \epsilon_{b-1}^2) = O(\epsilon^3)\ .
\end{multline}
We now consider the critical limit $\delta\to 0$ infinitesimal. 
The two terms in brackets vanishes at $\delta=0$, because of
Eqs.~(\ref{eq_orderparam_crit}) and (\ref{eq_orderparam_tangent}). 
Expanding also in powers of $\delta$, we obtain
\begin{equation}
B \delta = \lambda \epsilon_b^2 - 2\epsilon_b \epsilon_{b-1} + \epsilon_{b-1}^2
+O(\delta^2, \, \delta\epsilon,\, \epsilon^3)\ , \label{eq_minbarr_eps}
\end{equation}
where $B$ and $\lambda$ are as defined in 
App.~\ref{sec_app_minsize}, cf., for instance, Eq.~(\ref{eq:SizePlateauEq}).

Notice that Eq.~(\ref{eq_minbarr_eps}) is simpler than the analogous equation
for minimal size rearrangements, i.e.  Eq.~(\ref{eq:SizePlateauEq}).
In particular, it is not necessary to introduce generating
functions or Laplace transforms: Eq.~(\ref{eq_minbarr_eps}) is
local in the ``direct space" of  barriers.

Let us consider first the large $b$ behavior at the critical
point $\g_{\rm d}$. The left hand 
side of (\ref{eq_minbarr_eps}) vanishes (as well as all corrections 
proportional to $\delta$), and one can look for a solution 
such that $\epsilon_b \sim e^{-\omega_a b}$ as $b\to\infty$. 
Injecting this Ansatz in Eq.~(\ref{eq_minbarr_eps}) and
requiring the terms of order $e^{-2\omega_a b}$ to cancel
yields the condition $\lambda-2e^{\omega_a}+e^{2\omega_a}=0$,
which fixes $\omega_a$, as announced in Eq.~(\ref{eq_minbarr_oa}).

In order for Eq.~(\ref{eq_minbarr_eps}) to have a non-trivial limit
within the plateau regime, we must have $Q_b=\phi_{\rm d}+O(\delta^{1/2})$
in this regime.
We therefore assume the existence of a diverging scale
$b_0=b_0(\delta)$, and define 
$\bar{\epsilon}_m=\delta^{-1/2}\epsilon_{b=m+b_0}$. 
Substituting in Eq.~(\ref{eq_minbarr_eps}) and taking the $\delta\to 0$
limit, we get
\begin{equation}
B = \lambda \bar{\epsilon}_m^2 - 2\bar{\epsilon}_m \bar{\epsilon}_{m-1} + 
\bar{\epsilon}_{m-1}^2 \ .\label{eq:ScalingBarrEq}
\end{equation}
It is easy to work out the asymptotic behavior of the solution of 
this equation. We get
$\bar{\epsilon}_m\simeq C_- e^{-\omega_a m}$ as $m\to -\infty$, 
and  $\bar{\epsilon}_m\simeq -C_+e^{\omega_b m}$ as $m\to +\infty$, where 
$\omega_b$ is given by Eq.~(\ref{eq_minbarr_ob}) 
and $C_{\pm}$ are two positive constants.

\begin{figure}
\includegraphics[width=7cm]{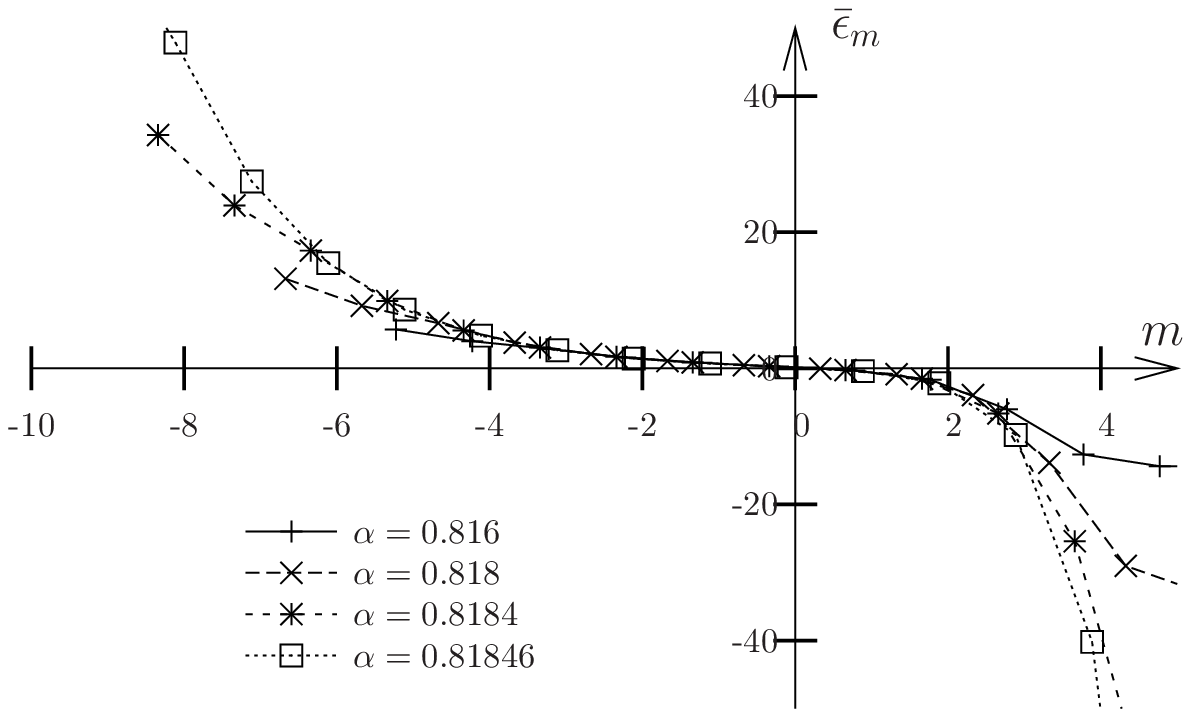} \hspace{1cm}
\includegraphics[width=7cm]{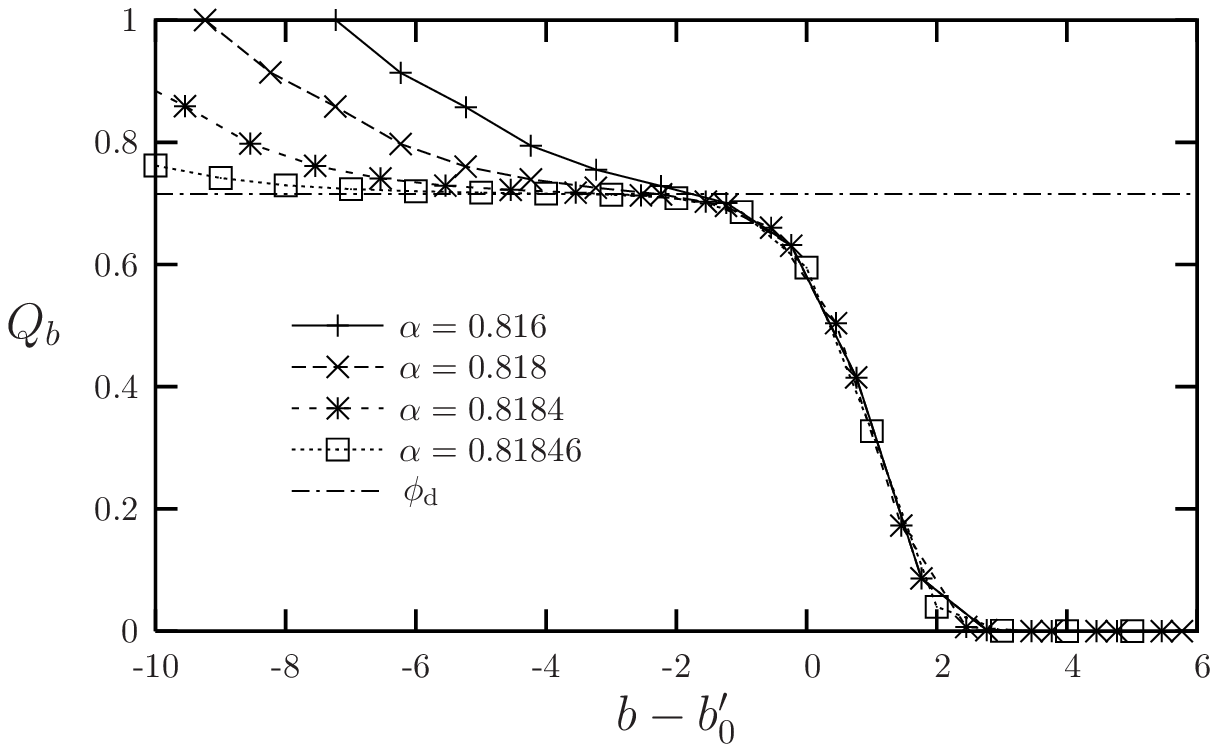}
\caption{Left: the scaling function $\bar{\epsilon}_m$ for the minimal
barrier rearrangements. Right: collapse on the scale of exit from the plateau.}
\label{fig_minbarr_as}
\end{figure}

The scale $b_0$ is fixed by matching the behaviour
of $\epsilon_b$ at $\g_{\rm d}$ for $b \to \infty$ with the one of
$\bar{\epsilon}_m$ when $m\to -\infty$. Consider a large (but independent of
$\delta$) value of $b$, and the limit $\g \to \g_{\rm d}$:
\begin{equation}
\epsilon_{b} \sim e^{-\omega_a b} \sim \delta^{1/2} \bar{\epsilon}_{m=b-b_0} 
\sim \delta^{1/2} e^{\omega_a b_0} e^{-\omega_a b} \ .
\end{equation}
The dependence on $\delta$ cancels only if
$b_0(\delta)\simeq \log(1/\delta)/(2\omega_a)$.
In Fig.~\ref{fig_minbarr_as}, left frame, we plot the outcome of a 
numerical resolution of the equations (\ref{eq_Qb}) along the lines
exposed in the previous Section. The data are rescaled in such a way
to exhibit the plateau scaling. More precisely, we plot
\begin{equation}
\delta^{-1/2}[Q_{b}-\phi_{\rm d}]\, , \label{eq:PlottedPlateau}
\end{equation}
for several values of $\delta$, 
as a function of $m=b-\log(1/\delta) /(2\omega_a)$.
The good collapse of the curves confirm the analysis
just presented.

Notice that the quantity (\ref{eq:PlottedPlateau}) is not expected to converge
to a unique curve as a function of $m$, even if $b_0(\delta)$
is chosen optimally (here we just derived its asymptotic behavior).
In other words, $\bar{\epsilon}_m$ is not a scaling function in the usual 
sense.
The reason is that  Eq.~(\ref{eq:ScalingBarrEq}) has more than one solution, 
and  two such solutions $\bar{\epsilon}^{(1)}$ and $\bar{\epsilon}^{(2)}$ 
cannot be superimposed through a shift of the $m$ axis (this is related to 
the fact that $m$ is discrete).
On the other hand, a shift can be found such that
$\bar{\epsilon}^{(1)}_m<\bar{\epsilon}^{(2)}_m<\bar{\epsilon}^{(1)}_{m+1}$.
Therefore the function  (\ref{eq:PlottedPlateau}) will converge, as a function
of $m=b-b_0(\delta)$, within a $O(1)$ distance around any solution
of Eq.~(\ref{eq:ScalingBarrEq}). This is enough for the analysis presented 
in this pages to be valid.

Finally, the scale $b'_0(\delta)$ for the decrease of $Q_b$ from 
$\phi_{\rm d}$ to $0$, can be defined as the smallest $b$ such that
$Q_b<Q_*$ with $Q_*$ any fixed number smaller than $\phi_{\rm d}$.
It can be estimated by requiring
$\epsilon_{b}$ to be of order 1, or, equivalently, $\bar{\epsilon}_b$ 
to be of order $\delta^{-1/2}$. This implies $\bar{\epsilon}_{b'_0-b_0}\sim \delta^{-1/2}$,
i.e. $e^{\omega_b (b'_0-b_0)} \sim \delta^{-1/2}$, hence 
\begin{equation}
b'_0(\delta) \simeq b_0(\delta) +\frac{1}{2\omega_b} \log(1/\delta)
\simeq \left(\frac{1}{2\omega_a}+\frac{1}{2\omega_b}\right)\, 
\log(1/\delta)   \ ,
\end{equation}
as announced in Eq.~(\ref{eq_minbarr_bp0}). 

Figure \ref{fig_minbarr_as}, right frame, shows $Q_{b}$ (obtained 
from the numerical resolution of Eqs.~(\ref{eq_Qb})), as a function
of $m'=b-b_0'(\delta)$. The curves for different values
of $\delta$ collapse between 0 and $\phi_{\rm d}$, thus confirming our
 analysis.

Contrary to the minimal size case, one cannot define a scaling function
for the last decay from the plateau value to 0. Indeed, this happens on an
interval of $b$ which remains of order 1 when $\delta \to 0$.  
This means that a finite fraction of the barriers is not just of order
$b_0'(\delta)$, but within a finite additive constant from $b_0'(\delta)$.

This asymptotic study has been performed on the distributions $Q$ and 
$\Qh$ of the messages on the links of the factor graph (anchored barriers). 
However, actual barriers to spin flips differ at most by one from
anchored barriers. The discussion presented so far is thus also valid 
for the actual barrier distribution $Q_b^{\rm site}$.
%
%
\subsection{Average size of minimal barrier rearrangement}
\label{sec_app_sizeminbarr}

Let us explain how to derive the results presented in 
Sec.~\ref{sec:SizeMinBarr} on the average size of minimal barrier 
rearrangements. To this aim one
has to solve Eqs.~(\ref{eq_sizebarr_1}) and (\ref{eq_sizebarr_2}). The first
of these equations involves a constrained sum on $l$ anchored barriers 
$b_1,\dots,b_l$, whose disjoint combination $\fh_l(b_1,\dots,b_l)$ 
leads to a prescribed barrier $b$. This is similar to 
Eq.~(\ref{eq_minbarr_distrib}) (first equation) that was solved in 
App.~\ref{sec_app_minbarr}.
Unlike in Eq.~(\ref{eq_minbarr_distrib}), one of the $l$ barriers is not 
drawn from the distribution $\hat{q}$, but is the argument of the unknown 
sequence $M$. Let us define
an integrated version of this sequence (as well as of the related $\Mh$),
\begin{equation}
N_b=\sum_{b'<b} M_{b'} \, , \;\;\;\;\;\;\;\;\;
 \Nh_b=\sum_{b'<b} \Mh_{b'} \, .
\end{equation}
Notice that, at odds with our usual conventions we integrated to the left. 
Equation (\ref{eq_sizebarr_1}) can thus
be rewritten as
\begin{equation}
N_b = 1-Q_b + \sum_{l=1}^{2b-3} p_l l\, N_b^{(l)} \ , \;\;\;\;\;\;\;\;\;\;
N_b^{(l)} = \sum_{b_1,\dots,b_l} \Mh_{b_1} \qh_{b_2} \dots
\qh_{b_l} \ind(\fh_l(b_1,\dots,b_l)<b) \ .
\label{eq_Nb}
\end{equation}
The sum in $N_b^{(l)}$ can be transformed following the strategy exposed in
App.~\ref{sec_app_minbarr}, treating the distinguished indice $b_1$ separately
from the others. A similar reasoning leads indeed to
\begin{multline}
N_b^{(l)} = \sum_{\{m_j,m'_j\}}\!\!\!\!{}'\;\; \frac{(l-1)!}
{\prod_{j=1}^{\left\lfloor l/2 \right\rfloor}(m_j-m_{j+1})! 
m_{\left\lfloor l/2 \right\rfloor+1}!}\cdot\\
\cdot\left(\prod_{j=1}^{\left\lfloor l/2 \right\rfloor}
\qh_{b-j}^{m_j-m_{j+1}} \Mh_{b-j}^{m'_j-m'_{j+1}} \right)
(1-\Qh_{b- \left\lfloor l/2 \right\rfloor})
^{m_{\left\lfloor l/2 \right\rfloor+1}}
(\Nh_{b- \left\lfloor l/2 \right\rfloor})
^{m'_{\left\lfloor l/2 \right\rfloor+1}} \ ,\label{eq:ExplicitSizeBarr}
\end{multline}
where the parameters of the sum obey the following constraints:
\begin{multline}
l-1=m_1 \ge \dots \ge m_{\left\lfloor l/2 \right\rfloor+1} \ , \quad 
1=m'_1 \ge \dots \ge m'_{\left\lfloor l/2 \right\rfloor+1} \ , \quad
m_j+m'_j \ge l+3-2j \ \ \forall j\in[2,\left\lfloor l/2 \right\rfloor+1] \ .
\end{multline}
We have encoded the value of the special indice $b_1$ in the sequence
of the $m'_j$, with $m'_j =1 \Leftrightarrow b_1 < b-j+1$.
The form (\ref{eq:ExplicitSizeBarr})
is suitable for numerical resolution of Eqs.~(\ref{eq_sizebarr_1}), 
(\ref{eq_sizebarr_2}), following one of the
two strategies (recursion or functional iteration) explained 
in Sec.~\ref{sec_app_minb_num}.

Let us now study the asymptotic regime $\g \to \g_{\rm d}$. Following
the notations of App.~\ref{sec_app_minbarr}, we can expand 
Eq.~ (\ref{eq:ExplicitSizeBarr}) in powers of
$\eph_b=\Qh_b - \phh_{\rm d}$:
\begin{equation}
N_b^{(l)} = (1-\phh_{\rm d})^{l-1} \Nh_b + 
(l-1) (1-\phh_{\rm d})^{l-2}[\Nh_{b-1} (\eph_{b-1}-\eph_b) - 
\Nh_b \eph_{b-1}] + O(\eph^2) \ .
\end{equation}
Inserting this expansion in Eq.~(\ref{eq_Nb}), and extending the sum on $l$ to 
infinity (as in Sec.~\ref{sec_app_minb_ana}, 
this produces a negligible error) leads to
\begin{equation}
N_b = 1-\phi_{\rm d} - \epsilon_b + \g p e^{-\g p \phh_{\rm d}}
\Nh_b + (\g p)^2 e^{-\g p \phh_{\rm d}} 
[\Nh_{b-1} (\hat{\epsilon}_{b-1}-\eph_b) - 
\Nh_b \hat{\epsilon}_{b-1}] + O(\eph^2) \ .\label{eq_sizebarr_asy01}
\end{equation}
Expanding also Eq.~(\ref{eq_sizebarr_2}) to the same order yields
\begin{eqnarray}
\Mh_b &=& M_b (p-1) \phi_{\rm d}^{p-2} \left[ 1 + 
\frac{p-2}{2 \phi_{\rm d}} (\epsilon_b + \epsilon_{b+1}) \right] 
+O(M\epsilon^2)\, .\label{eq_sizebarr_asy02}
\end{eqnarray}
Next we take finite differences of Eq.~(\ref{eq_sizebarr_asy01})
and trade $\Nh$ for $N$, thanks to (\ref{eq_sizebarr_asy02}), and
$\hat{\epsilon}$ for $\epsilon$, thanks to Eq.~(\ref{eq:MinSizeDistr2}).
We  finally get
\begin{multline}
[1-p(p-1)\g\phi_{\rm d}^{p-2}e^{-p\g\phh_{\rm d}}](N_{b+1}-N_b) = 
\frac{1}{2}p(p-1)(p-2)\g\phi_{\rm d}^{p-3}e^{-p\g\phh_{\rm d}}
(N_{b+1}-N_b)(\epsilon_b+\epsilon_{b+1})-\\
-p^2(p-1)^2\g^2\phi_{\rm d}^{2p-4}e^{-p\g\phh_{\rm d}}
[N_{b+1}\epsilon_b+N_b(\epsilon_{b+1}-\epsilon_b-\epsilon_{b-1})+
N_{b-1}(\epsilon_{b-1}-\epsilon_b)]+O(1,N\epsilon^2)\, .\label{eq_sizebarr_asy}
\end{multline}

As usual, we consider three distinct asymptotic regimes. Let us begin by
assuming $\g = \g_{\rm d}$ and $b$ large. The l.h.s. of 
Eq.~(\ref{eq_sizebarr_asy}) vanishes because of  
Eq.~(\ref{eq_orderparam_tangent}).
Moreover, we know from  App.~\ref{sec_app_minb_ana}
that $\epsilon_b\sim e^{-\omega_a b}$. Numerical
computations suggest the Ansatz $N_b \sim e^{\mu_a b}$, with 
$\mu_a > \omega_a >0$. Under this Ansatz terms of order $1$
and $N\epsilon^2$ can be safely neglected in
Eq.~(\ref{eq_sizebarr_asy}). Requiring the r.h.s. to vanish to the  
leading order $e^{(\mu_a-\omega_a) b}$, 
yields the condition
\begin{equation}
\frac{(e^{\mu_a -\omega_a}-1) 
(e^{\omega_a} + e^{-\mu_a} - e^{-(\mu_a - \omega_a)})}
{(e^{\mu_a}-1) (e^{-\omega_a}+1)} = \frac{\lambda}{2} \ .
\end{equation}
It is not hard to show that the above equation admits a unique solution larger
than $\omega_a$, thus completely determining the exponent  $\mu_a$.
For instance when $p=3$ one obtains 
$\mu_a \approx 0.81650 > \omega_a \approx 0.57432$.

Consider now the intermediate regime, $\delta  \to 0$,
with $b = m + b'_0(\delta)$, where  $b'_0(\delta)\simeq
\log(1/ \delta)/(2 \omega_a)$ and $m$ is finite. We found in this regime a 
scaling function $\bar{\epsilon}_m$, such that 
$\epsilon_b \simeq \delta^{1/2} \bar{\epsilon}_m$. Let us introduce
now a similar scaling function for the $N_b$'s,
$N_b \sim \delta^{-x} \overline{N}_m$, with $x$ a still unknown exponent that
we assume to be larger than $1/2$. With this assumption, the dominant terms in 
Eq.~(\ref{eq_sizebarr_asy}) are of order $\delta^{-x+1/2}$.  
Imposing Eq.~(\ref{eq_sizebarr_asy}) to hold at this order, we obtain
the following equation for the scaling function $\overline{N}_m$:
\begin{equation}
\overline{N}_{m+1} \overline{\epsilon}_m + 
\overline{N}_m (\overline{\epsilon}_{m+1} - \overline{\epsilon}_m - 
\overline{\epsilon}_{m-1}) +
\overline{N}_{m-1}(\overline{\epsilon}_{m-1} - \overline{\epsilon}_m)
 = \frac{\lambda}{2} (\overline{N}_{m+1} - \overline{N}_m)
(\overline{\epsilon}_{m+1} + \overline{\epsilon}_m) \ .
\end{equation}
We found in App.~\ref{sec_app_minb_ana} that 
$\overline{\epsilon}_m \sim e^{-\omega_a m}$ (resp. 
$\overline{\epsilon}_m \sim e^{\omega_b m}$)
in the $m \to -\infty$ (resp. $+\infty$) limit. 
Using these results, one can derive the asymptotic
behaviour of $\overline{N}_m$ in these two limits. 
When $m \to -\infty$, we  have $\overline{N}_m \sim e^{\mu_a m}$,
where $\mu_a$ is the exponent previously determined.
When $m \to +\infty$, $\overline{N}_m \sim e^{\mu_b m}$,
where $\mu_b$ is the unique positive solution of
\begin{equation}
\frac{(e^{\mu_b +\omega_b}-1) 
(e^{-\omega_b} + e^{-\mu_b} - e^{-\mu_b - \omega_b})}
{(e^{\mu_b}-1) (e^{\omega_b}+1)} = \frac{\lambda}{2} \ .
\end{equation}
For instance, we obtain $\mu_b \approx 3.90350$ for $p=3$.

The exponent $x$ is obtained by matching  the limit
$b \to \infty$ of the $\g=\g_{\rm d}$ solution, and the limit $m \to -\infty$
of the intermediate regime:
\begin{equation}
N_b \sim e^{\mu_a b} \sim \delta^{-x} e^{\mu_a (b-b_0)} = 
e^{\mu_a b} \delta^{-x + \frac{\mu_a}{2\omega_a}} \ .
\end{equation}
For the dependence on $\delta$ to cancel, we have to take 
$x=\mu_a /(2 \omega_a)$. Note that the assumption $x>1/2$ is a posteriori
confirmed, as $\mu_a > \omega_a$. 

\begin{figure}
\includegraphics[width=9cm]{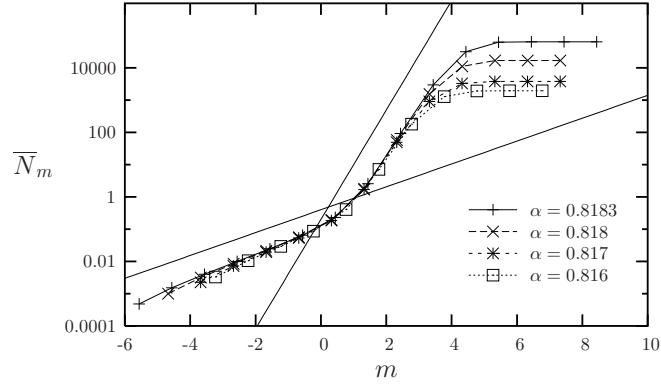} 
\caption{The intermediate regime scaling function for the study of
the size of minimal barrier rearrangements. The two solid straight lines have
the slope of the asymptotic behaviours, $\overline{N}_m \sim e^{\mu_{a/b}m}$
when $m \to \pm \infty$.}
\label{fig_sizebarr_int}
\end{figure}

This analysis of the intermediate regime is confirmed by 
Fig.~\ref{fig_sizebarr_int}. Here we plot
$\delta^x N_b$, versus $m=b -\log(1/\delta)/(2\omega_a)$, with $N_b$
computed numerically for several values of $\delta$, and $x$ determined 
analytically as explained above.

Finally, we look for the asymptotic behavior of $N_b$ when $b\sim b_0'(\delta)$
the typical height of large minimal barriers  
(recall that $b'_0(\delta) \simeq \Upsilon \log(1/\delta)$).
From the behaviour of $\overline{N}_m$ when $m \to + \infty$, we get
\begin{equation}
N_b \sim \delta^{-x} e^{\mu_b (b'_0-b_0)} \sim \delta^{-\nu_{\rm barr}} \ ,
\quad  \nu_{\rm barr} = \frac{\mu_a}{2 \omega_a} +\frac{\mu_b}{2 \omega_b} \ .
\end{equation}
As $q_b$ is of order 1 in this regime, the exponent $\nu_{\rm barr}$ is
indeed the one which controls the divergence of $M_b/q_b$,
the average size of large scale minimal barrier rearrangements.
%
%
\subsection{Average distance in minimal sizes rearrangements}
\label{sec_app_depth}

We study in this Appendix the asymptotic behaviour of the solution
$T_n$  of Eqs.~(\ref{eq_depth}). This allows to compute the average 
radius of minimal size rearrangements. Let us denote $g(x)$ and $\tilde{g}(x)$
the g.f.'s associated repectively to $T_n$ and $\Tt_n$. 
The first of equations (\ref{eq_depth}) can be rewritten as
\begin{equation}
g(x) = p\g\,  f(x) \tilde{g}(x) \ ,
\label{eq_depth_gf}
\end{equation}
where $f$ is the g.f. of the $q_n$'s.

Consider first the solution at $\g=\g_{\rm d}$. From the results of
Sec.~\ref{sec_app_minsize} we have the following asymptotic behaviors as
$n\to\infty$ ($s\to 0$):
\begin{eqnarray}
f(1-s) & = & 1-\phi_{\rm d} - A \Gamma(1-a) s^a +o(s^a)
\ , \label{eq_depth_f} \\
\frac{\qh_n}{q_n} & = & (p-1) \phi_{\rm d}^{p-2} + A (p-1)(p-2)
\phi_{\rm d}^{p-3} n^{-a} +o(n^{-a})\ ,\label{eq:Qrapp}\\
n\qh_n & = & A a (p-1) \phi_{\rm d}^{p-2} n^{-a} +o(n^{-a})\ .\label{eq:Qn}
\end{eqnarray}
Motivated by the numerical solution of Eqs.~(\ref{eq_depth}), we assume that, 
when $\g=\g_{\rm d}$, $T_n$ admits a finite limit $T_{\infty}$
as $n\to\infty$. This implies $g(1-s) = T_\infty s^{-1}+o(s^{-1})$
as $s\to 0$. Using the second of equations (\ref{eq_depth}), together with the 
formulae (\ref{eq:Qrapp}), (\ref{eq:Qn}), we get
\begin{eqnarray}
\gt(1-s) = (p-1) \phi_{\rm d}^{p-2} \, g(1-s)
+ A \Gamma(1-a) (p-1) \phi_{\rm d}^{p-3} ((p-2) T_\infty + a \phi_{\rm d})
s^{a-1}  +o(s^{-1+a})\ .
\end{eqnarray}
Plugging this Ansatz in Eq.~(\ref{eq_depth_gf}) along with the expansion
(\ref{eq_depth_f}), one realizes that terms of the type
${\rm cst} \cdot g(1-s)$ (with ${\rm cst}$ an $s$-independent constant)
cancel  because of Eq.~(\ref{eq_orderparam_tangent}).
The leading surviving terms are of order $s^{-1+a}$. Imposing them to vanish,
we get
\begin{equation}
T_\infty = \frac{a (1-\phi_{\rm d})}{1-\lambda} \ .
\end{equation}

Let us now turn to the intermediate regime, on the scale 
$n_0(\delta) = \delta^{-1/2a}$, and 
define the scaling variable $t=n/n_0(\delta)$. We  found in 
Sec.~\ref{app:IntermediateSizes} that
\begin{eqnarray}
f(x=1-y/n_0) & =& 1-\phi_{\rm d} - \delta^{1/2} \Lap[\epsilon](y) +
o(\delta^{1/2})\ ,\\
\frac{\qh_n}{q_n} & = & (p-1) \phi_{\rm d}^{p-2} + (p-1)(p-2)
\phi_{\rm d}^{p-3} \delta^{1/2} \epsilon(t) + o(\delta^{1/2})\ , \\
n\qh_n &=& - (p-1) \phi_{\rm d}^{p-2} \delta^{1/2} \, t \epsilon'(t)
+ o(\delta^{1/2}) \ ,
\end{eqnarray}
where $\epsilon(t)$ is the scaling function of $Q_n$ around the plateau,
introduced in Eq.~(\ref{eq_def_epsilon_t}).

Let us make the following Ansatz on this scale:
\begin{equation}
T_n \simeq F(t=n/n_0) \ , \;\;\;\;\;\;\;\;\; \Tt_n \simeq \Ft(t=n/n_0)\, .
\end{equation}
In this regime, the second of equations (\ref{eq_depth}) reads
\begin{equation}
\Tt_n = (p-1) \phi_{\rm d}^{p-2} T_n 
+ \delta^{1/2} (p-1) \phi_{\rm d}^{p-3} ((p-2) F(t) \epsilon(t) - 
\phi_{\rm d} t \epsilon'(t)) +o(\delta^{1/2})\ .
\end{equation}
Next we insert this formula in Eq.~(\ref{eq_depth_gf}) and expand
in $\delta$. Leading order terms are of the 
form ${\rm cst}\cdot T_n$ (of order $\delta^0$), and cancel as usual. 
The scaling function $F(t)$ is is determined by the vanishing of
$O(\delta^{1/2})$ terms, which leads to 
\begin{equation}
y \Lap[\epsilon] (y)\,\Lap[F](y)
- \lambda \Lap[\epsilon F](y)
= -(1-\phi_{\rm d}) \Lap[t \epsilon'](y) \, .
\end{equation}
Studying the limit $y \to \infty$ of this equation, one can check that
$F(t)$ reaches for small $t$ the value $T_\infty$ determined above, 
fulfilling the consistency condition between the end of the finite $n$ regime 
and the beginning of the ``intermediate'' one. Consider now the limit
$t \to \infty$, and assume $F(t) \sim t^x$, with
$x$ a yet unknown positive exponent. Recalling that $\epsilon(t) \sim t^b$ 
as  $t\to\infty$, the above equation yields:
\begin{equation}
\frac{\Gamma(1+x) \Gamma(1+b)}{\Gamma(1+x+b)} = \lambda \ .
\label{eq_depth_x}
\end{equation}
It follows from the convexity of $\log\Gamma(1+x)$ for $x\ge 0$ that
this equation has a unique positive solution, which, because of
Eq.~(\ref{eq_MCT1_b}), is $x=b$.

Finally, let us turn to the large rearrangements regime, $n = t n'_0(\delta)$, 
with $n'_0(\delta) \sim \delta^{- \nu}$. 
We define scaling functions on this last
scale by
\begin{equation}
T_n \simeq \delta^{- \zeta} G(t=n/n'_0) \ , \;\;\;\;\;\;\;\;\; 
\Tt_n \simeq \delta^{- \zeta} \Gt(t=n/n'_0) \ .
\end{equation}
We defined in App.~\ref{sec_app_minsize_slow} the scaling function
$Q(t)$ for the minimal size distribution $Q_n$ in this
regime. From this definition, we have
\begin{eqnarray}
f(x=1-y/n'_0) &\simeq& \psi(y) =1- y \int_0^\infty  Q(t) e^{-y t} \, \de t\ ,\\
\frac{\qh_n}{q_n} &\simeq& (p-1) Q(t)^{p-2} \ , \\
n\hat{q}_n &\simeq& - (p-1) t Q'(t) Q(t)^{p-2} \ .
\end{eqnarray}
These last two formulae allow to relate $G(t)$ and $\Gt(t)$,
using the second of equations (\ref{eq_depth}). Assuming $\zeta>0$
(which is consistent with the outcome of numerical computations), we
get
\begin{equation}
\Gt(t) = (p-1) Q(t)^{p-2} G(t)  \ .
\end{equation}
Injecting this expression in Eq.~(\ref{eq_depth_gf}),
we finally  obtain an equation determining the scaling function $G(t)$,
\begin{equation}
\Lap[G](y) = p (p-1)  \g_{\rm d}\psi(y)
\Lap[Q^{p-2} G](y) \ .
\end{equation}
Using Eq.~(\ref{eq:FinalScaling}), it is not hard to show that 
$G(t) = -\theta_* \,t\, Q'(t)$ is, for any $\theta_*>0$, a solution of 
this equation.
This solution yields $G(t) \sim t^b$ for $t \to 0$, consistently
with the above study of the intermediate regime.

A matching argument allows to fix the exponent $\zeta$.
Consider indeed $n =t n_0(\delta)$, with $t \gg 1$ but 
independent of $\delta$. From the limit $t \to \infty$ of the intermediate
regime we have $T_n \sim t^b$.
Consider instead the limit $t \to 0$ of the final regime, we get
\begin{equation}
T_n \sim \delta^{-\zeta}\, \left( tn_0(\delta)/ n_0'(\delta) 
\right)^b 
\sim t^b\, \delta^{-\zeta + \frac{1}{2}} \ .
\end{equation}
Matching the two asymptotics, we obtain the universal exponent $\zeta =1/2$. 

Let us finally compute the  
quantity $\theta_n = T_n /(n q_n)$, i.e. the average radius of a 
rearrangement of size $n$. Using the above solution we find,
on the scale of large minimal size rearrangements
$n\sim n_0'(\delta)$:
\begin{equation}
\theta_n \simeq \delta^{-\zeta} \frac{G(t)}{-t Q'(t)}  = \theta_*\, 
\delta^{-\zeta}\ .
\end{equation}
%
%
\subsection{Geometrical susceptibility}

In this Appendix we work out the critical behavior of the 
geometrical susceptibility defined in Sec.~\ref{sec:GeometricalSusceptibility}.
%
%
\subsubsection{Minimal size rearrangements}
\label{sec_app_geomsuc_sizes}

The susceptibility for minimal size rearrangements is determined
by solving Eqs.~(\ref{eq:FinalSuscSize}). It is helpful to rewrite the 
first of these equations as
\begin{equation}
S_n = 1-Q_n+(1-Q_n)\Sh_n+\sum_{m=1}^{n-1} q_m(\Sh_{n-m}-\Sh_n)\, .
\label{eq:SubtractedSusc}
\end{equation}
As usual, we start by considering the large $n$ behavior at $\g=\g_{\rm d}$.
Numerical computations suggest the Ansatz $S_n\simeq S_{*} n^w$,
with $w>0$ an exponent to be determined. From the second of 
Eqs.~(\ref{eq:FinalSuscSize}), and recalling
that $Q_n=\phi_{\rm d}+An^{-a}+o(n^{-a})$, we get
\begin{equation}
\Sh_n = p(p-1)\g_{\rm d}\phi_{\rm d}^{p-2}S_n 
+p(p-1)(p-2)\g_{\rm d}\phi_{\rm d}^{p-3}AS_{*}n^{w-a}+o(n^{w-a})\, .
\end{equation}
This implies in particular 
$\Sh_n \simeq p(p-1)\g_{\rm d}\phi_{\rm d}^{p-2}S_*n^w$. 
This expression can be substituted in Eq.~(\ref{eq:SubtractedSusc}).
The leading terms are of the type ${\rm cst}\, S_n$ (and of order 
$n^w$) and cancel because of Eq.~(\ref{eq_orderparam_tangent}).
We get an equation of the form
\begin{equation}
0 = 1-\phi_{\rm d}+AS_*C\, n^{w-a}+o(1,n^{w-a})\, ,\label{eq:FastSusc}
\end{equation}
where $C$ is $n$-independent and given by
\begin{equation}
C= p(p-1)\g_{\rm d} \phi_{\rm d}^{p-2}\left\{\lambda-
\frac{\Gamma(1-a)\Gamma(1+w)}{\Gamma(1-a+w)}\right\}\, ,\label{eq:ConstSusc}
\end{equation}
and $\lambda$ is the constant already introduced in Sec.~\ref{sec:AsymSizes}.
Notice that Eq.~(\ref{eq:FastSusc}) cannot be satisfied for 
$w<a$ (in this case, it would imply $0=1-\phi_{\rm d}$), nor
for $w>a$ (it can be shown\footnote{Consider the 
expression in curly brackets in Eq.~(\ref{eq:ConstSusc})
(the prefactor is positive). Proving that it is negative is equivalent to
proving that $F_a(1+w-a)>F_a(1-2a)$, where
$F_a(x) =\log\Gamma(x+a)-\log\Gamma(x)$. This in turns follows from the
fact that $\log\Gamma(x)$ is strictly convex for $x>0$.} 
that $C<0$ for $w>0$). Therefore we have necessarily $w=a$, and 
Eq.~(\ref{eq:FastSusc}) determines the prefactor $S_*$.
\begin{figure}
\includegraphics[width=8cm]{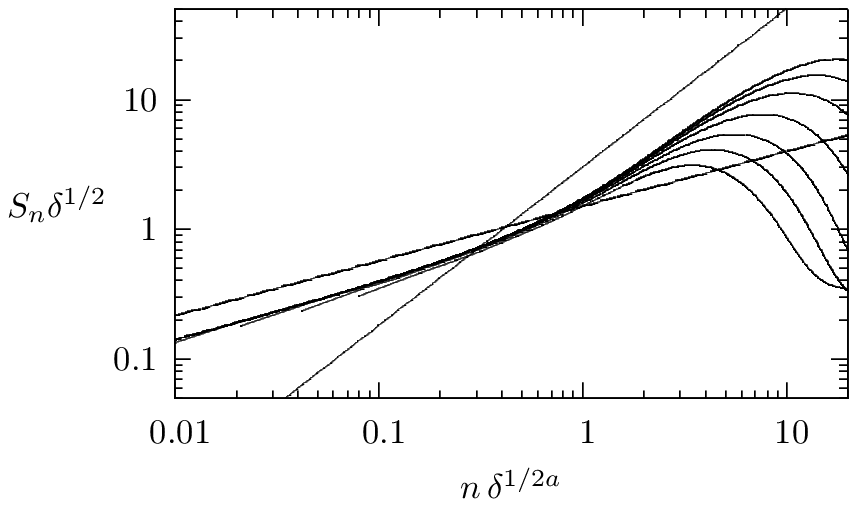} 
\includegraphics[width=8cm]{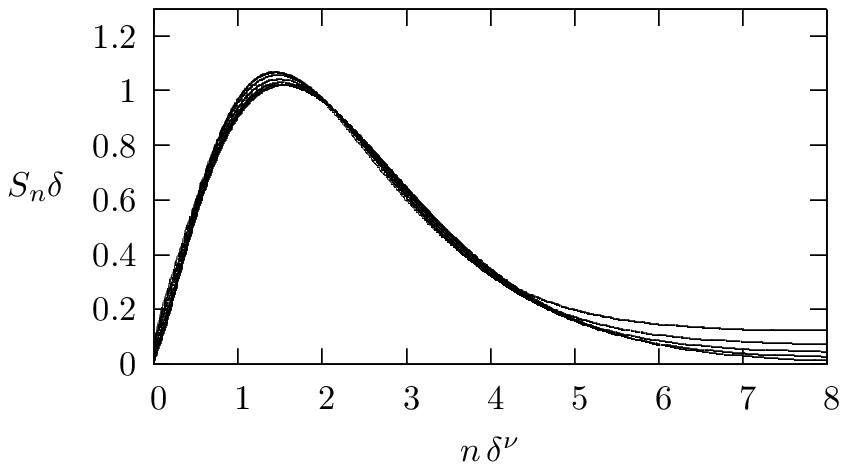}
\caption{Critical scaling of the geometrical susceptibility for
minimal size earrangements. Different curves refer to $p=3$ and 
$\g = 0.7$, $0.75$, $0.78$, $0.8$, $0.81$, $0.814$, $0.816$. 
Left: intermediate regime. The straigth lines correspond to the 
asymptotic behaviors $t^a$ (as $t\to 0$) and $t^b$ (as $t\to\infty$).
Right: large scale rearrangements regime.}
\label{fig:ChiSize}
\end{figure}

Next we consider  the intermediate scale $n_0(\delta)=\delta^{-1/2a}$.
Since $S_n\sim n^{a}$ at $\g = \g_{\rm d}$, we expect $S_n\sim \delta^{-1/2}$
for $\g<\g_{\rm d}$ and $n\sim n_0(\delta)$.
We assume the scaling behaviour
\begin{equation}
S_n\simeq n^{-x}\, R(t=n/n_0(\delta))\, ,
\end{equation}
with an analogous Ansatz for $\Sh_n$ (with scaling function $\Rh(t)$),
and $x>0$ an exponent yet to be determined.
Equations (\ref{eq:FinalSuscSize}) yield
(cf. also the form (\ref{eq:SubtractedSusc}))
\begin{eqnarray}
\Rh(t) & = & p(p-1)\g_{\rm d}\phi_{\rm d}^{p-2}R(t)+
p(p-1)(p-2)\g_{\rm d}\phi_{\rm d}^{p-3}\epsilon(t)R(t)\,\delta^{1/2}+
o(\delta^{1/2})\, ,\\
R(t) & = & (1-\phi_{\rm d})\delta^x+(1-\phi_{\rm d})\Rh(t)
-\epsilon(t) \Rh(t)\,\delta^{1/2} -\delta^{1/2}
\int_0^t\epsilon'(s)[\Rh(t-s)-\Rh(t)]\,\de s + o(\delta^{1/2})\, .
\end{eqnarray}
Eliminating $\Rh(t)$, we obtain an equation determining the scaling
function $R(t)$. For $x=1/2$ we get:
\begin{equation}
(1-\phi_{\rm d})^2 + (\lambda-1)\epsilon(t) R(t)-
\int_0^t\epsilon'(s)[R(t-s)-R(t)]\,\de s = 0\, .\label{eq:IntermediateSuscSize}
\end{equation}
Other values of $x$ can be excluded via an analysis of the 
resulting equations: we shall stick to $x=1/2$ hereafter.
The asymptotic behavior of the solution of Eq.~(\ref{eq:IntermediateSuscSize})
can be determined by recalling that $\epsilon(t)\sim t^{-a}$ as $t\to 0$,
and  $\epsilon(t)\sim -t^{b}$ as $t\to \infty$. We get
$R(t)\sim t^a$ for $t\to 0$ and $R(t)\sim t^b$ for $t\to\infty$.
The choice $x=1/2$ is confirmed by matching the $t\to 0$ 
behavior with the analysis of the finite $n$ regime, explained above.

In Fig.~\ref{fig:ChiSize}, left frame, we plot $\delta^{1/2}S_n$
versus $\delta^{1/2a}n$, with $S_n$ determined via a numerical solution of
Eqs.~(\ref{eq:FinalSuscSize}). The good collapse confirms the validity 
of the above analysis.

Let us finally consider the scale of large rearrangements
$n_0'(\delta)= \delta^{-\nu}$, and assume in this regime
\begin{eqnarray}
S_n \simeq \delta^{-\eta}\, S(t=n/n_0(\delta))\, ,
\end{eqnarray}
with an analogous Ansatz for $\Sh_n$, and $\eta>0$ a new critical  exponent. 
Injecting this form in Eqs.~(\ref{eq:FinalSuscSize}), and
recalling that $Q_n\simeq Q( t=n/n_0(\delta))$ on the same scale, 
we get:
\begin{eqnarray}
S(t) &=& (1-\phi_{\rm d})\Sh(t)-\int_0^tQ'(s)\Sh(t-s)\, \de s\, ,\\
\Sh(t) & = &  p(p-1)\g_{\rm d} Q(t)^{p-2}S(t)\, .
\end{eqnarray}
Studying these equations in the $t\to 0$ limit allows to determine
the asymptotic behaviour $S(t) \sim t^b$ as $t\to 0$. This allows 
to determine the exponent $\eta$ through a matching procedure. 
If we let $n=t\, n_0(\delta)$, with $t$  large but $\delta$--independent,
from our study, respectively, of the intermediate and final regimes,
we get
\begin{eqnarray}
S_n\sim \delta^{-1/2}\, t^{b} \sim \delta^{-\eta} \left(t\, n_0(\delta)/
n_0'(\delta)\right)^b\, .
\end{eqnarray}
This in turn implies $\eta=1$.

In Fig.~\ref{fig:ChiSize}, right frame, we check the scaling hypothesis
in the final regime (as well as the exponent $\eta=1$) against
a numerical solution of Eqs.~(\ref{eq:FinalSuscSize}).
%
%
\subsubsection{Minimal barrier rearrangements}
\label{sec_app_geomsuc_barriers}

This part of the Appendix is devoted to the resolution of 
Eqs.~(\ref{eq_suscbarr1},\ref{eq_suscbarr2}) for the geometric susceptibility
of minimal barrier rearrangements. Using the fact that $\sum_b \sh_b =0$,
Eq.~(\ref{eq_suscbarr1}) can be rewritten as
\begin{equation}
S_b = 1 - Q_b - \sum_{l=0}^{2b-4} p_l S_b^{(l+1)} \ , \quad
S_b^{(l)} = \sum_{b_1,\dots,b_l}\sh_{b_1} \qh_{b_2} \dots  
\qh_{b_l} \ind(\hat{f}_l(b_1,\dots,b_l)<b ) \ .
\end{equation}
The quantity $S_b^{(l)}$ is very similar to $N_b^{(l)}$ defined in
Eq.~(\ref{eq_Nb}),
with the sequence $\sh_b$ replacing $\Mh_b$. In particular $S_b^{(l)}$
admits a representation of the form (\ref{eq:ExplicitSizeBarr}),
where $\Nh_b$ is replaced by $-\Sh_b$. The numerical computation
of $S_b$  proceeds along the lines exposed in Sec.~\ref{sec_app_sizeminbarr}. 

The expansion of  
Eqs.~(\ref{eq_suscbarr1}), (\ref{eq_suscbarr2}) in powers of 
$\epsilon_b = Q_b -\phi_{\rm d}$ is also analogous to the one of
 Sec.~\ref{sec_app_sizeminbarr}. We get
\begin{eqnarray}
S_b &=& 1 - \phi_{\rm d} - \epsilon_b + e^{-\g p \phi_d^{p-1}} \Sh_b
+ \g p (p-1) \phi_{\rm d}^{p-2} e^{-\g p \phi_d^{p-1}} 
[\Sh_{b-1} (\epsilon_{b-1}-\epsilon_b) - \Sh_b \epsilon_{b-1}]
+ O(\epsilon^2S) \ , \label{eq:SuscBarrAs1}\\
\Sh_b &=& \g p (p-1) \phi_{\rm d}^{p-2} S_b 
+ \g p (p-1) (p-2) \phi_{\rm d}^{p-3} \epsilon_b S_b +O(\epsilon^2S)\ .\
\label{eq:SuscBarrAs2}
\end{eqnarray}
Next we eliminate $\Sh_b$ from the above equations and
assume the asymptotic behavior $S_b\simeq S_* e^{wb}$, with $w>0$
to be determined. Recalling that $\epsilon_{b}\simeq A\, e^{-\omega_ab}$,
we get
\begin{eqnarray}
0 = (1-\phi_{\rm d}) + C S_*\, e^{(w-\omega_a)b} +o(1,e^{(w-\omega_a)b})\, ,
\label{eq:SuscBarrFast}
\end{eqnarray}
where
\begin{eqnarray}
C = p(p-1)\g_{\rm d}\phi_{\rm d}^{p-2}\left\{
\lambda-[e^{\omega_a}- (e^{\omega_a}-1)e^{-w}]\right\}\, ,
\end{eqnarray}
and $\lambda$ is defined in Sec.~\ref{sec:AsymSizes}. Using
an argument similar to the one in the previous Section, one can show that
Eq.~(\ref{eq:SuscBarrFast}) implies $w=\omega_a$.

\begin{figure}
\includegraphics[width=8cm]{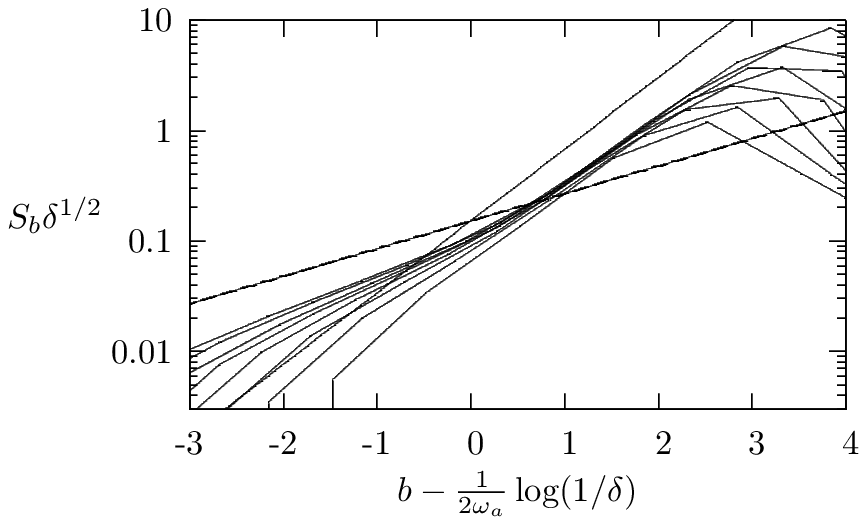} 
\includegraphics[width=8cm]{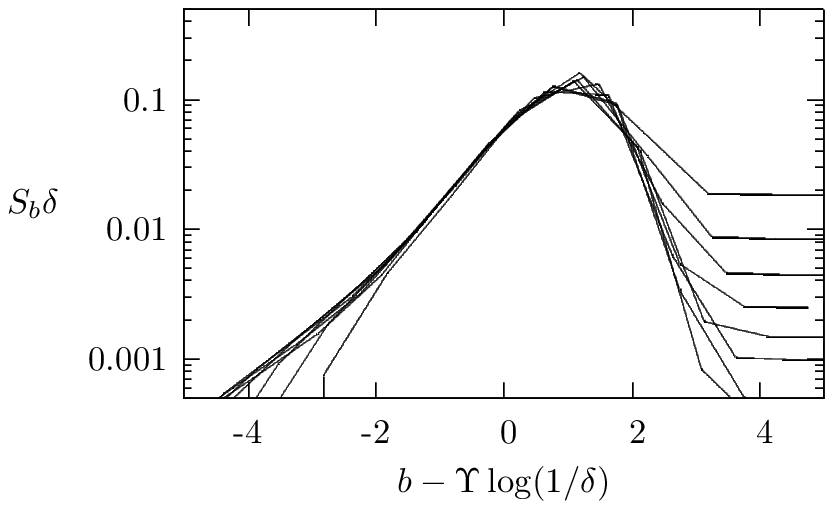}
\caption{Scaling of the geometrical susceptibility for
minimal barrier rearrangements. Different curves refer to $p=3$ and
$\g = 0.8$, $0.81$, $0.814$, $0.816$, $0.817$, $0.8175$, $0.818$, $0.8182$. 
Left: intermediate regime. The straigth lines correspond to the 
asymptotic behaviors $e^{\omega_am}$ (as $m\to -\infty$) and 
$e^{\omega_bm}$ (as $m\to\infty$).
Right: large scale rearrangements regime.}
\label{fig:ChiBarr}
\end{figure}
Let us consider now the intermediate regime $b=m+b_0(\delta)$, where 
$b_0(\delta)=\log(1/ \delta)/(2\omega_a)$ and 
$\epsilon_b \sim \delta^{1/2} \overline{\epsilon}_m$. Inserting the Ansatz 
$S_b = \delta^{-x} \overline{S}_m$ in Eqs.~(\ref{eq:SuscBarrAs1}), 
(\ref{eq:SuscBarrAs2}), dominant terms (of order $\delta^{-x}$) cancel.
For next-to-leading terms to cancel, one has to choose $x=1/2$. The
scaling function $\overline{S}_m$ is then found to obey
\begin{equation}
\lambda \overline{S}_m \overline{\epsilon}_m + 
\overline{S}_{m-1} (\overline{\epsilon}_{m-1} - \overline{\epsilon}_m)
- \overline{S}_m \overline{\epsilon}_{m-1} + (1-\phi_{\rm d})^2 = 0 \ .
\end{equation}
The asymptotic behaviour of $\overline{S}_m$  can be deduced from 
this equation, using the results on $\bar{\epsilon}_m$,
cf.~Sec.~\ref{sec_app_minb_ana}. 
One finds $\overline{S}_m \sim e^{\omega_a m}$ as $m\to -\infty$ 
(thus matching the behavior in the first regime) and 
$\overline{S}_m \sim e^{\omega_b m}$ as $m\to \infty$. 

In Fig.~\ref{fig:ChiBarr}, left frame, we compare 
the postulated scaling in the 
intermediate regime with a numerical solution of 
Eqs.~(\ref{eq_suscbarr1}), (\ref{eq_suscbarr2}).

Finally, on the large rearrangement scale 
$b_0'(\delta)\sim \Upsilon\log(1/\delta)$, we expect $S_b\sim\delta^{-\eta}$.
The exponent $\eta$ can be determined  by matching this behavior
with the large $m$ asymptotics in the intermediate scale:
\begin{equation}
S_{b'_0} \sim \delta^{-1/2} \overline{S}_{m=b'_0-b_0}
\sim \delta^{-1/2} \exp\left\{\omega_b \frac{1}{2\omega_b}\log(1/ \delta)
\right\} \ .
\end{equation}
We thus have $\eta=1$ as claimed in Sec.~\ref{sec_geomsuc_barr}.

%
%
\section{Low temperature behavior of the dynamical line}
\label{app:LowTDline}

In Sec.~\ref{ScalingSection}, we claimed that the dynamical transition
line behaves as described in Eq.~(\ref{eq:LowTDynamical}) 
in the low temperature limit. A proof of this statement 
requires a particularly involved asymptotic calculation, and goes beyond
the scope of this paper. On the other hand, a simple argument can 
be constructed as follows. 

Consider $\g>\g_{\rm d}$ and expand the solution
of the 1RSB equations (\ref{eq:1RSB1}), (\ref{eq:1RSB2}) as $T\to 0$,
with $\g$ and $m<1$ fixed. At $T=0$, the order parameters
$\Qc(\rho)$, $\Qch(\rh)$ concentrate over local field distributions
$\rho(h)$, $\rh(u)$ supported  on integer fields 
$h, u\in {\mathbb Z}$. More precisely, 
$\rh(u) = \frac{1}{2}\delta(u-1)+\frac{1}{2}\delta(u+1)$ with probability
$\phh$ (corresponding to hyperedges in the backbone), and 
$\rh(u) = \delta(u)$ with probability $1-\phh$ (hyperedges
outside the backbone).  Analogous expressions hold for $\Qc(\rho)$.

A small non-zero temperature has two effects on $\Qch(\rh)$. First of all,
any local field distribution $\rh(u)$ in the support of  $\Qch$
aquires a peak for $u\approx 0$. Second, the peaks on
integer fields acquire a small width. In general, this width is of order
$T$ and the last effect dominates. For XORSAT,
the width is of order $e^{-2\beta}$ and can be neglected
to first order (this can be checked through usual population dynamics
algorithms and yields indeed a consistent low $T$ expansion).
Therefore, the leading correction is captured by  
the form $\rh(u) \approx \frac{1}{2}(1-\rh_0)\delta(u-1)+ \rh_0\delta(u)+
 \frac{1}{2}(1-\rh_0)\delta(u+1)$. The distribution $\Qch(\rh)$ 
is concentrated over $\rh(u)$ of this form and is completely determined
as 
\begin{eqnarray}
\rh_0=\left\{
\begin{array}{ll}
n2^me^{-2\beta m}&\mbox{with probability $\Qh_n$,}\\
1 & \mbox{with probability $1-\phh$.}
\end{array}
\right.
\end{eqnarray}
The distribution $\Qh_n$ has  normalization $\sum_n \Qh_n = \phh$ , 
and satisfies the equations
\begin{eqnarray}
\Qh_n &= &\sum_{n_1\dots n_{p-1}}Q_{n_1}\cdots Q_{n_{p-1}}\,
\delta_{n-n_1-\cdots-n_{p-1}}\, ,\label{eq:LowT1}\\
Q_n & = & [1-\bar{p}_0-\bar{p}_1-\bar{p}_2]\,\delta_{n,0}+
\bar{p}_2\,\delta_{n,1}+\bar{p}_1\,\Qh_n\, .\label{eq:LowT2}
\end{eqnarray}
where $\bar{p}_l = (p\g\phh)^le^{-p\g\phh}/l!$, and $Q_n$ parametrizes
the functional distribution $\Qc(\rho)$. At any fixed 
$\g>\g_{\rm d}$, the leading correction to $\rho(h)$, $\rh(u)$ is of order
$e^{-2\beta m}$. Since $m<1$, it is consistent to neglect the 
$O(e^{-2\beta})$ width of the peaks on integer $u$'s.

By studying Eqs.~(\ref{eq:LowT1}), (\ref{eq:LowT2})
as $\g\downarrow \g_{\rm d}$, one can show that the typical scale of
$n$ diverges as $(\g-\g_{\rm d})^{-1}$. The above low-$T$ 
expansion breaks down when $\rh_0\sim (\g-\g_{\rm d})^{-1}
e^{-2\beta m} = O(1)$, i.e. for $\g = \g_{\rm d} + O(e^{-2\beta m})$.
The DPT corresponds to a singularity in the solution
of Eqs.~(\ref{eq:1RSB1}), (\ref{eq:1RSB2}) at $m=1-$. If we identify this
singularity with the low-$T$ expansion breakdown, we get the
scaling (\ref{eq:LowTDynamical}).

\begin{figure}
\begin{center}
\includegraphics[width=7cm]{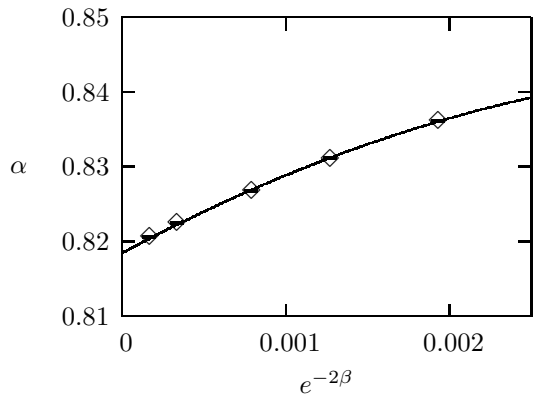} 
\end{center}
\caption{Low temperature behavior of the dynamical transition line.
Symbols (with error bars) correspond to a numerical determination of
$\g_{\rm d}(\beta)$ along the lines of App.~\ref{app:PointToSet}.
The line is a fit to the form $\g_{\rm d}(\beta) = \g_{\rm d}+
x_{\rm d}e^{-2\beta}+y_{\rm d}e^{-4\beta}$.}
\label{dline}
\end{figure}
The above argument does not fix the coefficient $x_{\rm d}$.
In order to determine it numerically (for $p=3$), 
we computed $\g_{\rm d}(\beta)$
numerically for several values of $\beta$, and fitted the results to the
form  (\ref{eq:LowTDynamical}). At each value of $\beta$,
$\g_{\rm d}$ was determined by computing the length $\ell_*(\ve)$
defined in Sec.~\ref{sec:PointToSet} using the method of
App.~\ref{app:PointToSet} and fitting the result using the form
$\ell_*(\ve) \approx A(\g_{\rm d}(\beta)-\g)^{-1/2}$. 
The resulting critical points are reported in Fig.~\ref{dline}.

Next, the values for $T\le 0.32$ were fitted using the form
$\g_{\rm d}(\beta) = \g_{\rm d}+x_{\rm d}e^{-2\beta}+y_{\rm d}e^{-4\beta}$.
This gives a very good description of the data if
$x_{\rm d} = 11.76$, and $y_{\rm d} = 1378$. 
The final result (with a somehow subjective
estimate of the error) is
\begin{eqnarray}
x_{\rm d} = 11.7(5)\, .
\end{eqnarray}

We conclude by stressing that, in general mean-field models 
on sparse graphs, the low-temperature behavior of the dynamical 
transition line can be different from Eq.~(\ref{eq:LowTDynamical}), because 
of $O(T)$ cavity fields.
%
%

\end{document}